\documentclass[trackchanges,twocolumn]{aastex701}
\usepackage{makecell}
\usepackage{graphicx}  
\usepackage{amsmath}
\usepackage{float}
\usepackage{comment}
\usepackage{txfonts}
\newcommand{\ko}[1]{#1}

\usepackage[rightcaption]{sidecap}
\usepackage{tablefootnote}
\usepackage{natbib}
\usepackage{booktabs}
\usepackage{multirow}
\usepackage{comment}
\usepackage{graphicx}
\usepackage{txfonts}

\begin{document} 
\title{Radiatively Controlled Thermal Stability of High-Altitude Clouds in Exoplanetary Atmospheres}

\author{Taro M. Shefferson-Nagata}
\affiliation{Meikei Gakuen High School, 1-1 Inarimae, Tsukuba, Ibaraki, 305-8502, Japan}
\email[show]{tarofractal@gmail.com} 

\correspondingauthor{Kazumasa Ohno}
\author[orcid=0000-0003-3290-6758]{Kazumasa Ohno}
\affiliation{Division of Science, National Astronomical Observatory of Japan, 2-21-1 Osawa, Mitaka, Tokyo, 1818588, Japan}
\affiliation{Graduate Institute for Advanced Studies, SOKENDAI, 2-21-1 Osawa, Mitaka, Tokyo 181-8588, Japan}
\email[show]{ohno.k.ab.715@gmail.com}

\begin{abstract}
One of the most striking findings of exoplanetary science is the ubiquity of clouds. A conventional approach to infer the cloud compositions is to utilize the thermochemical equilibrium model assuming the same temperature shared with clouds and ambient gases; however, this assumption is actually not always valid, especially in the tenuous upper atmospheres where particle-gas collisions are infrequent. In this study, we investigate the radiative equilibrium temperatures of exoplanetary clouds to assess the thermal stability of aerosols in the low atmospheric pressure limit. For eight cloud-forming condensates (KCl, ZnS, Na$_2$S, MnS, SiO$_2$, Mg$_2$SiO$_4$, Fe, and Al$_2$O$_3$), we solve the energy balance between stellar radiative heating and infrared cooling to calculate the particle temperature, which is then compared with their condensation temperatures. We find three composition-dependent groups in the particle-temperature behavior, and show that silicate condensates (SiO$_2$ and Mg$_2$SiO$_4$) maintain a cool enough temperature in a wide range of stellar and planetary conditions. Conversely, sulfide condensates (ZnS, Na$_2$S, and MnS) are readily heated to their sublimation temperatures even on temperate planets. WASP-17b and HD 189733b fall within the thermodynamically stable regime for SiO$_2$ clouds, consistent with recent JWST observations. We also examine the vertical profiles of cloud temperatures on WASP-17b, showing that Fe clouds cannot exist on the dayside, and Al$_2$O$_3$ clouds can exist only in a confined region, even though atmospheric temperature appears to allow the formation of these clouds. This study provides novel insights on cloud compositions in upper exoplanetary atmospheres, testable by upcoming atmospheric surveys of JWST and Ariel.
\end{abstract}

\keywords{\uat{Exoplanet astronomy}{486} --- \uat{Exoplanet atmospheres}{487} ---  \uat{Atmospheric clouds}{2180} --- \uat{Hot Jupiters}{753} --- \uat{Hot Neptunes}{754} --- \uat{Mini Neptunes}{1063} }

\section{Introduction}

The characterization of exoplanetary atmospheres has advanced rapidly over the last decade.
One of the most striking findings is the ubiquity of clouds and hazes. Transmission spectra often show muted molecular features and/or continuum spectral slopes at optical wavelengths \citep[e.g.,][]{2014Natur.505...69K,2016Natur.529...59S}, both of which can be explained by the presence of aerosol layers that scatter and absorb starlight.
Clouds and hazes are not merely observational complications; they directly shape atmospheric temperature structures by changing how stellar radiation is absorbed and scattered \citep[e.g.,][]{Heng+12,Morley+15,Lavvas2021,Ohno24}, thereby affecting atmospheric chemistry \citep{Molaverdikhani+20} and dynamics \citep[e.g.,][]{Steinrueck+23,Steinrueck+25}.
Understanding exoplanetary clouds and hazes is one of the central topics of exoplanetary atmospheric science \citep[][]{Gao+21}.

To assess the impacts of clouds on atmospheric observations, it is crucial to know what they are made of.
A conventional approach is to utilize the thermochemical equilibrium model to identify which condensates are thermodynamically stable for a given temperature and pressure \citep[e.g.,][]{Burrows&Sharp99,Visscher+06,Visscher+10,Mbarek&Kempton16}. 
In this approach, a condensate is postulated to be stable wherever the local gas temperature falls below the thermodynamic condensation temperature. 
This method predicts the presence of various cloud species, such as salt and sulfide condensates \citep[KCl, ZnS, Na$_2$S, MnS,][]{Visscher+06,Morley12}, silicate and iron condensates \citep[e.g., Mg$_2$SiO$_4$, MgSiO$_3$, Fe,][]{Visscher+10}, and Al- and Ti-bearing condensates \citep[e.g., Al$_2$O$_3$, CaTiO$_3$,][]{Wakeford+17}.
Many cloud microphysical models \citep[e.g.,][]{Powell+18,Ohno&Okuzumi18,Gao&Benneke18,Ormel&Min19,Gao+20,Lee+25,Kiefer+26} and global circulation models of cloudy atmospheres \citep[e.g.,][]{Parmentier+16,Roman&Rauscher19,Mehta+25} determine cloud compositions based on the condensation curve derived by the thermochemical model.

The conventional approach postulates that condensates and ambient gases establish thermal equilibrium to share the same temperature; however, this assumption is actually not always valid, especially in the tenuous upper atmosphere.
In the context of the Solar System, it has been suggested that the temperatures of cloud particles in low-pressure regions ($\lesssim {10}^{-4}~{\rm bar}$) diverge greatly from the ambient gas temperature on Earth \citep{Fiocco+75,Fiocco+76} and Mars \citep{Goldenson+08,Haberle+25}.
For hot Jupiters, \citet{Lavvas2021} introduced a radiative-convective model that accounts for the temperature decoupling between photochemical hazes and ambient gases. 
They showed that, in low-pressure regions of $\lesssim 10^{-3}$~bar, the temperature of haze particles can substantially diverge from that of the surrounding gas because the collision rate between gas molecules and haze particles is too low to ensure thermal equilibrium.
In the calculation of \citet{Lavvas2021} for the hot Jupiter HD 189733b, the temperature difference between haze particles and surrounding gases reaches $\sim100~{\rm K}$ for the soot composition and $\sim400~{\rm K}$ for the composition of the Titan-haze analog.
Their studies demonstrated that radiative processes can dominate the energy balance to control the aerosol temperature in low-pressure regions, which is relevant to the transmission spectroscopy.

The possible thermal decoupling raises an important question: Can cloud species of interest remain thermodynamically stable in the upper atmosphere?
In this radiative-equilibrium regime, aerosol temperature is controlled by the competition between stellar radiative heating and infrared cooling by thermal emission from aerosol particles.
Depending on the relative efficiency of heating and cooling, the aerosol temperature may rise to its sublimation temperature even if the ambient gas temperature is cool enough to condense the aerosols. 
A relevant topic was studied in the context of circumstellar disks: \citet{Jones22} investigated the thermal stability of nano-diamond particles in circumstellar disks by calculating the radiative-equilibrium temperature.
\citet{Jones22} showed that the temperature of nano-diamonds could rise to graphitization and sublimation temperature even at $\gtrsim 10~{\rm AU}$ around the Herbig Ae/Be stars with effective temperature of $T_{\rm eff}\sim 7500$--$10500~{\rm K}$.
The superheated temperature is attributed to the inefficient infrared cooling of diamonds, which underscores the critical role of material-dependent optical constants in controlling the particle temperature under radiative equilibrium.
In the context of exoplanetary atmospheres, no previous studies have investigated the thermal stability of aerosol particles under radiative equilibrium.


In this study, we investigate the radiative equilibrium temperature of exoplanetary clouds, which may dictate whether the cloud can stably exist in upper atmospheres.
The goal of this study is to provide a theoretical framework for assessing the thermal stability of aerosols in the limit of low atmospheric pressure, where the conventional assumption of gas-aerosol thermal equilibrium does not hold. 
The organization of this paper is as follows.
Section \ref{sec:method} introduces the methodology of this paper.
In Section \ref{sec:results}, we investigate how the radiative-equilibrium temperature of aerosol particles varies with particle compositions, sizes, and stellar spectral type.
We then explore under what stellar and planetary conditions the aerosol particles remain thermally stable under radiative equilibrium.
In Section \ref{sec:discussion}, we discuss implications of our results to exoplanetary cloud compositions and caveats of this study.
We also examine the vertical profiles of cloud temperatures on WASP-17b to discuss plausible cloud compositions in the planet's upper atmosphere.
In Section \ref{sec:summary}, we summarize the findings of this paper.

\section{Methods}\label{sec:method}

\subsection{Equilibrium Temperature of Clouds}
\begin{figure*}[t]
\centering
\includegraphics[width=0.48\textwidth]{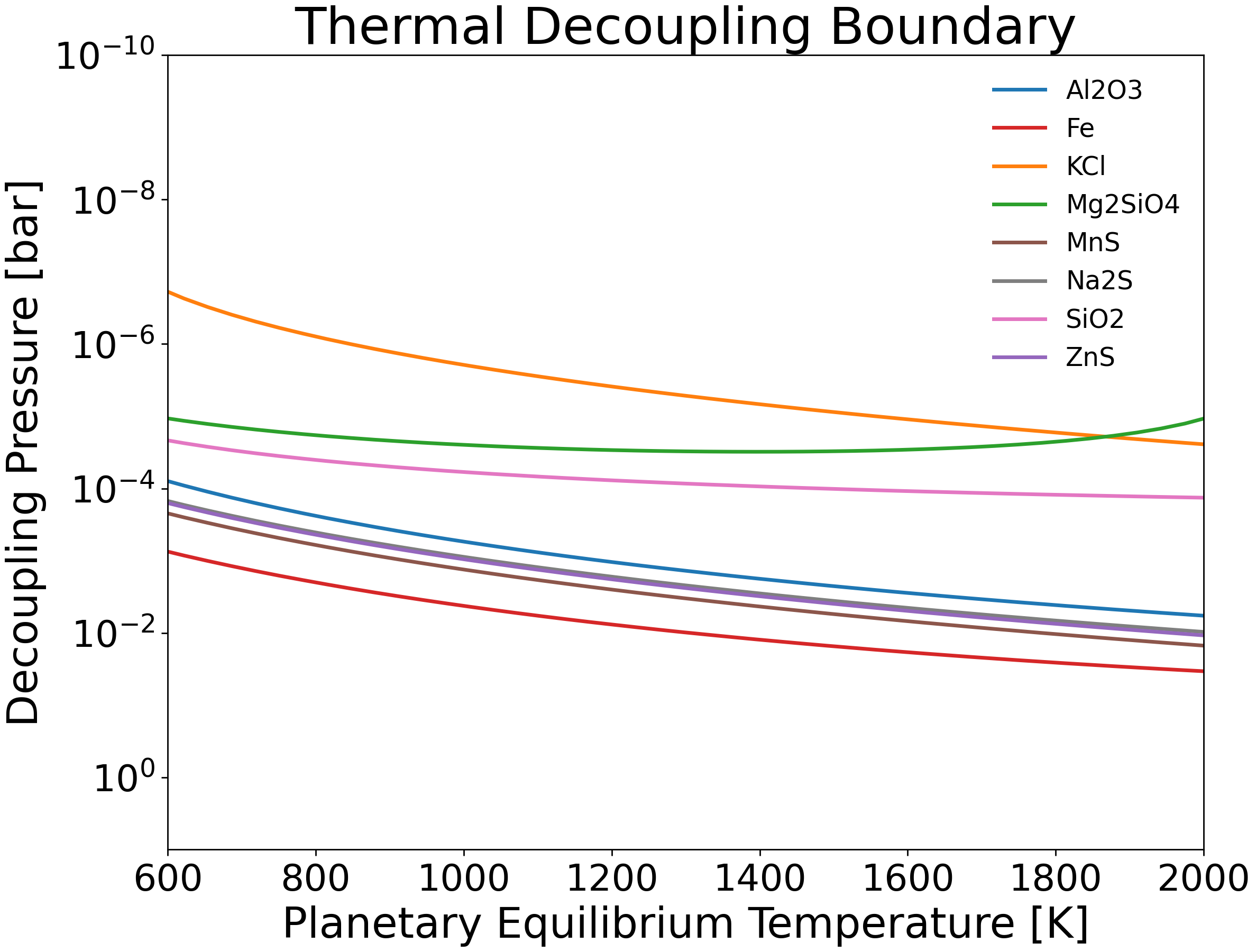}
\hfill
\includegraphics[width=0.512\textwidth]{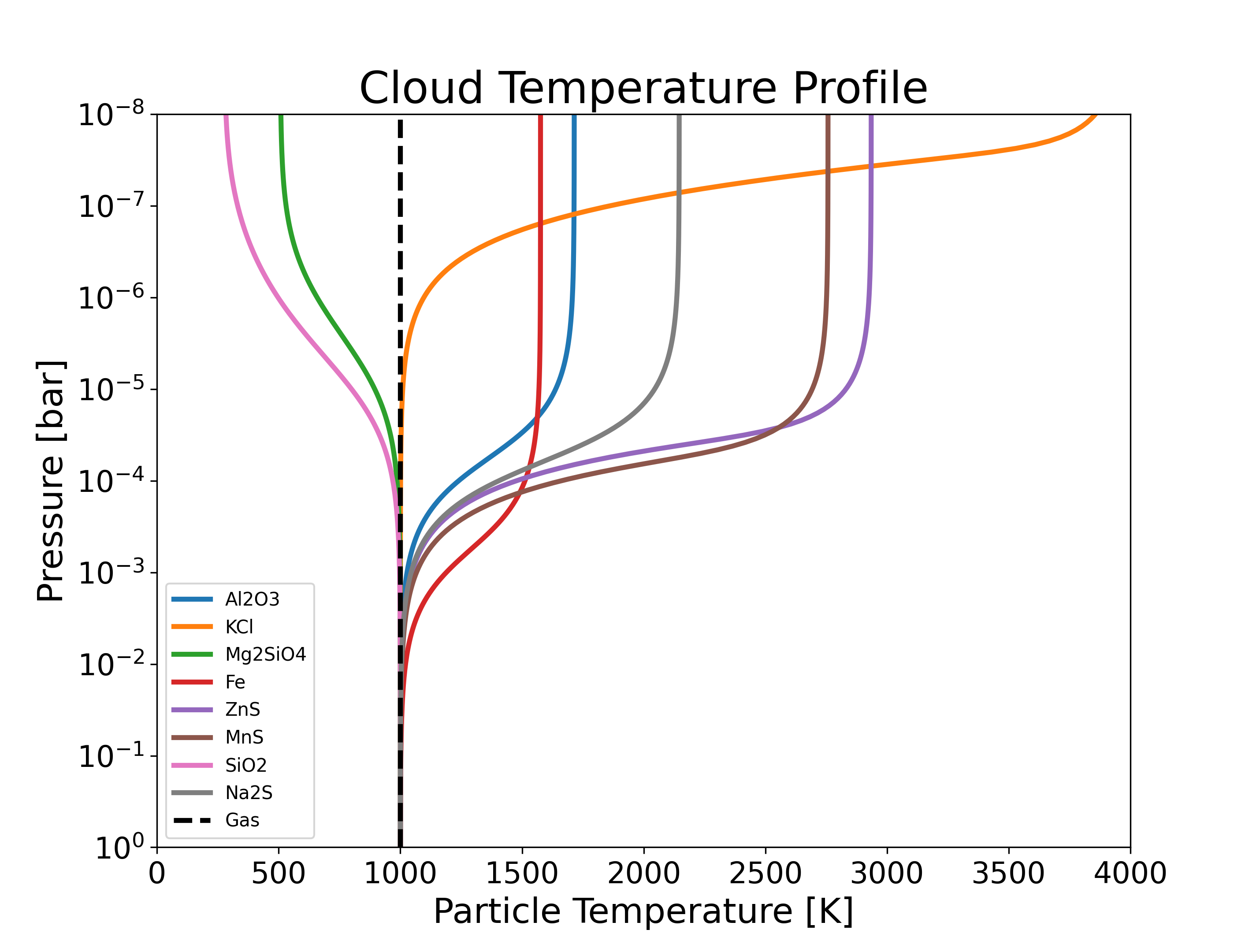}
\caption{
Onset of particle--gas thermal decoupling.
Left: Decoupling pressure $P_{\rm dec}$ as a function of planetary equilibrium temperature $T_{\rm eq}$ for the eight condensates considered in this study. Here, $P_{\rm dec}$ is defined by $|T_{\rm p}-T_{\rm gas}|/T_{\rm gas}=0.05$, and we assume an isothermal atmosphere of $T_{\rm gas}=T_{\rm eq}$, cloud particle radius of $0.1~{\rm {\mu}m}$ and $6000~{\rm K}$ black body radiation as an incident stellar spectrum.
Right: Pressure-dependent particle temperatures for a representative case with $T_{\rm eq}=1000$ K.}
\label{fig:decoupling}
\end{figure*}
In general, steady-state particle temperature obeys the energy conservation equation given by

\begin{align}\label{eq:Energy_balance2}
\frac{R_{\rm *}^2}{d^2}\int_{0}^{\infty} Q_{\mathrm{abs}}(a,\lambda)\,
B_{\lambda}(T_{\mathrm{eff}})\, d\lambda
- 4\int_{0}^{\infty} Q_{\mathrm{abs}}(a,\lambda)\,
B_{\lambda}(T_{\mathrm{p}})\, d\lambda \notag\\
+2\alpha_{\rm acc}v_{\rm th}n_{\rm gas}k_{\rm B}(T_{\rm gas}-T_{\rm p})=0
\end{align}
where $Q_{\mathrm{abs}}(a,\lambda)$ is the wavelength-dependent absorption efficiency, 
$B_{\lambda}(T)$ is the Planck function, $R_\star$ is the stellar radius, $d$ is the particle’s distance from the central star that is equivalent to the planetary orbital distance, $\sigma$ is the Stefan-Boltzmann constant, $T_{\mathrm{eff}}$ is the stellar effective temperature that characterizes the total radiative flux emitted per unit surface area of the star, $\alpha_{\rm acc}\approx0.5$ is the heat accommodation coefficient \citep{Draine+11}, $v_{\rm th}=\sqrt{8k_{\rm B}T_{\rm gas}/\pi m_{\rm gas}}$ is the mean thermal velocity of ambient gas, $n_{\rm gas}$ is the number density of gas particles, and $T_{\rm gas}$ is the gas temperature. 
In Equation \eqref{eq:Energy_balance2}, the first term stands for the stellar radiative heating, the second term stands for the cooling of the particle through thermal emission, and the third term stands for the thermal relaxation with ambient gas temperature through gas-particle collisions \citep{Draine+11}.
We have approximated a stellar spectrum as a black body radiation for simplicity.

It is useful to estimate under what conditions the cloud particle temperature deviates from ambient gas temperature.
To this end, we define the decoupling pressure, $P_{\rm dec}$, as the pressure at which the fractional temperature difference between the particle and ambient gas first reaches $5\%$ \footnote{The choice of $5\%$ is somewhat arbitrary but hardly affects the argument presented here, since the particle temperature quickly departs from the gas temperature once the decoupling occurs, as seen in Figure \ref{fig:decoupling}.} as the pressure decreases, namely,
\begin{equation}
\frac{|T_{\rm p}-T_{\rm gas}|}{T_{\rm gas}} = 0.05.
\end{equation}
To evaluate $P_{\rm dec}$, we solve Equation (1) assuming a vertically isothermal atmosphere with $T_{\rm gas} = T_{\rm eq}$, where $T_{\rm eq}$ is the planetary equilibrium temperature with zero Bond albedo and full heat redistribution, given by
\begin{equation}\label{eq:Teq}
    T_{\rm eq}=T_{\rm eff}\left( \frac{R_{\rm *}}{2d}\right)^{1/2}.
\end{equation}
The left panel of Figure~\ref{fig:decoupling} shows the resulting decoupling pressure as a function of planetary equilibrium temperature for a particle radius of $a = 0.1~\mu$m and a stellar effective temperature of $T_{\rm eff}=6000~{\rm K}$, while the right panel presents the corresponding pressure-dependent particle temperature profiles for a representative case with $T_{\rm eq} = 1000$~K.
In general, particle temperatures remain close to the gas temperature at high pressure, whereas they deviate from it at $\lesssim{10}^{-2}$--${10}^{-4}~{\rm bar}$, depending on particle compositions and planetary equilibrium temperature.
As shown in the right panel of Figure~\ref{fig:decoupling}, the particle temperature eventually converges to a certain species-dependent temperature at low pressure limit, which is controlled by the balance between radiative heating and cooling.


In this study, we mainly focus on the radiative-equilibrium particle temperature applicable to the low pressure limit to obtain a unified picture of the particle's stability at upper atmospheres.
Setting $n_{\rm gas}\rightarrow0$ in Equation \eqref{eq:Energy_balance2}, we calculate $T_{\mathrm{p}}$ by solving the energy balance equation given by

\begin{equation}\label{eq:Energy_balance}
\frac{R_\star^{2}}{d^{2}}
   \int_{0}^{\infty} Q_{\mathrm{abs}}(a,\lambda)\,
   B_{\lambda}(T_{\mathrm{eff}})\, d\lambda
   = 4\int_{0}^{\infty} Q_{\mathrm{abs}}(a,\lambda)\,
   B_{\lambda}(T_{\mathrm{p}})\, d\lambda.
\end{equation}
Equation~\eqref{eq:Energy_balance} is identical to the master equation adopted by \citet{Jones22}.
In this study, we approximate the incoming stellar radiation by blackbody radiation.
Our approach has the advantage of computing the particle temperature without knowing the details of the atmospheric temperature-pressure profile.
We investigate the pressure-dependent vertical profile of the particle temperature by directly solving Equation \eqref{eq:Energy_balance2} for a specific hot Jupiter WASP-17b in Section \ref{sec:PT_particle}.

Note that, for wavelength-independent $Q_{\rm abs}$, Equation \eqref{eq:Energy_balance} yields a particle temperature that is identical to the conventional planetary equilibrium temperature of Equation \eqref{eq:Teq}.
Combining Equations \eqref{eq:Energy_balance} and \eqref{eq:Teq}, we can express the particle temperature in terms of the planetary equilibrium temperature as
\begin{equation}
    T_{\rm p}=T_{\rm eq}\left( \frac{\left\langle Q_{\mathrm{abs}}(a,\lambda,T_{\rm eff}) \right\rangle}{\left\langle Q_{\mathrm{abs}}(a,\lambda,T_{\mathrm{p}}) \right\rangle}\right)^{1/4},
\end{equation}
where the Planck-mean absorption efficiency $\langle Q_{\mathrm{abs}}(a,\lambda,T) \rangle$ at temperature $T$ is defined as
\begin{equation}\label{eq:Planck_mean_Qabs}
\left\langle Q_{\mathrm{abs}}(a,\lambda,T) \right\rangle
\equiv 
\frac{\int_{0}^{\infty} Q_{\mathrm{abs}}(a,\lambda)\, 
    B_{\lambda}(T)\, d\lambda}
   {\int_{0}^{\infty} B_{\lambda}(T)\, d\lambda}. 
\end{equation}
Thus, the deviation of particle temperature from the equilibrium (blackbody) temperature is determined by the ratio of the Planck-averaged absorption efficiency at stellar effective temperature to that at particle temperature, which stands for the relative strength of stellar radiative heating and particle's infrared cooling.
We solve Equation \eqref{eq:Energy_balance} with respect to $T_{\rm p}$ using the bisection method, where the trapezoidal rule is adopted to conduct the numerical integration.

\begin{table*}[t]
\centering
\caption{Condensation temperature relations and
calculated condensation temperatures at $P=10^{-3}$~bar and $[\mathrm{Fe}/\mathrm{H}]=+1.0$. Condensation temperature is in Kelvin, and $P_T\equiv P/{\rm bar}$ is the atmospheric pressure in bar. Each formula is quoted from \citet{Morley12}, \citet{Visscher+10}, \citet{Grant+23}, and \citet{Wakeford+17}.}
\label{tab:Tcond}
\begin{tabular}{l | l l} \hline
Condensate & Relation & $T_{\mathrm{cond}}$ [K] \\
\hline
MnS &
$10^{4}/T_{\mathrm{cond}} \approx 7.447 - 0.42\log_{10} P_{T}-0.84[\mathrm{Fe/H}]$ &
1271 \\

KCl &$10^{4}/T_{\mathrm{cond}} \approx 12.479 - 0.879\log_{10} P_{T}-0.879[\mathrm{Fe/H}]$&
702\\

Na$_2$S &$10^{4}/T_{\mathrm{cond}} \approx 10.045 - 0.72\log_{10} P_{T}-1.08[\mathrm{Fe/H}]$ & 
898\\

ZnS &
$10^{4}/T_{\mathrm{cond}} \approx 12.527 - 0.63\log_{10} P_{T}$
$-1.26[\mathrm{Fe/H}]$ &
760\\

SiO$_2$ &
$10^{4}/T_{\mathrm{cond}} \approx 6.14 - 0.351\log_{10} P_{T}-0.70[\mathrm{Fe/H}]$ &
1540\\

Mg$_2$SiO$_4$ &
$10^{4}/T_{\mathrm{cond}} \approx 5.89 - 0.37\log_{10} P_{T}-0.73[\mathrm{Fe/H}]$ &
1594\\

Fe &
$10^{4}/T_{\mathrm{cond}} \approx 5.44 - 0.48\log_{10} P_{T}-0.48[\mathrm{Fe/H}]$ &
1562\\

Al$_2$O$_3$ &
${10}^{4}/T_{\mathrm{cond}} \approx 5.014 - 0.2179\log_{10} P_{T} +2.264\times10^{-3}(\log_{10} P_{T})^{2} - 0.580[\mathrm{Fe/H}]$  &
1957\\

\end{tabular}
\end{table*}
\subsection{Absorption Efficiency}
\begin{figure}[h]
\centering
\includegraphics[width=0.45\textwidth]{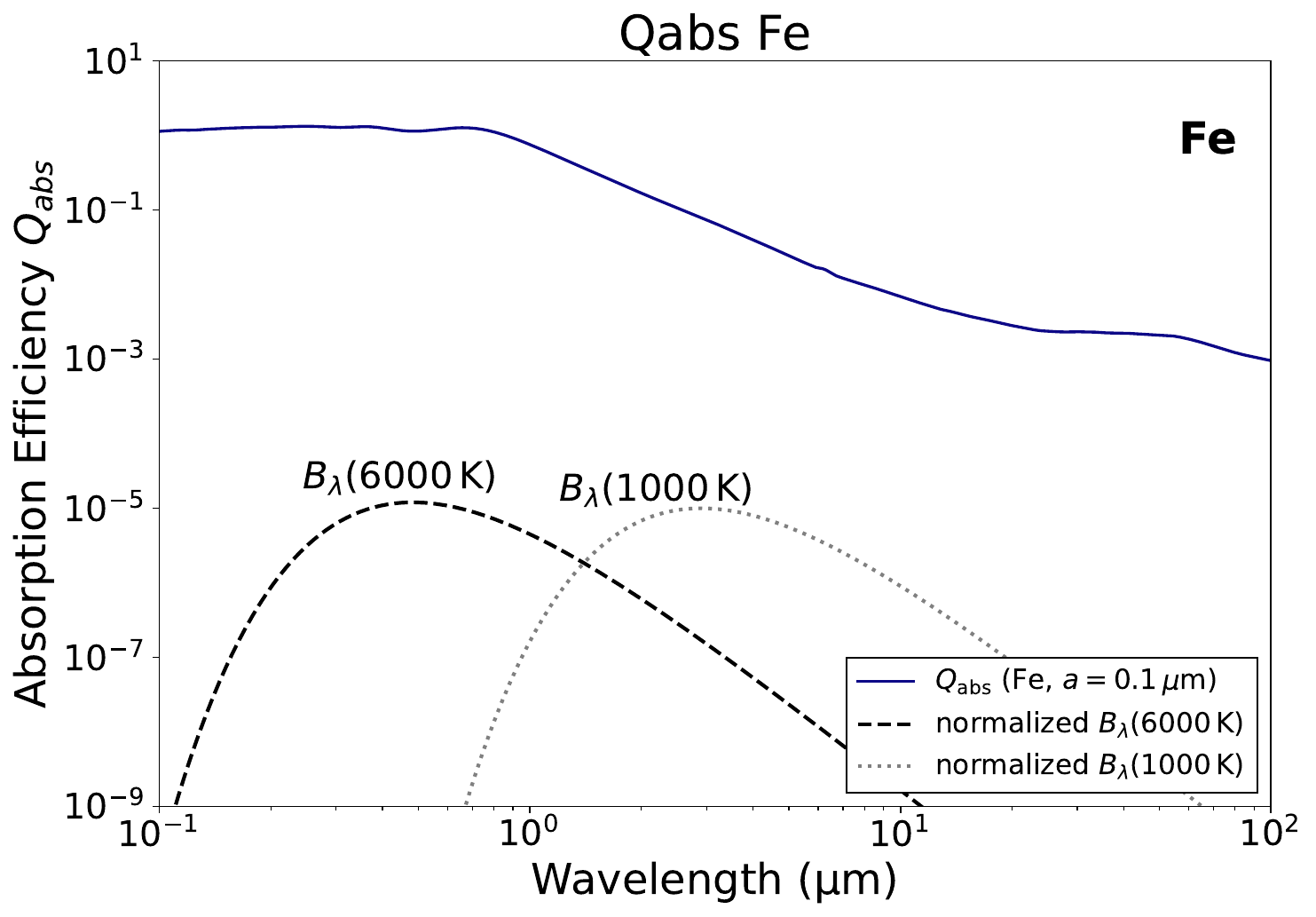}
\includegraphics[width=0.45\textwidth]{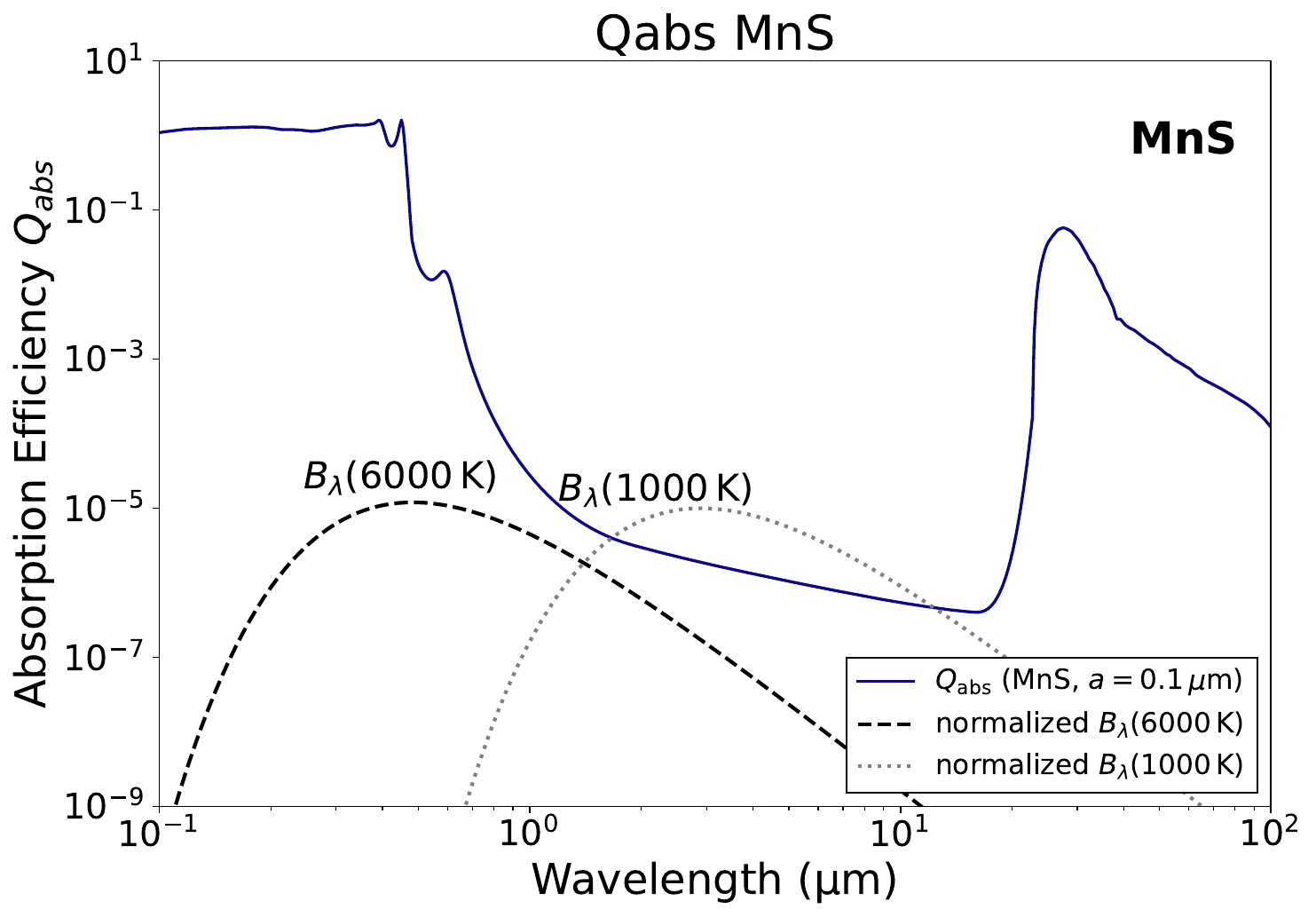}
\includegraphics[width=0.45\textwidth]{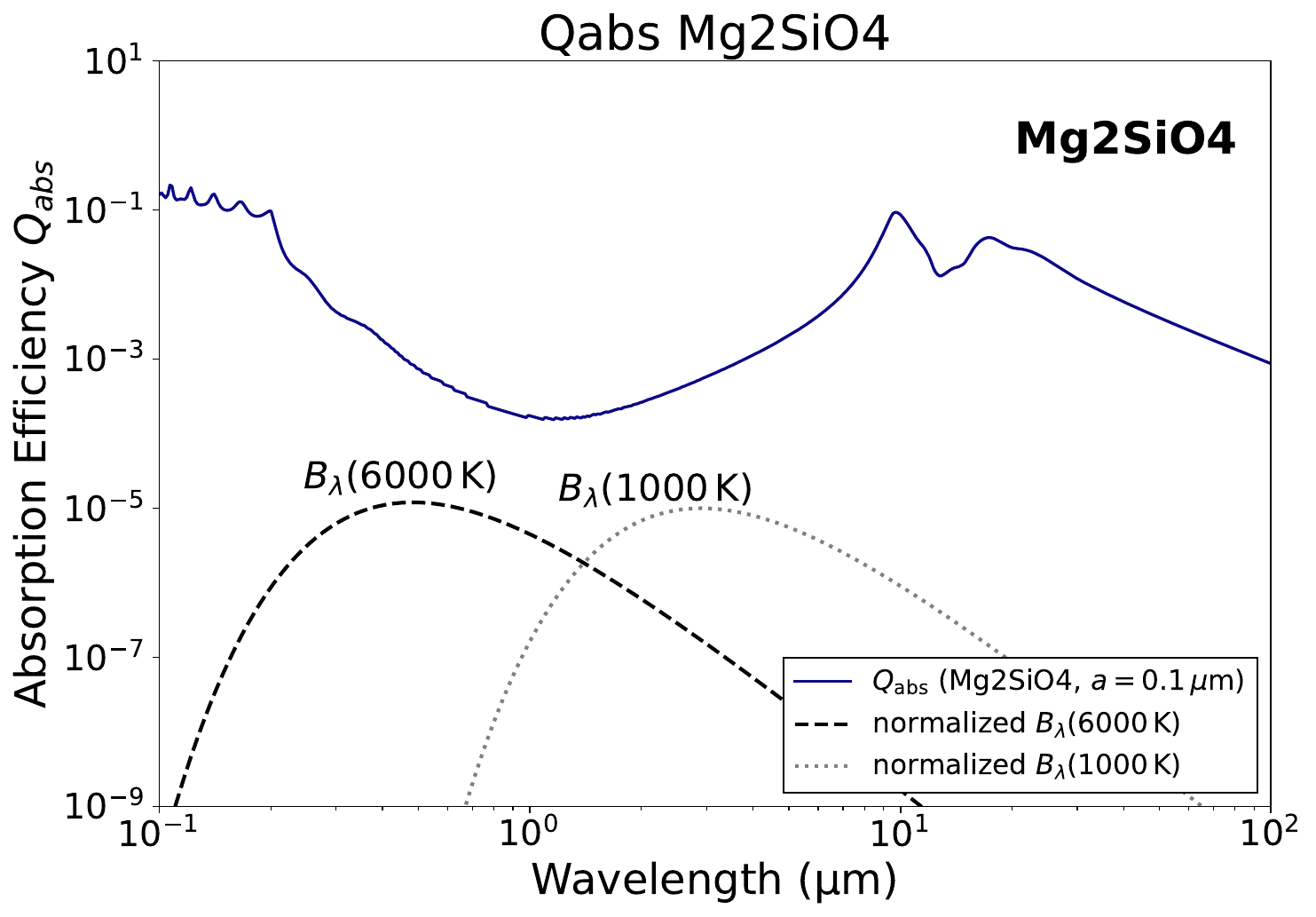}
\caption{
Representative examples of the wavelength-dependent absorption efficiency $Q_{\rm abs}(a,\lambda)$
for Fe, MnS, and Mg$_2$SiO$_4$, overlaid with normalized blackbody curves at $6000$~K and $1000$~K.
We set the particle radius to $a=0.1~\mu$m.
}
\label{fig:Qabs_BB}
\end{figure}

We compute the absorption efficiency of cloud particles $Q_{\text{abs}}(a,\lambda)$ based on Mie theory using the open-source Python package miepython \citep{prahl_miepython_2026}.
We carry out the calculations for various substances expected to form as clouds in exoplanets: {MnS}, {KCl}, {SiO$_2$ ($\alpha$-quartz)}, {amorphous Mg$_2$SiO$_4$}, {Fe}, {Na$_2$S}, {Al$_2$O$_3$}, {ZnS} \citep[e.g.,][]{Gao+21}. 
We adopt wavelength-dependent complex refractive indices $m(\lambda)=n(\lambda)+ik(\lambda)$
compiled by \citet{Kitzmann&Heng18}.
The data set covers the wavelength range of $\lambda \approx 0.1$--$100~{\rm \mu m}$ for most species, covering a wavelength range at which stellar radiative heating and infrared cooling take place. 
For {Mg$_2$SiO$_4$} and {Al$_2$O$_3$}, refractive-index data are unavailable below 
$\lambda = 0.2~\mu\mathrm{m}$; therefore, \ko{we assumed constant values of $n$ and $k$ from $0.2$ to $0.1~\mu\mathrm{m}$ equal to  that at $0.2~\mu\mathrm{m}$}.



The spectral shapes of $Q_{\rm abs}$ differ substantially among condensates.
Figure \ref{fig:Qabs_BB} shows $Q_{\rm abs}$ for Fe, MnS, and Mg$_2$SiO$_4$ as a representative case.
To illustrate the connection between $Q_{\rm abs}$ and radiative heating/cooling more directly, Figure~\ref{fig:Qabs_BB} also overlays representative $Q_{\rm abs}$ curves with normalized blackbody spectra at $6000$ K and $1000$ K, corresponding to typical stellar irradiation and cloud thermal emission, respectively. Fe maintains relatively high $Q_{\rm abs}$ over both wavelength ranges, allowing efficient heating and cooling. MnS shows strong absorption at short wavelengths overlapping with stellar irradiation, but relatively weak absorption over much of the thermal infrared, leading to efficient heating but inefficient cooling. In contrast, Mg$_2$SiO$_4$ has weak absorption at stellar wavelengths and relatively stronger absorption in the infrared, favoring weak heating and efficient cooling. 
Such inter-species diversity in $Q_{\rm abs}$ is crucial for determining the balance between radiative heating and cooling and thus the resulting particle equilibrium temperature.

Figure~\ref{fig:Qabs_all_full} in Appendix \ref{app:qabs} summarizes the
wavelength-dependent absorption efficiency, $Q_{\rm abs}(a,\lambda)$,
for all condensates considered in this study at several particle radii.
In general, $Q_{\rm abs}$ is larger at shorter wavelengths and decreases for smaller particles,
reflecting the reduced absorption efficiency in the small-particle regime.
As the particle size increases, $Q_{\rm abs}$ tends to approach the geometric-optics limit,
$Q_{\rm abs}{\sim}1$, most clearly for strongly absorptive materials such as Fe.
Note that we focus on $Q_{\rm abs}$ for given particle sizes without considering size distributions because our present interest lies in the temperature of an individual cloud particle.
The Mie calculation produces spikes in $Q_{\rm abs}$ profile due to the assumption of a perfect spherical shape. 
We tested the particle temperature calculation using a $Q_{\rm abs}$ profile smoothed by the method of \citet{Batalha+26} and found that the Mie spikes hardly affect the results.
Therefore, we adopt $Q_{\rm abs}$ without smoothing in the rest of this paper.


\subsection{Criterion of Particle's Thermal Stability}
To evaluate thermal stability, we compare the particle temperature $T_{\rm p}$ with the condensation temperature $T_{\rm cond}$ derived by previous studies.
For each species, particles are supposed to be thermally stable if $T_{\rm p}<T_{\rm cond}$ and unstable otherwise. 
We compute $T_{\rm cond}$ (in K) from published equilibrium condensation relations compiled from \citet{Morley12} (MnS, KCl, Na$_2$S, ZnS), \citet{Visscher+10} (Mg$_2$SiO$_4$, Fe), \citet{Grant+23} (SiO$_2$), and \citet{Wakeford+17} (Al$_2$O$_3$). 
The condensation temperature in the literature depends on atmospheric pressure $P$ and metallicity [Fe/H].
Throughout this work, we adopt $P=10^{-3}$~bar and $[\mathrm{Fe}/\mathrm{H}]=+1.0$ as representative of a region probed by transmission spectroscopy. 
We choose $[\mathrm{Fe}/\mathrm{H}]=+1.0$ to mimic the metal–enriched atmospheres reported for several close–in giants \citep[e.g.,][]{Feinstein+23,Grant+23,Fu+25}. 
Table~\ref{tab:Tcond} summarizes these relations together with the resulting condensation temperatures ($T_{\rm cond}$) under the adopted conditions.

\section{Results}\label{sec:results}

\begin{figure*}[t]
\centering
\includegraphics[width=0.75\textwidth]{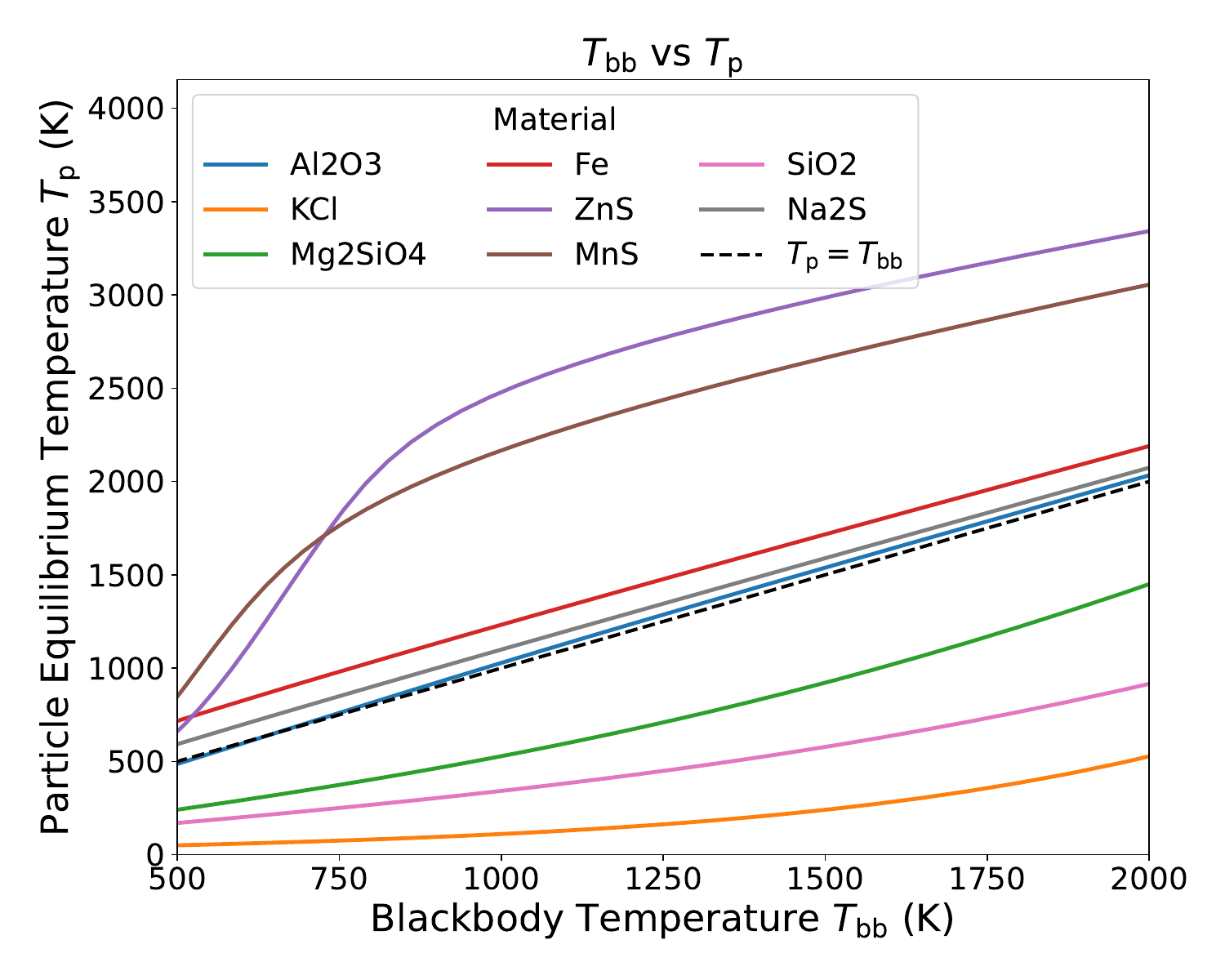}
\caption{
Particle equilibrium temperature $T_{\rm p}$ as a function of blackbody temperature $T_{\rm bb}$.
Different colored lines show the particle temperature for different condensates. We set the particle radius to $10~{\rm {\mu}m}$ and stellar effective temperature to $T_{\rm eff}=5000K$. 
}
\label{fig:Tbb_vs_Tp}
\end{figure*}

\subsection{General trend of particle temperature}

Particle temperature is sensitive to particle compositions. 
To illustrate this dependence, Figure~\ref{fig:Tbb_vs_Tp} shows the calculated particle equilibrium temperature as a function of the corresponding blackbody equilibrium temperature $T_{\rm bb}$, i.e., the conventional planetary equilibrium temperature $T_{\rm eq}$ (Equation \ref{eq:Teq}). 
Since we assume that cloud particles are isothermal due to efficient heat redistribution by conduction, $T_{\rm eq}$ can serve as a reference, as it similarly assumes full heat redistribution over the planet.
Here, we set the particle radius to $a=10~{\rm {\mu}m}$ and the stellar effective temperature to $T_{\rm eff}=5000K$, and assume that cloud particles are thermally stable at any temperature to illustrate the behavior of particle temperature.
Overall, the behavior of $T_{\rm p}$ varies markedly among condensate species, showing composition-dependent deviations from the blackbody trend. 
A closer inspection reveals a clear separation: several condensates (Fe, Al$_2$O$_3$, Na$_2$S, MnS, ZnS) remain consistently hotter than the blackbody value, others (Mg$_2$SiO$_4$, SiO$_2$, KCl) remain cooler, and a small subset (MnS, ZnS) exhibits a distinct transitional behavior. 
Motivated by these tendencies, we classified the condensates into three groups.

We refer to the first group as the broadband absorber group, which consists of Fe, Al$_2$O$_3$, and Na$_2$S.
Notably, the temperatures of these clouds closely follow the blackbody temperature.
These trends can be understood from the $Q_{\rm abs}$ spectra of each condensate. 
The broadband absorber group (Fe, Na$_2$S, Al$_2$O$_3$) maintain a relatively high $Q_{\rm abs}$ in both optical-NIR wavelength band (0.3--1.0 \micron) and the thermal-IR band (1--10 \micron). 
The heating by stellar radiation occurs predominantly in the optical to near-infrared (NIR) wavelengths, whereas the cooling by thermal emission occurs predominantly at longer infrared wavelengths. 
Since the broadband absorber group condensates have high $Q_{\rm abs}$ in both bands, these condensates have the ability to heat and cool both effectively. 
Because blackbodies have a similar characteristic, in this plane, the temperatures of Na$_2$S, Fe, and Al$_2$O$_3$ behave similarly to the temperature of the blackbody, though these condensates are systematically warmer than the blackbody due to gradual decrease of $Q_{\rm abs}$ toward long wavelengths (see Figure \ref{fig:Qabs_all_full}).

For the \ko{second}, optical-heating-dominated group (MnS, ZnS), the clouds have high $Q_{\rm abs}$ in 0.3--1.0 {\micron} wavelength, thus resulting in efficient heating, whereas their radiative cooling is inefficient due to the low $Q_{\rm abs}$ in the 1--10 {\micron} band (see Figures \ref{fig:Qabs_BB} and \ref{fig:Qabs_all_full}). 
Due to this inequality of heating and cooling, particles are strongly influenced by the heating from the stellar irradiation and struggle to release the heat for cooling. 
To compensate for the low ability of radiative cooling, the particle temperature increases rapidly with increasing planetary equilibrium temperature.
At $T_{\rm p}\gtrsim2000~{\rm K}$, the rate of increase in particle temperature \ko{becomes more} moderate.
This transition behavior occurs because the optical-to-NIR band, where $Q_{\rm abs}$ increases toward short wavelengths, begins to play a role in particle cooling \footnote{We note that, in practice, this behavior is unlikely to be seen in real optical-heating-dominated group, as the transition occurs at particle temperature much hotter than their sublimation temperature.}.
These species exhibit temperatures much hotter than the blackbody temperature, which significantly affects their thermal stability as discussed later.

The \ko{third}, IR-cooling-dominated group,
 (SiO$_2$, Mg$_2$SiO$_4$, KCl) has an opposite characteristic to that of the optical-heating-dominated group. 
$Q_{\rm abs}$ is low across the stellar optical--NIR heating band (0.3--1.0 \micron), making radiative heating inefficient (see Figures \ref{fig:Qabs_BB} and \ref{fig:Qabs_all_full}).
Meanwhile, $Q_{\rm abs}$ remains relatively high over the thermal-IR cooling band (1.0--10 \micron) compared to the heating band, enabling efficient radiative cooling.
Because the cooling ability dominates over the heating ability, their particle temperatures lie systematically below the blackbody reference across a wide range of planetary equilibrium temperature.

\begin{figure}[t] 
\centering
\includegraphics[clip,width=\hsize]{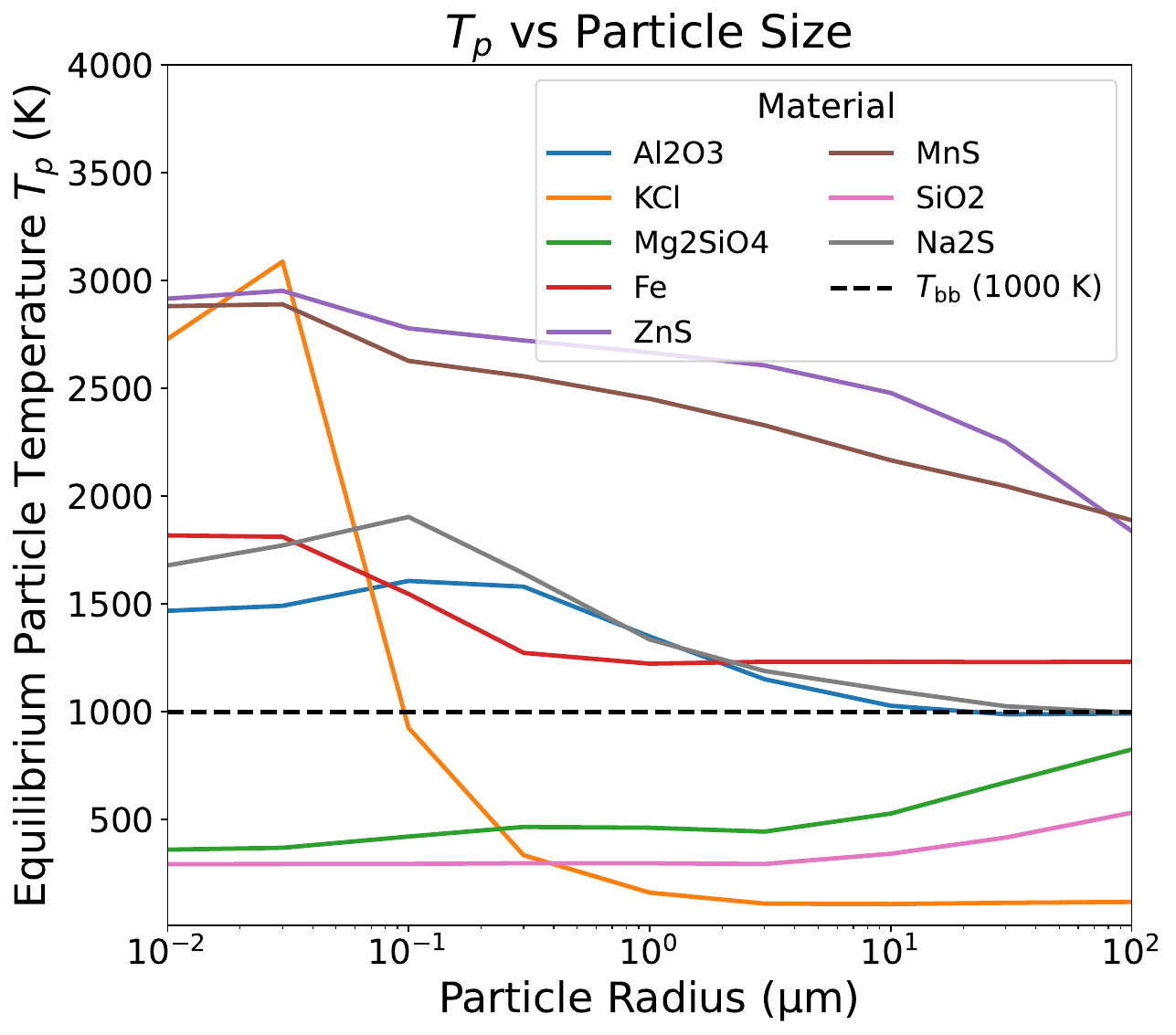}
\caption{
Particle equilibrium temperature $T_{\rm p}$ versus particle radius $a$ at selected $T_{\rm bb}$ for the eight condensates considered in this study. We fix the stellar effective temperature to $T_{\rm eff}=5000$ K and the orbital distance to yield $T_{\text{bb}}=1000~ \text{K}$.
}
\label{fig:Tp_vs_size}
\end{figure}

\subsection{Particle Size Dependence}
It is worthwhile to examine the size dependence of the temperature, as microphysical and dynamical processes would yield cloud particles with diverse sizes.
Figure~\ref{fig:Tp_vs_size} summarizes the particle temperature as a function of a particle size.
The dependence on particle size follows a simple law: smaller particles become warmer, and temperatures of larger particles tend to approach a blackbody temperature.
Smaller particles tend to decrease their $Q_{\rm abs}$ in longer wavelengths (see Figures \ref{fig:Qabs_BB} and \ref{fig:Qabs_all_full}), making their cooling rather inefficient. This simple rule is evident in Fe, Na$_2$S, Al$_2$O$_3$, MnS, KCl, and ZnS.

Although we introduced the three category of broadband absorber, optical-heating-dominated, and IR-cooling-dominated groups in Section~3.1, the actual behavior of particle temperature can depend on the particle size.
For example, for sufficiently small particles, the broadband absorber condensates introduced in Section~3.1 deviate significantly from the blackbody temperature. This behavior arises because their absorption efficiency $Q_{\rm abs}$ is strongly suppressed at the longer wavelengths relevant for radiative cooling.

More complex trend can be seen in KCl clouds: their temperature remains cooler than the blackbody at the particle radii of $\gtrsim1~{\rm {\mu}m}$ but rapidly rises to $\sim3000~{\rm K}$ at $\sim0.03~{\rm {\mu}m}$ and again decrease toward smaller particle sizes, although we classified KCl as IR-cooling-dominated condensate.
This behavior is attributed to the enhanced $Q_{\rm abs}$ at visible to UV wavelengths at smaller particle sizes, which lead the stellar heating to surpass the IR cooling for tiny KCl particles.
The reason why particle temperature peaks at the radius of $\lesssim0.03~{\rm {\mu}m}$ is attributed to the fact that the heating ability of KCl saturates at $a\gtrsim0.03~{\rm {\mu}m}$.
That is, $Q_{\rm abs}$ values at wavelengths of $<0.2~{\rm {\mu}m}$ increase with increasing particle size at $a\lesssim0.03~{\rm {\mu}m}$ (see Figure \ref{fig:Qabs_all_full}), leading to hotter temperature.
At $a\gtrsim0.03~{\rm {\mu}m}$, KCl particles retain $Q_{\rm abs}\sim1$ at wavelengths of $<0.2~{\rm {\mu}m}$, while $Q_{\rm abs}$ at longer wavelengths continue to increase with size, yielding stronger cooling ability for larger particles.
Consequently, particle temperature decreases with particle size at $a\gtrsim0.03~{\rm {\mu}m}$ where the heating ability is saturated.

\begin{figure}[t]
\centering
\includegraphics[clip,width=\hsize]{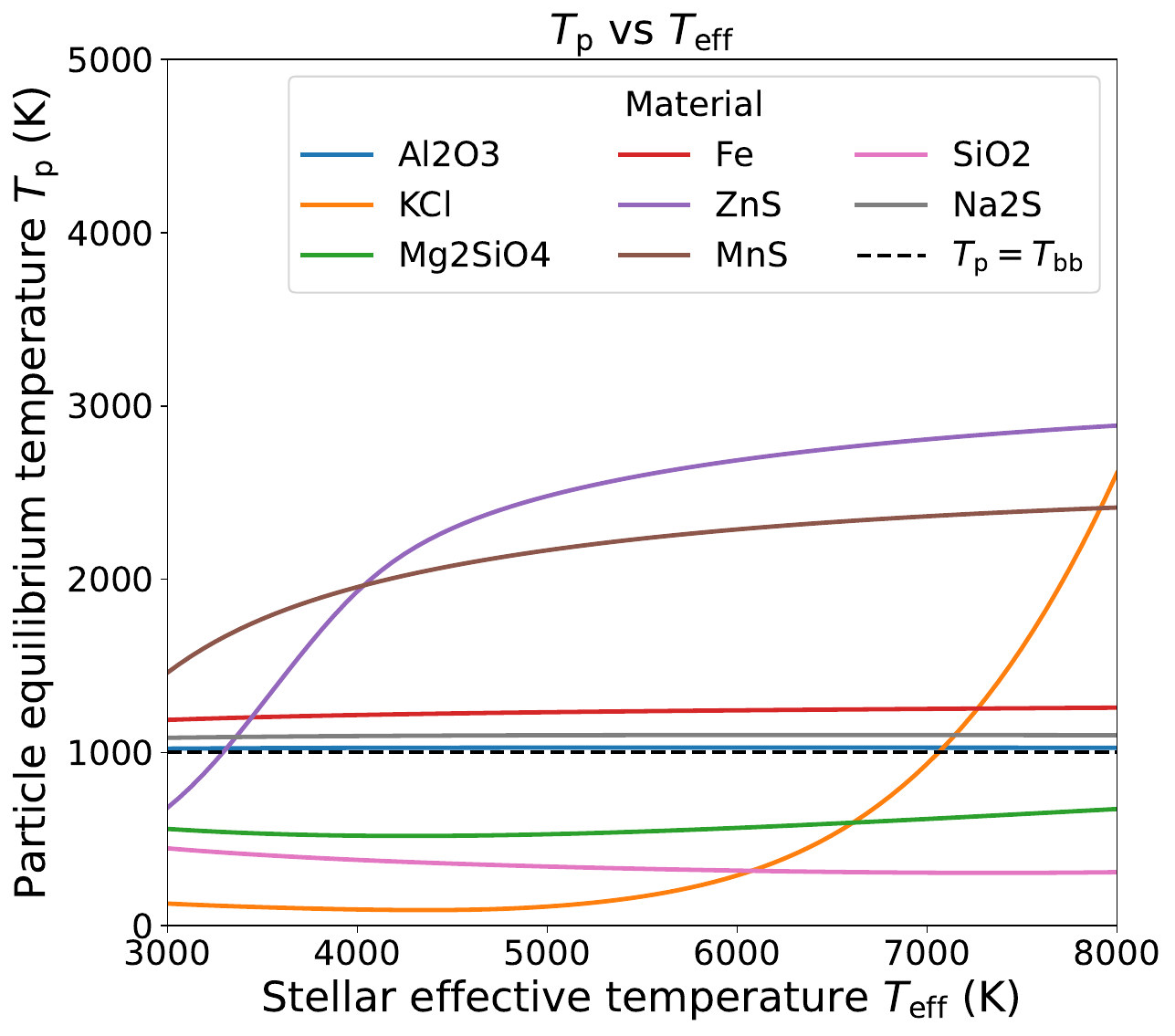}

\caption{Particle equilibrium temperature $T_{\rm p}$ as a function of stellar effective temperature $T_{\rm eff}$.
Different colored lines represent different condensate species. The particle radius is fixed at $a = 10~\mu{\rm m}$, the stellar radius at $1~ R_{\odot}$, and the blackbody temperature at $T_{\rm bb} = 1000~ {\rm K}$.}
\label{fig:Teff_vs_Tp}
\end{figure}

\begin{figure*}[ht]
\centering

\includegraphics[width=0.395\textwidth]{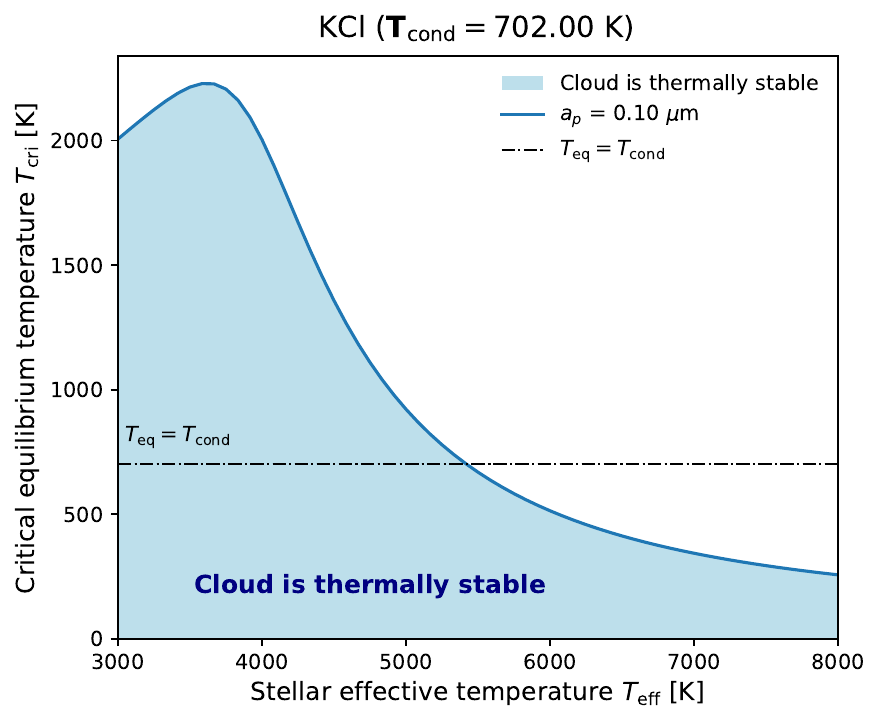}
\includegraphics[width=0.395\textwidth]{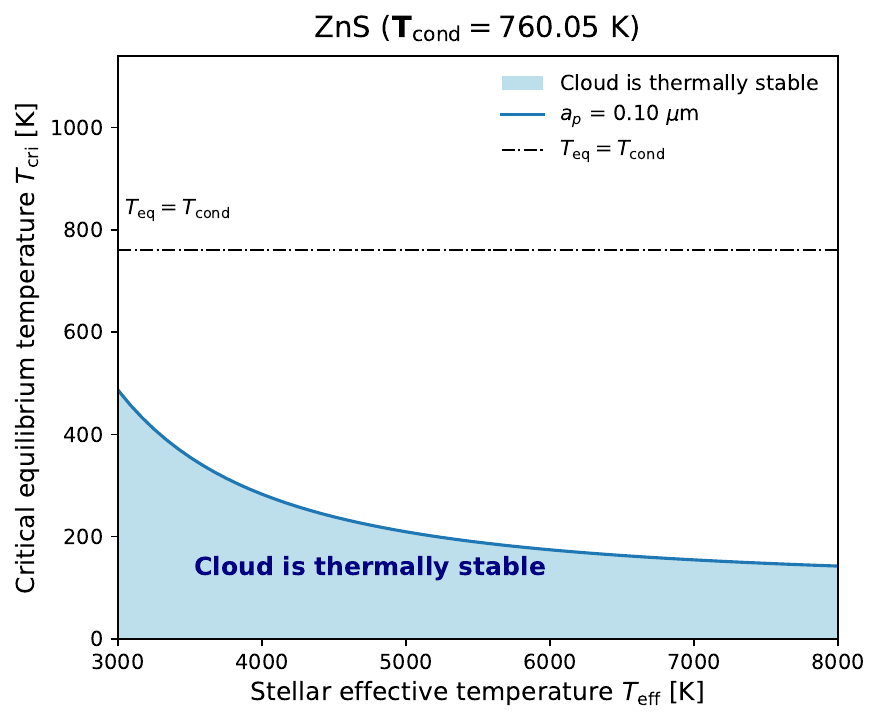}
\includegraphics[width=0.395\textwidth]{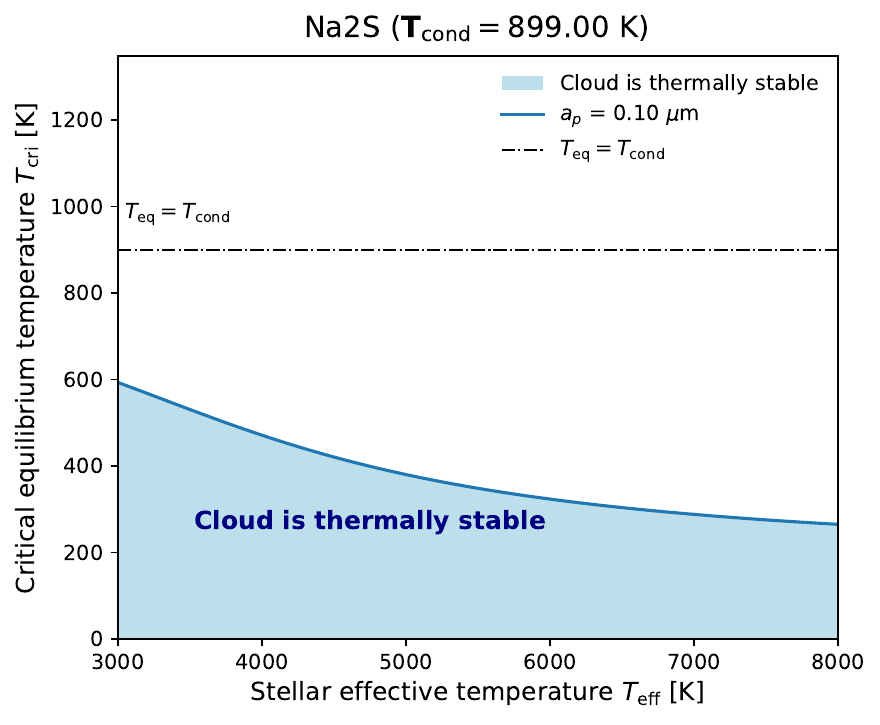}
\includegraphics[width=0.395\textwidth]{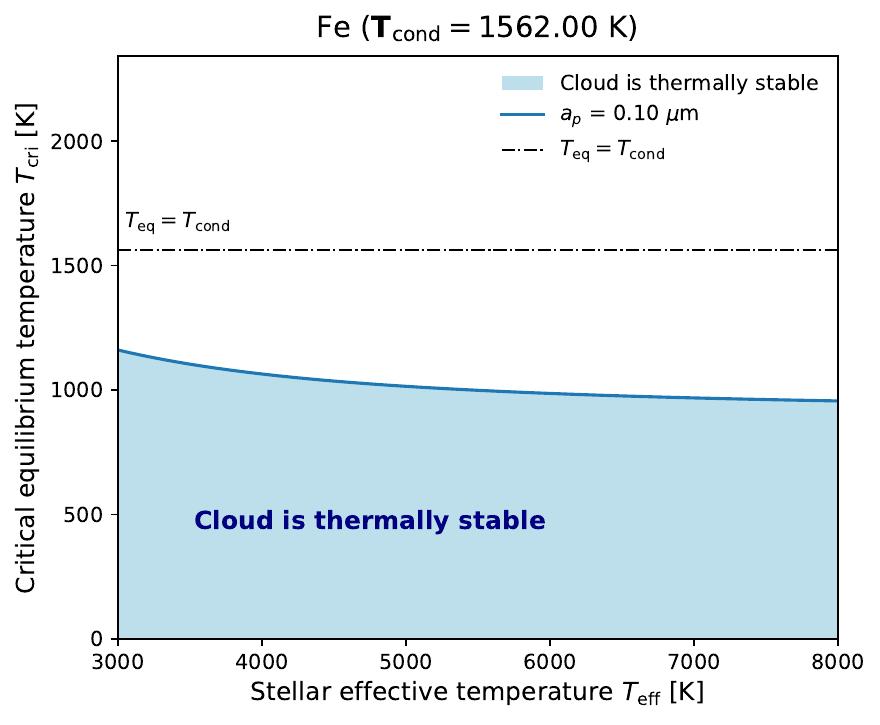}
\includegraphics[width=0.395\textwidth]{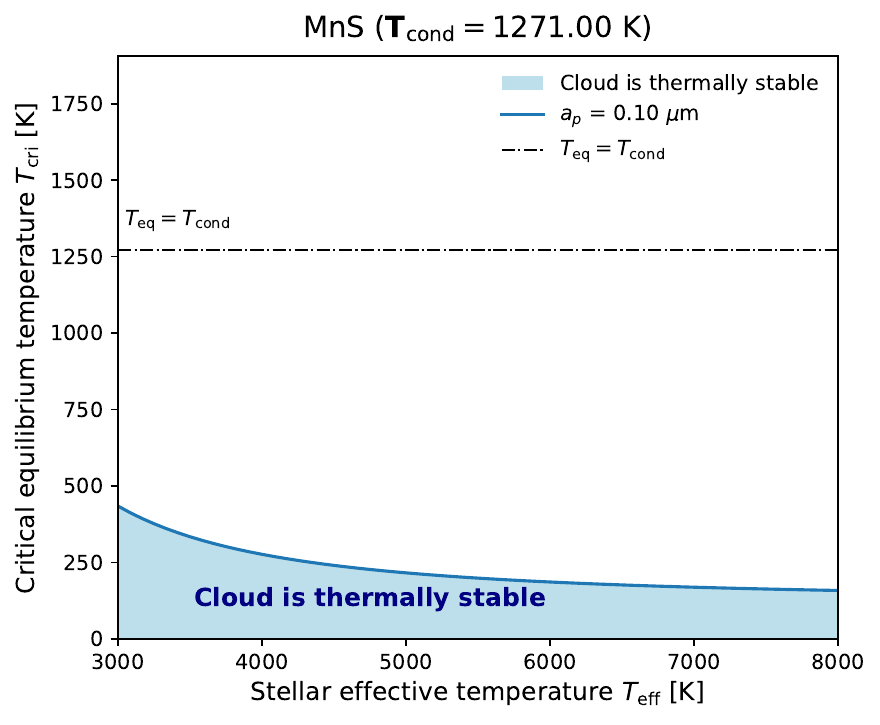}
\includegraphics[width=0.395\textwidth]{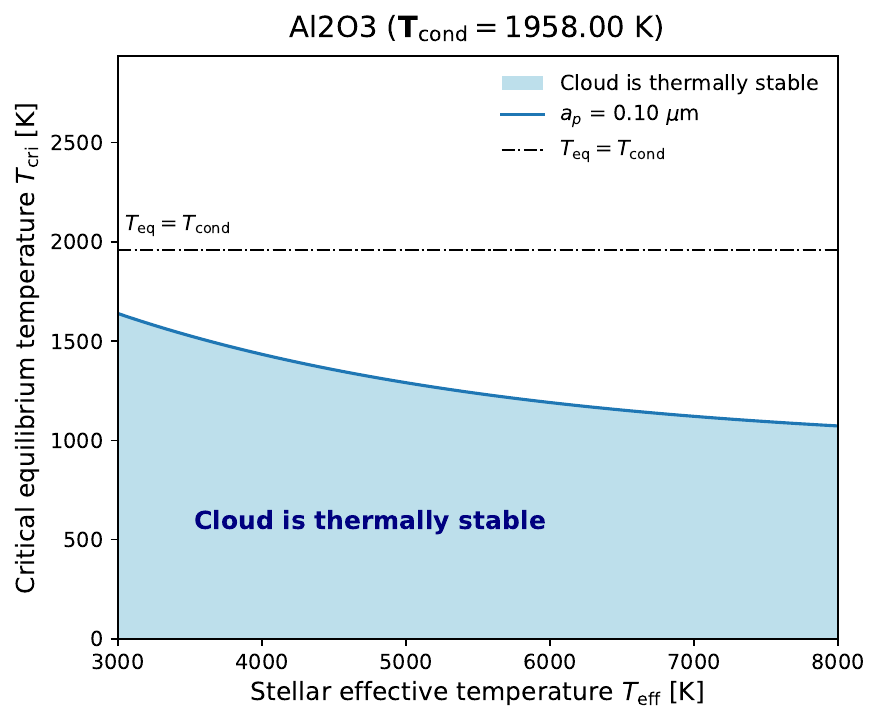}
\caption{
The critical planetary equilibrium temperature $T_{\rm cri}$ as a function of stellar effective temperature $T_{\rm eff}$ for KCl, ZnS, Na$_2$S, MnS, Fe, and Al$_2$O$_3$. The shaded regions denote the $T_{\rm eq}$---$T_{\rm eff}$ spaces at which the particle temperatures are cool enough to avoid sublimation. The dash-dot lines show the $T_{\rm eq}=T_{\rm cond}$ relations for reference. We set the particle radius to $0.1~{\rm {\mu}m}$. The calculations assume $P = 10^{-3}~{\rm bar}$ and $[\mathrm{Fe}/\mathrm{H}] = +1.0$, representative of the region probed by transmission spectroscopy.}

\label{fig:Teq_vs_Teff}
\end{figure*}

\subsection{Stellar effective temperature dependence}\label{sec:Teff_dependence}

Since stellar radiation controls the heating rate, it is of great interest to examine how the particle temperature depends on the stellar spectral type.
Figure~\ref{fig:Teff_vs_Tp} shows the equilibrium temperature of cloud particles,
$T_{\rm p}$, as a function of the stellar effective temperature $T_{\rm eff}$
for all condensates considered in this study, assuming a particle radius of
$a = 10~\mu$m. The blackbody temperature $T_{\rm bb}$ is also shown for reference.
By comparing this figure with the $T_{\rm p}=T_{\rm bb}$ relation,
it is evident that the condensates can again be classified into the three
categories introduced in Section~3.1, based on their degree of deviation
from the blackbody temperature.

One notable exception is the behavior of KCl.
As shown in Figure~\ref{fig:Teff_vs_Tp}, KCl exhibits a rapid increase in
particle temperature above $T_{\rm eff} \sim 6500$~K,
resulting in a pronounced temperature jump comparable to those seen for
MnS and ZnS.
This behavior can be attributed to the wavelength dependence of radiative
heating and cooling.
As $T_{\rm eff}$ increases, the stellar irradiation shifts toward shorter
wavelengths, where the absorption efficiency $Q_{\rm abs}$ of KCl becomes
significantly larger.
At the same time, the characteristic wavelength of thermal emission also
moves to shorter mid-infrared wavelengths where $Q_{\rm abs}$ is significantly low.
Consequently, stellar heating becomes increasingly efficient while
radiative cooling remains relatively inefficient, leading to extremely hot temperature for KCl around hot stars.

\subsection{Forbidden Zones for Cloud Stability}\label{sec:cond_boundary}

\begin{figure}[t]
    \centering
    \includegraphics[width=0.48\textwidth]
    {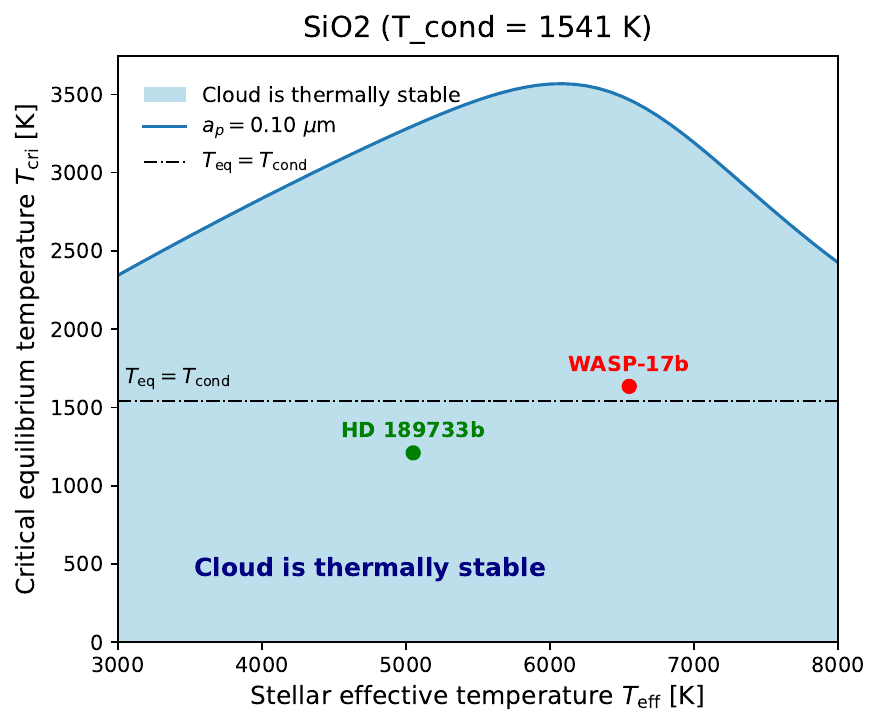}
    \includegraphics[width=0.48\textwidth]{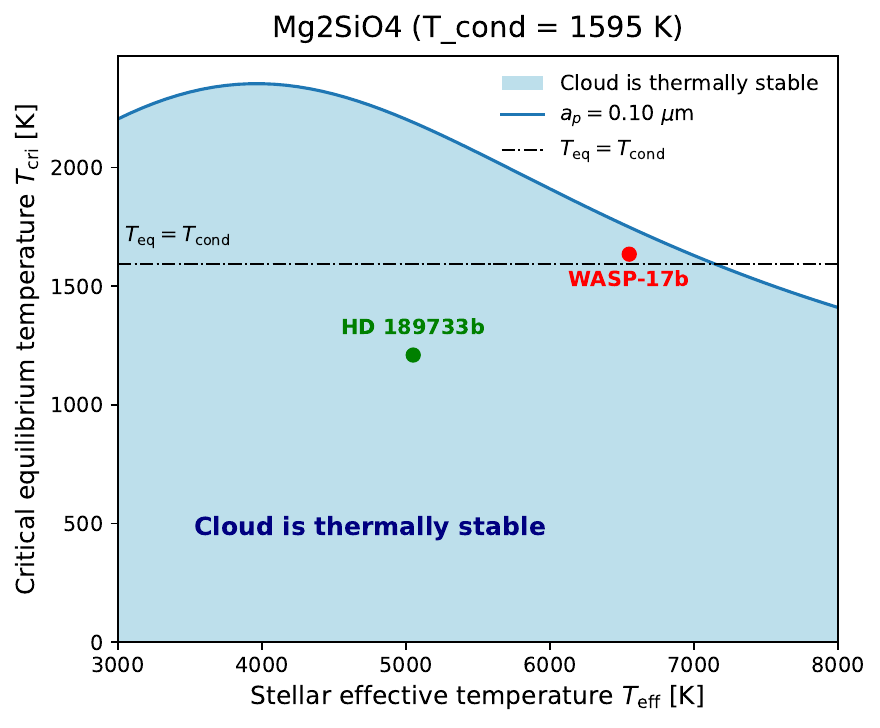}
    \caption{Same as Figure \ref{fig:Teq_vs_Teff} but for SiO$_2$ and Mg$_2$SiO$_4$. The red and green dot denote the $T_{\rm eq}$--$T_{\rm eff}$ relations of WASP-17b and HD 189733b for which SiO$_2$ cloud has been discovered.}\label{fig:boundary_silicates}
\end{figure}

Since the radiative–equilibrium particle temperature $T_{\rm p}$ can greatly deviate from the planetary equilibrium temperature $T_{\rm eq}$, it is useful to determine under which planetary properties each condensate can stably exist in the upper atmosphere. 
Under radiative-equilibrium, for a given particle composition, a cloud particle temperature is uniquely determined by planetary equilibrium temperature $T_{\rm eq}$, stellar effective temperature $T_{\rm eff}$, and particle size $a$.
For each condensate and particle size, we therefore define a critical equilibrium temperature $T_{\rm cri}(T_{\rm eff},a)$ as the value of $T_{\rm eq}$ at which the particle temperature $T_{\rm p}(T_{\rm eff},\,T_{\rm eq},\,a)$ reaches the condensation temperature $T_{\rm cond}$ listed in Table~\ref{tab:Tcond}:
\begin{equation}
    T_{\rm p}(T_{\rm eff},\,T_{\rm cri},\,a) = T_{\rm cond}.
    \label{eq:Tcri_def}
\end{equation}
We compute $T_{\rm cri}$ on a grid of stellar effective temperatures $T_{\rm eff}$ by numerically solving the implicit equation of $T_{\rm p}(T_{\rm eff},\,T_{\rm eq},\,a) = T_{\rm cond}$ with respect to $T_{\rm eq}$ using the bisection method, where the particle radius is fixed to $a=0.1~{\rm {\mu}m}$ to represent tiny particles that can stay in upper tenuous atmospheres. 
In the $(T_{\rm eff},T_{\rm eq})$ plane, the resulting curves delineate “forbidden zones’’ for cloud stability: planets located above a species’ boundary yield particle temperature hotter than condensation temperature (i.e., $T_{\rm p}>T_{\rm cond}$) and are therefore predicted to lack that condensate in their upper atmospheres, whereas those below the boundary are thermodynamically allowed to host it.
Figures~\ref{fig:Teq_vs_Teff} and \ref{fig:boundary_silicates} summarize the resulting condensation boundaries for all eight condensates.

Planet's critical temperature $T_{\rm cri }$ generally decreases monotonically with increasing $T_{\rm eff}$. 
The cause of this trend can be understood from Figure~\ref{fig:Teff_vs_Tp}.
Because the radiation from the hotter host star shifts towards shorter wavelengths at which the absorption efficiency $Q_{\rm abs}$ tends to be higher, hotter stars cause stronger heating, leading to a hotter particle temperature at a given planetary equilibrium temperature $T_{\mathrm{eq}}$, as introduced in Section \ref{sec:Teff_dependence} and Figure~\ref{fig:Teff_vs_Tp}.
As a result, the $T_{\mathrm{eq}}$ required to reach the evaporation temperature becomes lower around hotter stars. 

Among our condensates, KCl, SiO$_2$, and Mg$_2$SiO$_4$ deviate from the monotonic trend: their critical equilibrium temperatures exhibit a local maximum around $T_{\rm eff}\sim 4000$–$5000$ K.  For $T_{\rm eff}\gtrsim 5000$ K, the stellar spectrum peaks in the optical band and the short-wavelength absorption is efficient. 
As $T_{\rm eff}$ increases, the optical heating of the particles becomes stronger, reducing planetary equilibrium temperature $T_{\rm eq}$ to achieve a particle temperature hotter than $T_{\rm cond}$.
As a result, the critical $T_{\rm eq}$ decreases with $T_{\rm eff}$, as for the other species.  In contrast, at $T_{\rm eff}\lesssim 5000$ K stellar radiation peak falls into optical-to-near infrared wavelengths where these condensates have very low $Q_{\rm abs}$, leading to negligible heating in this wavelength band.  
Their heating is instead dominated by absorption in the thermal-infrared ($\lambda\gtrsim 10~\mu{\rm m}$), where $Q_{\rm abs}$ has a prominent peak.
Since stellar incident energy is deposited to longer wavelengths at lower $T_{\rm eff}$, the critical $T_{\rm eq}$ again decreases as $T_{\rm eff}$ decreases in this cool-star regime.  The local maximum therefore arises at the transition between these two regimes, where the dominant heating band switches from the optical to the infrared.

We find that optical-heating-dominated condensates (ZnS, MnS) and broadband absorber condensates (Na$_2$S, Fe, Al$_2$O$_3$) can undergo sublimation even if planetary equilibrium temperatures are considerably cooler than the condensation temperature of each species.
In particular, sulfide condensates can become thermally unstable even on relatively cool planets with $T_{\rm eq}\sim 300$~K.
While a microphysical model predicted the difficulty in forming sulfide clouds due to inefficient nucleation \citep{Gao+20}, this study further demonstrates that sulfide clouds, even if they form somehow, tend to be thermally unstable due to inefficient infrared cooling.

Our results also demonstrate that silicate clouds (SiO$_2$, Mg$_2$SiO$_4$) can remain thermally stable even on extremely hot planets with $T_{\rm eq}\gtrsim 2000$~K.  In particular, G and K type host stars ($T_{\rm eff}\sim 4000$–$6000$~K) provide a favorable environment for maintaining their thermal stability, representing a potential ``sweet spot’’ for the presence of silicate clouds at upper atmospheres.  
Salt clouds (KCl) exhibit a partly similar trend: they are thermally unstable even on cool planets of $T_{\rm eq}\sim400~{\rm K}$ around hot stars of $T_{\rm eff}>6000~{\rm K}$, whereas they are stable even on hot planets of $T_{\rm eq}>1000~{\rm K}$ around cool stars of $T_{\rm eff}<5000~{\rm K}$.
This result suggests that KCl clouds may preferentially exist only on planets around cool stars.
We note that our current analysis supposes an upper atmosphere where particle temperatures are set purely by radiative processes.
Section \ref{sec:PT_particle} presents a stability assessment for several condensates on a specific hot Jupiter WASP-17b by taking into account thermal relaxation with ambient gases.

\subsection{Application to WASP-17b and HD 189733b}
Recent JWST observations discovered spectral features of SiO$_2$ clouds in hot Jupiter WASP-17b \citep{Grant+23}.
It is of great interest to see whether WASP-17b falls within the thermodynamically allowed regime for SiO$_2$ clouds in the stability diagram established in Section \ref{sec:cond_boundary}.
For WASP-17b, we adopt a stellar effective temperature of $T_\star=6550$~K \citep{Stassun+17}, a stellar radius of $R_\star=1.38\,R_\odot$ \citep{Anderson+10}, and a planetary semi-major axis of $a=0.0515$~AU \citep{Bonomo+17}. 
These parameters yield planetary equilibrium temperature of $T_{\rm eq} = 1635~{\rm K}$, which lies within the thermally stable region of SiO$_2$ (Figure \ref{fig:boundary_silicates}).
Thus, our result is consistent with the SiO$_2$ clouds on WASP-17b reported by \citet{Grant+23}.
We note that our calculation assesses the thermal stability of SiO$_2$ clouds after they form somehow.
The microphysical pathways leading to SiO$_2$ cloud formation, which is suggested to involve non-equilibrium condensation \citep{Helling+06} and/or consumption of Mg vapors through the formation of Mg$_2$SiO$_4$ clouds in deeper layers of the atmosphere \citep{Huang+24}, remain to be explored.

The same argument can be applied to HD 189733b for which \citet{Inglis+24} recently reported the detection of SiO$_2$ clouds.
For HD~189733b, we adopt a stellar effective temperature of
$T_\star = 5050$~K \citep{Addison2019}, a stellar radius of $R_\star = 0.765R_\odot$ \citep{Addison2019}, and a semi-major axis of $a = 0.031$~AU \citep{Paredes2021}, yielding the equilibrium temperature of $T_{\rm eq} = 1210~{\rm K}$.
Again, HD 189733b lies within the region where SiO$_2$ cloud is thermodynamically stable in our stability diagram (Figure \ref{fig:boundary_silicates}).
Thus, our result is compatible with the presence of high-altitude SiO$_2$ clouds on HD~189733b reported by \citet{Inglis+24}.

One of the conundrums of WASP-17b and HD 189733b is the non-detection of Mg$_2$SiO$_4$ clouds, as the equilibrium condensation sequence predicts the formation of Mg$_2$SiO$_4$ clouds before the formation of SiO$_2$ clouds \citep{Visscher+10}.
Although WASP-17b lies close to the boundary of thermal stability for Mg$_2$SiO$_4$ in Figure \ref{fig:boundary_silicates}, both planets lie within the thermally stable region of Mg$_2$SiO$_4$ in our stability diagram.
Thus, other processes should be responsible for the non-detection of Mg$_2$SiO$_4$ clouds in WASP-17b and HD 189733b, such as the sequestration of Mg$_2$SiO$_4$ clouds below the observable regions or unrecognized processes that hinder Mg$_2$SiO$_4$ cloud formation.

\section{Discussion}\label{sec:discussion}

\subsection{Vertical Profiles of Cloud Particle Temperatures on WASP-17b}\label{sec:PT_particle}
\begin{figure*}[!htbp]
    \centering
    \includegraphics[width=0.8\linewidth]{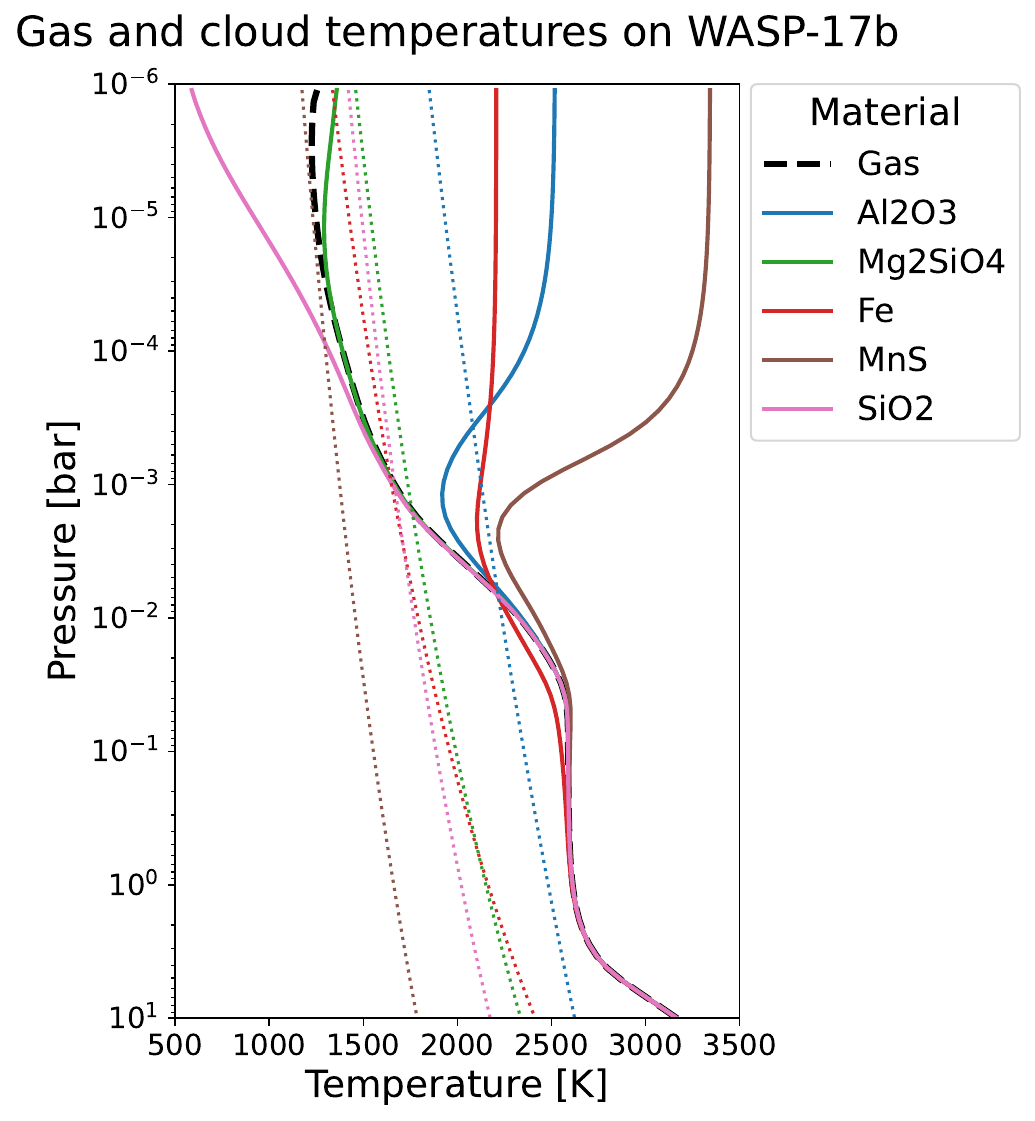}
    \caption{Gas temperature profile (black dashed) and equilibrium particle temperatures of Al$_2$O$_3$, Mg$_2$SiO$_4$, Fe, MnS, SiO$_2$ (solid lines) that potentially form as clouds on WASP-17b. The dotted lines show the condensation curves with [Fe/H] = +1.7 for each condensate. 
    We fix a particle radius to $a=0.1~\mu$m.
    }
    \label{fig:tp_wasp17}
\end{figure*}

\begin{figure*}[!htbp]
    \centering
    \includegraphics[width=\hsize]{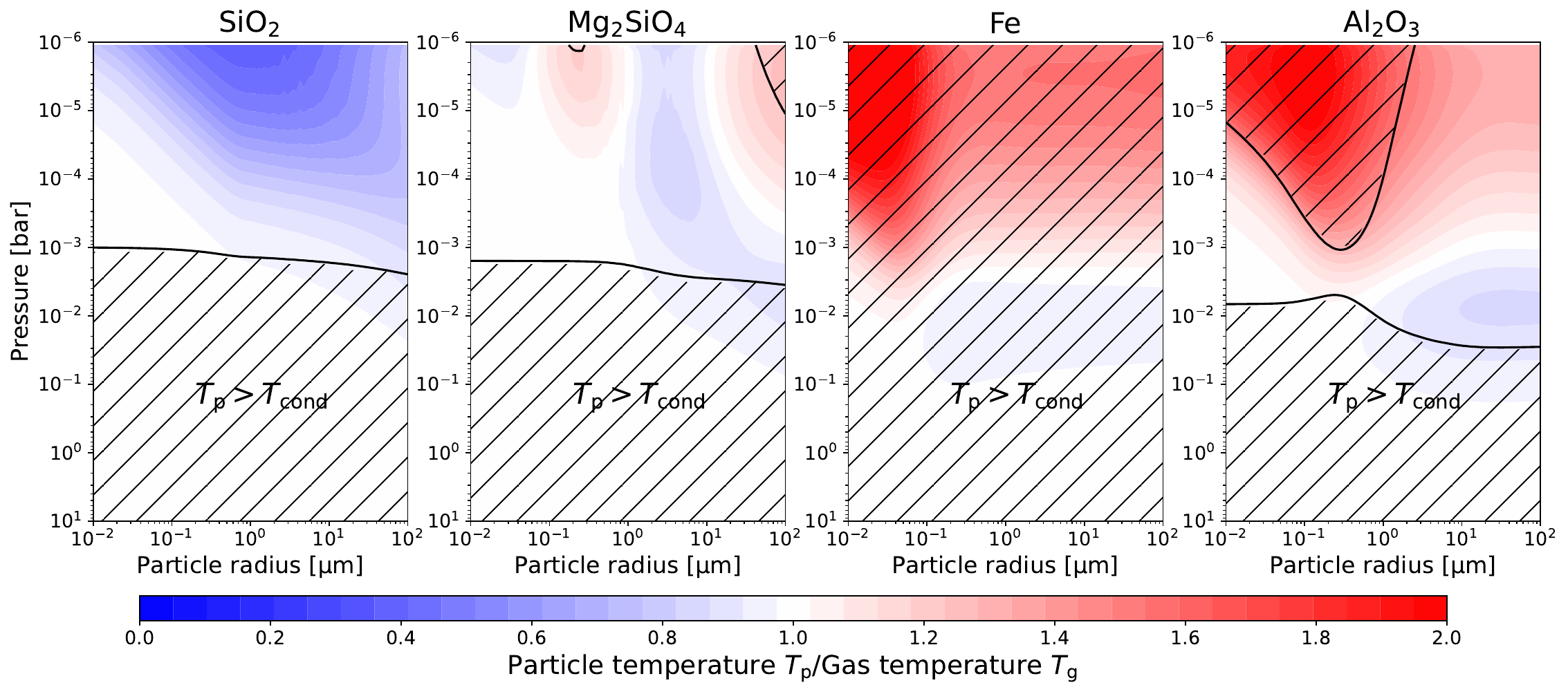}
    \caption{Particle temperature normalized by the gas temperature as a function of pressure and particle radius for SiO$_2$, Mg$_2$SiO$_4$, Fe, and Al$_2$O$_3$. The hatched region denotes the phase space where the particle temperature exceeds the condensation temperature, leading to the sublimation of clouds.
    }
    \label{fig:tp_wasp17_2}
\end{figure*}

The preceding sections focus on the radiative-equilibrium temperature of cloud particles to obtain a unified picture of particle stability that is independent of atmospheric temperature structure.
In reality, however, the particle temperature would follow the ambient gas temperature at lower dense atmospheres and switches to the radiative-equilibrium temperature as cloud particles ascend to upper tenuous atmospheres.
In this subsection, we investigate the vertical structure of cloud particle temperature on WASP-17b.

WASP-17b is the first hot Jupiter for which an absorption feature of SiO$_2$ clouds was discovered \citep{Grant+23}.
\citet{Moran+24} also suggested that Al$_2$O$_3$ clouds may exist at lower atmosphere to explain the optical transmission spectrum.
According to the atmospheric temperature structure obtained by the best-fit PICASO model of \citet{Grant+23}, WASP-17b potentially hosts MnS, SiO$_2$, Mg$_2$SiO$_4$, Fe, and Al$_2$O$_3$ clouds (Figure \ref{fig:tp_wasp17}).
Here, we calculate the vertical profiles of particle temperature for these condensates by solving Equation \eqref{eq:Energy_balance2}, where the atmospheric mean molecular mass is assumed to be $m_{\rm gas}=2.3~{\rm amu}$.
For computing $T_{\rm cond}$ to assess the thermal stability of each cloud, we assume the atmospheric metallicity of [Fe/H]$=1.7$ following the latest constraint derived by JWST NIRISS and MIRI transmission spectra \citep{Louie+24}.

As shown in Figure \ref{fig:tp_wasp17}, the particle temperature greatly diverges from the gas temperature at pressure regions of $P\lesssim10^{-4}$--${10}^{-2}~{\rm bar}$ for $a=0.1~{\rm {\mu}m}$, depending on particle compositions.
The temperature of MnS, Fe, and Al$_2$O$_3$ clouds start to diverge from the ambient gas temperature around $\sim10^{-2}~{\rm bar}$, while the temperature decoupling takes place at $P<{10}^{-4}~{\rm bar}$ for SiO$_2$ and Mg$_2$SiO$_4$.
The difference originates from the fact that MnS, Fe, and Al$_2$O$_3$ have high $Q_{\rm abs}$ values in visible wavelength, allowing the radiative processes to dominate over thermal relaxation with gas-particle collisions at relatively high pressure regions. 
Since the radiative heating is inefficient for SiO$_2$ and Mg$_2$SiO$_4$, the gas-particle collisions control the particle's energy balance until the ambient gas pressure drops sufficiently.

Intriguingly, our calculation demonstrates that many cloud species cannot maintain thermal stability on WASP-17b, except for SiO$_2$ and Mg$_2$SiO$_4$.
For example, the temperatures of MnS, Fe, and Al$_2$O$_3$ condensates significantly exceed their own condensation temperatures at $\lesssim 10^{-3}~{\rm bar}$, even though the gas temperature is cool enough to allow the formation of these clouds.
The temperature of Mg$_2$SiO$_4$ condensate is also about to reach the condensation temperature at $\lesssim{10}^{-6}~{\rm bar}$.
Among the condensates tested here, SiO$_2$ is the most stable cloud species that can exist in upper atmospheres on WASP-17b.

Since the particle temperature also depends on particle size, Figure \ref{fig:tp_wasp17_2} shows the particle temperature as a function of pressure and particle radius, where the temperature is normalized by the gas temperature.
Figure \ref{fig:tp_wasp17_2} demonstrates that the particle temperature tends to follow the gas temperature at higher pressure with smaller particle sizes.
We find that Fe cloud is unstable everywhere in the atmosphere for particle sizes of $0.01$--$100~{\rm {\mu}m}$.
Al$_2$O$_3$ clouds could stably exist if the particle size exceeds ${\sim}1~{\rm {\mu}m}$, whereas Al$_2$O$_3$ can stably exist only in a confined region of ${\sim}10^{-2}$--${10}^{-3}~{\rm bar}$ for a particle radius of $\sim0.1~{\rm {\mu}m}$ and ${\sim}10^{-2}$--${10}^{-5}~{\rm bar}$ for $\sim0.01~{\rm {\mu}m}$.
In contrast, SiO$_2$ and Mg$_2$SiO$_4$ clouds maintain thermal stability at wide range of particle sizes at $<{10}^{-3}~{\rm bar}$, though the temperature of $\sim0.1~{\rm {\mu}m}$ sized Mg$_2$SiO$_4$ particles reaches the sublimation temperature at $\sim{10}^{-6}~{\rm bar}$.
Our result aligns with the predominant SiO$_2$ composition for the high-altitude clouds on WASP-17b suggested by \citet{Grant+23}.
We also note that the present result agrees with the prediction of our stability diagram in Section \ref{sec:cond_boundary}: Figure \ref{fig:Teq_vs_Teff} does predict that MnS, Fe, and Al$_2$O$_3$ clouds are not stable in the phase space of WASP-17b, demonstrating the ability of our stability diagram to assess the cloud stability at upper atmospheres.

\subsection{Relevance to Previous Studies of Exoplanetary Clouds}

Several studies attempted to constrain cloud compositions on hot Jupiters \ko{based on phase curve observations} \citep[e.g.,][]{Oreshenko+16,Parmentier+16,Morris+24}.
Using a global circulation model with equilibrium cloud framework, \citet{Parmentier+16} suggested that not all cloud species should exist on dayside hemispheres of hot Jupiters to explain their low geometric albedo, which the authors attributed to the presence of deep cold trap.
Thermal decoupling may appear to provide an alternative explanation at first glance, since it prohibits the formation of various mineral clouds including Na$_2$S, MnS and Fe at $T_{\rm eq}>1000~{\rm K}$ (Figure \ref{fig:Teq_vs_Teff}).
However, \citet{Parmentier+16} showed that the light curve barely depends on cloud top pressure unless it lies in $>10^{-2}~{\rm bar}$, meaning that clouds at $>10^{-2}~{\rm bar}$ are typically responsible for the phase curve observations.
Since particle-gas thermal decoupling is significant only at $\lesssim 10^{-2}~{\rm bar}$ as shown in Figure \ref{fig:decoupling}, thermal decoupling would not be the major cause of the cloudless daysides on hot Jupiters.

Our results should be more relevant to the interpretations of transmission spectra that probe low-pressure upper atmospheres.
\citet{Gao+20} argued, on the basis of an aerosol microphysics model constrained by trends seen in the amplitude of H$_2$O absorption band near $1.4~{\rm {\mu}m}$ \citep{Stevenson16,Fu+17}, that the aerosol opacity of giant exoplanets is generally dominated by silicates above $T_{\rm eq} \sim 950~\mathrm{K}$ and by hydrocarbon hazes below that temperature.
\citet{Gao+20} attributed the silicate dominance primarily to high nucleation barriers of other condensable species such as Fe and MnS.
Our results reinforce the conclusion of \citet{Gao+20} that silicate aerosols should play a major role in transmission spectra of hot giant exoplanets, but for a different physical reason.
That is, sulfide condensates such as MnS, ZnS, and Na$_2$S, as well as Fe, become thermally unstable due to inefficient IR cooling in the tenuous upper atmospheres at $T_{\rm eq}>1000~{\rm K}$ (see Figure \ref{fig:Teq_vs_Teff}), leaving only silicate condensates as stable cloud species.

\ko{Thermal decoupling is in particular relevant to the interpretation regarding sulfide clouds, especially MnS clouds.}
\citet{Gao+20} \ko{used a cloud microphysical model to show that mineral clouds dominated by silicate compositions can explain the trend seen in transmission spectra of hot Jupiters. 
They invoked the ``nucleation energy barrier'' to hinder the formation of other clouds including MnS.}
\ko{However,} \citet{Parmentier+16} suggested \ko{that MnS clouds are required to explain the Kepler light curve observations for} planets with $T_{\rm eq}<1600~{\rm K}$ \citep[see also Sect. 4.2 of][]{Gao+21}.
\ko{If MnS clouds can somehow overcome the nucleation barrier, as implied by phase-curve observations, it is natural to ask whether their presence remains compatible with the transmission spectrum trend. 
Our results suggest that MnS clouds unlikely contribute to transmission spectra significantly, even if they do form and play an important role in explaining the observed phase curves.
This is because
}
MnS clouds can readily be sublimated \ko{under thermal decoupling} due to inefficient IR cooling at upper tenuous atmospheres probed by transmission spectroscopy, whereas efficient gas-cloud temperature coupling can stabilize MnS clouds at high pressure region of $\gtrsim10^{-2}~{\rm bar}$ (see Figure \ref{fig:decoupling}) probed by the Kepler light curve \citep{Parmentier+16}.
\ko{In other words, sulfide clouds like MnS may affect atmospheric observations differently at different viewing geometries.}
Future multidimensional modeling with gas-cloud temperature decoupling is warranted to revisit the spatial distribution of MnS clouds and its impacts on atmospheric observations.


\ko{
}

\ko{In addition to hot Jupiters, many sub-Neptunes are also thought to host abundant aerosols due to their featureless transmission spectra \citep[e.g.,][]{Knutson+14,Crossfield&Kreidberg17,Benneke+19,Dymont+22,Brande+24,Kahle+25,Gordon+26}.
One such example is GJ1214 b that is suggested to host a thick aerosol layer high up in the atmosphere \citep[e.g.,][]{Kreidberg+14,Kempton+23,Schlawin+24}.}
While a number of studies argued the presence of thick organic hazes \citep[e.g.,][]{Morley+15,Kawashima&Ikoma18,Adams+19,Lavvas+19,Gao+23,Ohno+25,Steinrueck+25}, the presence of condensed salt clouds, namely KCl, ZnS, and Na$_2$S, remains a possible explanation as well \citep[e.g.,][]{Kempton+12,Charnay+15,Ohno&Okuzumi18,Gao&Benneke18,Ohno+20,Christie+22,Huang+24,Malsky+25}.
Intriguingly, GJ1214's effective temperature $T_{\rm eff}=3101~{\rm K}$ and planetary equilibrium temperature $T_{\rm eq}=567~{\rm K}$ \citep{Mahajan+24} are close to the forbidden zones of Na$_2$S and ZnS clouds in Figure \ref{fig:Teq_vs_Teff}, motivating a close inspection on the cloud stability on GJ1214 b under thermal decoupling.

\begin{figure}[t]
    \centering
    \includegraphics[width=\hsize]{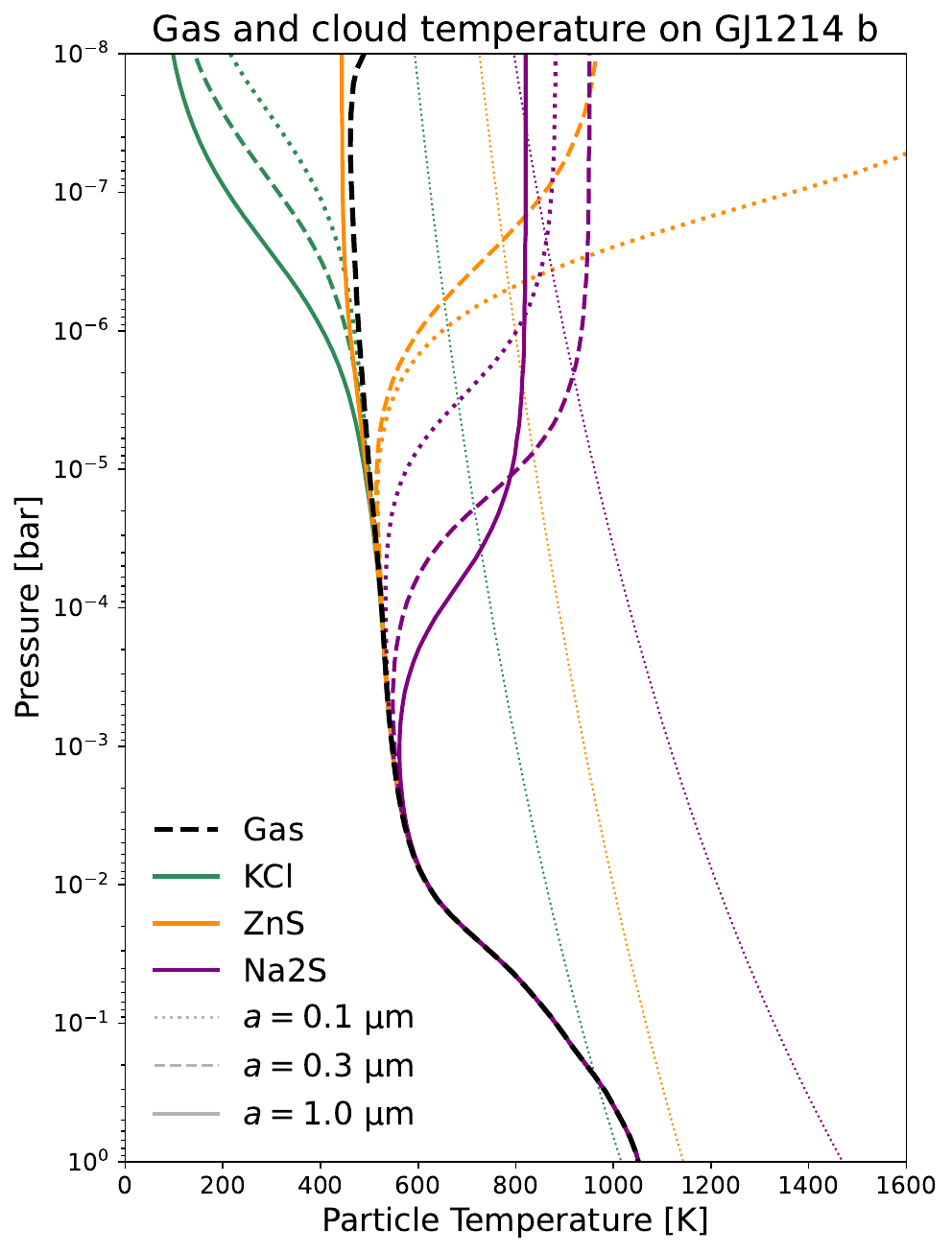}
    \caption{
    Vertical profiles of salt cloud temperatures on the sub-Neptune GJ1214 b.
    The green, orange, and purple lines show the temperatures of KCl, ZnS, and Na$_2$S clouds, respectively, along with their condensation temperatures for [M/H]$=3.0$.
    The dashed, dash-dot, and solid colored lines show the temperatures for particle radii of $a=0.1$, $0.3$ and $1.0~{\rm {\mu}m}$, respectively.
    }
    \label{fig:tp_GJ1214b}
\end{figure}

We calculate the vertical profiles of particle temperatures for KCl, ZnS and Na$_2$S clouds on GJ1214 b in Figure \ref{fig:tp_GJ1214b}, where we adopt the gas temperature profile for [M/H]$=3.0$ from \citet{Ohno+25} and mean molecular weight of $18~{\rm amu}$ to mimic extreme atmospheric metallicity suggested by JWST \citep{Kempton+23,Gao+23,Schlawin+24,Ohno+25}.
While KCl clouds exhibit temperatures cooler than the gas temperature, ZnS and Na$_2$S clouds exhibit temperatures systematically hotter than the gas temperature at $P\lesssim10^{-4}~{\rm bar}$.
In particular, temperatures of ZnS and Na$_2$S clouds exceed their condensation temperatures at $P\lesssim3$--$0.3\times10^{-6}~{\rm bar}$ for particle radii of $a=0.1$ and $0.3~{\rm {\mu}m}$, suggesting that submicron ZnS and Na$_2$S cloud particles may be unable to exist in such low pressure regions.
ZnS and Na$_2$S cloud particles with $a\gtrsim1~{\rm {\mu}m}$ could still stably exist, though extreme vertical mixing would be needed to sustain micron-sized particles in low pressure regions \citep{Gao&Benneke18,Huang+24}. 
Further analysis would be needed to assess cloud stability if aerosols on GJ1214 b are the salt clouds condensed on organic hazes \citep{Lavvas+24,Malsky+25}, though it is beyond the scope of this study.

\subsection{Caveats and Model Assumptions}



In this study, we have introduced several simplifications to make the calculations tractable.
Although the following assumptions deliberately neglect a number of phenomena capable of occurring in exoplanetary atmospheres, they maintain a framework for a valid comparison across different condensates without delving into detailed atmospheric properties of a specific planet.
In what follows, we discuss the validity of the simplifications made in this study.

\subsubsection*{Shielding effects}
Our calculations assume that cloud particles in the upper atmosphere are directly exposed to stellar irradiation.  
This assumption is appropriate for optically thin regions where stellar radiation can reach individual particles without significant shielding.
When multiple optically-thick cloud layers exist in an atmosphere, the uppermost radiatively-thermally stable clouds may stabilize the lower unstable clouds by shielding stellar light, though detailed radiative transfer calculation will be needed to assess this possibility.

\subsubsection*{Multidimensionality}
While our present analysis focuses on 0D (Section \ref{sec:cond_boundary}) and 1D (Section \ref{sec:PT_particle}) atmospheres, the real atmospheres are 3D in nature.
Because stellar radiation never heats cloud particles on the nightside hemisphere, radiatively unstable species such as MnS and ZnS can condense back to cloud particles even if the inefficient infrared cooling prohibits their presence on the dayside.
The recondensed clouds would travel the west part of the dayside to some extent and eventually sublimate through the radiative process investigated in this study.

\begin{figure}[t]
    \centering
    \includegraphics[width=\hsize]{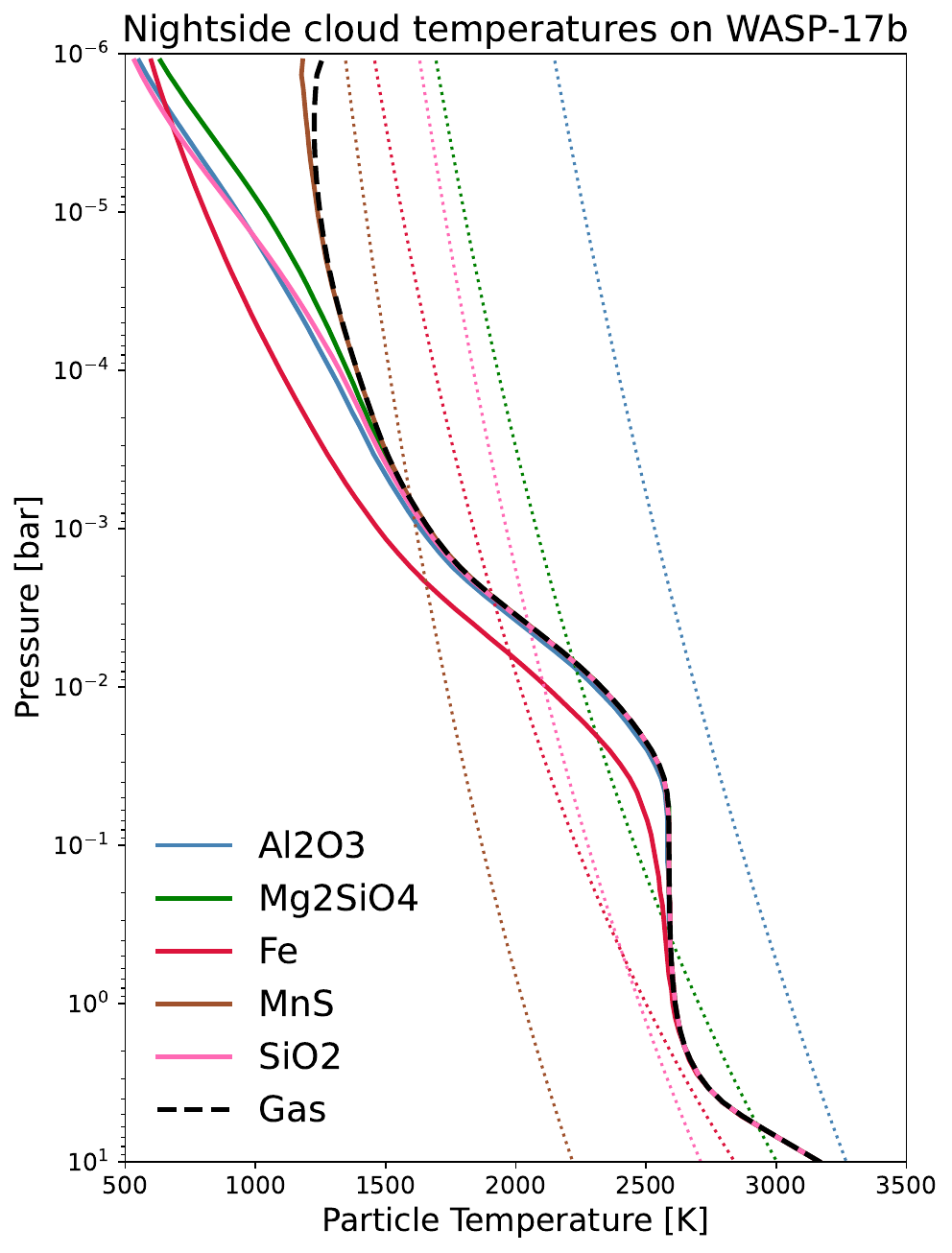}
    \caption{
    Same as Figure~\ref{fig:tp_wasp17}, but for an idealized nightside limit of WASP-17b where the stellar radiation heating is excluded.
    The particle radius is fixed to $a=0.1~\mu$m.
    }
    \label{fig:tp_wasp17_nightside}
\end{figure}

To illustrate this nightside effect, we performed an idealized nightside-limit calculation for WASP-17b, as shown in Figure~\ref{fig:tp_wasp17_nightside}. 
We solve the same energy balance equation (Equation \ref{eq:Energy_balance}) as in Figure~\ref{fig:tp_wasp17} using the same temperature--pressure profile but omit the stellar irradiation heating term.
Although cloud particles would still be heated by planetary radiation in reality, we ignore its contribution to assess the possible coldest temperature of cloud particles.
In this limit, particles in the low-pressure upper atmosphere become much cooler than in the dayside case, suggesting that the nightside could provide cold surfaces for the re-condensation of species that are thermally unstable on the irradiated dayside.

\ko{In the case of WASP-17b, for instance, Al$_2$O$_3$ clouds can form on the nightside, and may persist on the dayside when transported \ko{there by} the global circulation if those cloud particles grew to sufficiently large sizes of $>1~{\rm {\mu}m}$ on the nightside (see Figure \ref{fig:tp_wasp17_2}).}
Alternatively, the size distribution of Al$_2$O$_3$ cloud particles may evolve to a bimodal distribution once they enter the dayside from the nightside, since only particles with a certain size range are unstable in case of Al$_2$O$_3$.
Our results motivate future multidimensional studies combining atmospheric dynamics with cloud microphysics \citep[e.g.,][]{Lee+16,Lines+18,Lee23,Powell&Zhang24,Lee&Ohno25} to incorporate gas-cloud temperature decoupling.

\subsubsection*{Particle shape}
    We have assumed perfectly spherical cloud particles. 
    Actual condensate particles in exoplanetary atmospheres may exhibit irregular or aggregate morphologies \citep{Ohno+20,Samra+20,Hamil+24,Hamil+25,Vahidinia+24,Lodge+24,Lodge+25,Moran+25}. 
    However, we expect that the present results would hold for aggregate particles because their absorption efficiency $Q_{\rm abs}$ retains the same spectral behaviors as that of a spherical particle \citep{Kataoka+14}.
    Since irregularly shaped particles also yield a similar spectral behavior in $Q_{\rm abs}$ \citep{Lin+25}, we expect that our main results hold for non-spherical aerosols.

\subsubsection*{Optical constants}
    A complex refractive index was adopted for the calculations, and its dependence on temperature was not considered.
    Several laboratory studies reported that the peak positions of silicate features at $\gtrsim 9~{\rm {\mu}m}$ depend on temperature \citep[e.g.,][]{Koike+06,Zeidler+13,Zeidler+15}.
    However, the peak shifts are only submicron scale in the temperature range of $10$--$928~{\rm K}$, which are unlikely to greatly change the results of the present study.
    Different polymorphs alter the refractive indices of crystalline condensates as argued for SiO$_2$ in \citet{Moran+24}, but it also only causes small shifts in the infrared features from vibrational modes and unlikely affects the particle temperature significantly. 

\begin{figure}[t]
    \centering
    \includegraphics[width=\hsize]{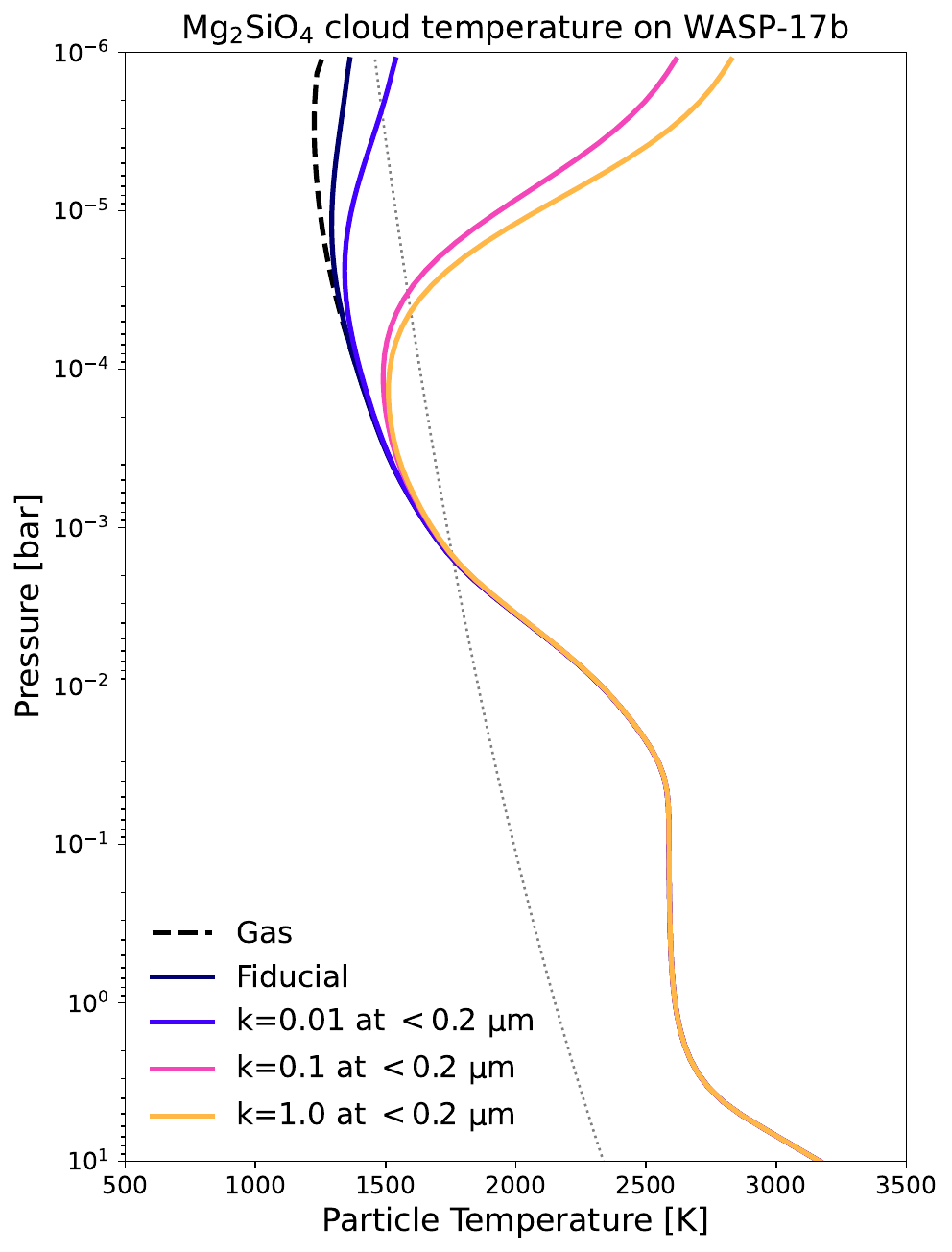}
    \caption{Same as Figure \ref{fig:tp_wasp17}, but for Mg$_2$SiO$_4$ clouds. Different colored lines show the Mg$_2$SiO$_4$ cloud temperature for different $k$ values at $<0.2~{\rm {\mu}m}$.
    }
    \label{fig:test_Mg2SiO4}
\end{figure}

    We assumed constant values of the refractive indices $n$ and $k$ for Mg$_2$SiO$_4$ and Al$_2$O$_3$ from $0.2~{\rm {\mu}m}$ down to $0.1~{\rm {\mu}m}$, equal to their values at $0.2~{\rm {\mu}m}$, due to the lack of experimental data.
    This treatment hardly affects the result of Al$_2$O$_3$ because stellar heating occurs at visible wavelengths, whereas the result of Mg$_2$SiO$_4$ depends on the actual value of refractive index at $<0.2~{\rm {\mu}m}$.
    Figure \ref{fig:test_Mg2SiO4} shows the vertical temperature structure of Mg$_2$SiO$_4$ clouds on WASP-17b for different values of the imaginary part of the refractive index $k$ at wavelengths of $<0.2~{\rm {\mu}m}$.
    Because Mg$_2$SiO$_4$ has relatively low $Q_{\rm abs}$ values in infrared wavelengths, higher $k$ values at $<0.2~{\rm {\mu}m}$ can cause significant heating over cooling, leading Mg$_2$SiO$_4$ clouds to sublimate at $\lesssim3\times{10}^{-5}~{\rm bar}$ for $k\gtrsim0.1$.
    Thus, it is vital to experimentally measure the refractive index in UV wavelengths for correctly estimating Mg$_2$SiO$_4$ particle temperature in upper tenuous atmospheres.


\subsubsection*{Sublimation kinetics}
    In evaluating the thermal stability of cloud particles, we have assumed that particles evaporate instantaneously once the particle temperature exceeds the thermodynamic condensation temperature. 
    In reality, evaporation and condensation proceed on finite microphysical timescales that depend on local supersaturation, particle size, and transport. 
    However, once the local temperature exceeds the equilibrium condensation temperature, the evaporation timescale of micron-sized condensates is typically extremely short compared to dynamical (advective or mixing) timescales \citep[see e.g., Figure~2 of][]{Powell+18}. 
    In this regime, the condensate abundance adjusts so rapidly that the system effectively behaves as if evaporation were instantaneous, justifying our use of a sharp stability boundary in the $(T_{\rm eff},T_{\rm eq})$ plane. 

\subsubsection*{Stellar spectrum}
    Although we have approximated the stellar spectrum by blackbody, we caution that the actual stellar spectrum contains a much stronger UV portion.
    This potentially causes a hotter temperature than our calculation for some condensates, especially SiO$_2$, Mg$_2$SiO$_4$, and KCl for which stellar radiation heating in UV band substantially contributes to the net heating rate.
    Meanwhile, stellar UV photons are greatly attenuated by photodissociation of gas phase molecules such as H$_2$O, CO, and CH$_4$.
    Detailed knowledge on stellar spectrum and atmospheric UV environments would be needed to better understand the temperature of those condensates, which is left to future studies.

\subsubsection*{FEEDBACK ON AMBIENT GAS Temperature}
In this study, we have focused on the thermal stability of cloud particles by solving for the particle temperature under a prescribed radiation field and gas temperature structure. This treatment does not self-consistently include the thermal feedback of cloud particles on the ambient gas. In reality, particles affect ambient gas temperature in multiple ways.
Cloud particles absorb or scatter the incident stellar \ko{flux} and outgoing planetary emission, which produces distinct gas temperature profiles from those with clear atmospheres \citep[e.g.,][]{Heng+12,Morley+15,Lavvas2021,Ohno24}.
Particles whose temperatures differ from that of the surrounding gas can exchange energy with the atmosphere through collisions and radiative transfer, potentially contributing to atmospheric heating or cooling. A striking example is provided by Pluto, where haze particles play a dominant role in the atmospheric radiative energy balance through its efficient infrared cooling \citep{Zhang+17_Pluto,Bertland+25}. 
It remains unclear to what extent the similar feedback affects the gas temperature profiles on hot exoplanets.
A self-consistent treatment coupling the thermal balance of cloud particles and the radiative-convective structure of the atmosphere will be important for future studies.

\section{Conclusion}\label{sec:summary}

We have presented a radiative–thermodynamic framework to assess the temperature and thermal stability of cloud-forming condensates in the tenuous upper atmospheres of exoplanets. Instead of assuming that cloud particles share the local gas temperature, we have calculated the particle temperature by solving the full radiative energy balance for individual grains using wavelength dependent absorption efficiencies $Q_{\rm abs}(a,\lambda)$ computed by the Mie theory.
Comparing these particle temperatures with condensation temperatures derived by equilibrium chemistry models, we have investigated critical planetary equilibrium temperatures above which clouds cannot maintain cool enough temperatures for eight commonly discussed condensates (MnS, KCl, SiO$_2$, Mg$_2$SiO$_4$, Fe, Na$_2$S, Al$_2$O$_3$, ZnS).
Our calculations yield the following principal physical insights:

\begin{enumerate}
    \item The condensates separate into three classes in terms of the behavior of particle temperatures: (i) the high--temperature condensates Fe and Al$_2$O$_3$ (with Na$_2$S also closely following this behavior) behave similarly to moderately superheated blackbodies as a consequence of efficient heating and cooling at both optical and infrared wavelengths; (ii) the optical-heating-dominated group MnS and ZnS are heating--dominated and exhibit a significantly hot temperature compared to blackbody due to efficient optical heating along with inefficient infrared cooling; and (iii) the IR-cooling-dominated group clouds SiO$_2$ and Mg$_2$SiO$_4$ remain systematically cooler than the blackbody temperature $T_{\rm bb}$, reflecting their reduced absorption efficiency to stellar irradiation.

    \item Our results show a clear size dependence: smaller particles generally reach higher temperatures, while larger grains remain cooler. This trend arises because $Q_{\rm abs}$ of small grains is strongly suppressed at the longer wavelengths relevant for radiative cooling, making their cooling inefficient and causing $T_{\rm p}$ to deviate from $T_{\rm bb}$, particularly for the high--temperature condensates. An exception is KCl, which exhibits a local temperature maximum around $a \simeq 0.03~\mu$m due to enhanced absorption in the heating band at $\lambda \sim 0.2~\mu$m, highlighting that species--specific features in $Q_{\rm abs}$ can locally modify this general rule.

    \item In general, particle temperature increases with increasing stellar effective temperature even if the planetary equilibrium temperature is the same. This is because the stellar radiation peak shifts to shorter wavelengths at which high absorption efficiency allows cloud particles to efficiently absorb stellar light and heat themselves.

    \item Under certain planetary equilibrium temperature $T_{\rm eq}$ and stellar effective temperature $T_{\rm eff}$, some condensates cannot maintain cool enough temperature to avoid sublimation irrespective of ambient gas temperature. 
    This phenomenon introduces a ''forbidden zone`` in $T_{\rm eq}$---$T_{\rm eff}$ space that dictates the thermal stability of cloud particles in upper tenuous atmospheres.  

    \item Optical-heating-dominated group condensates, especially ZnS and MnS, tend to be thermally unstable due to inefficient infrared cooling. Their radiative-equilibrium temperatures readily exceed sublimation temperatures even on temperate planets of $T_{\rm eq}\sim300~{\rm K}$.
    This result suggests the potential absence of optical-heating-dominated group clouds in upper atmospheres on hot exoplanets even if the atmospheric temperature allows the formation of those clouds.

    \item IR-cooling-dominated group clouds, especially SiO$_2$, can maintain cool enough temperature even on ultra-hot planets of $T_{\rm eq}\gtrsim2000~{\rm K}$ under radiative equilibrium. The high stability of IR-cooling-dominated group clouds originates from extremely low absorption efficiency in visible wavelengths, which suppresses the stellar radiation heating. This result suggests that, in upper atmospheres, these clouds can prevalently exist in a wide range of stellar radiation conditions compared to other mineral clouds.
    
    \item For several species (notably KCl, SiO$_2$, Mg$_2$SiO$_4$) the forbidden zones develop non–monotonic structure—local maxima in the threshold $T_{\rm eq}$ as a function of $T_{\rm eff}$, implying that these clouds preferentially exist on planets around certain types of  stars. In particular, the stability zone of KCl clouds greatly expands around the stars of $T_{\rm eff}<5000~{\rm K}$ and peak at $T_{\rm eff}\sim3500~{\rm K}$, indicating that KCl clouds may preferentially exist on planets around K- and M-type stars. 


    \item 
    WASP-17b and HD 189733b fall below the SiO$_2$ stability boundary, consistent with their observed silica spectral features. In addition, we have calculated the vertical profiles of cloud particle temperatures on WASP-17b, confirming the stability of SiO$_2$ clouds for a wide range of pressure and particle sizes. Furthermore, we find that Fe clouds cannot exist on WASP-17b for particle sizes of $0.01$--$100~{\rm {\mu}m}$, and Al$_2$O$_3$ clouds can exist only in a confined pressure range of $\gtrsim{10}^{-3}$--${10}^{-5}~{\rm bar}$ for particle sizes of $\lesssim1~{\rm {\mu}m}$.
\end{enumerate}

For irradiated exoplanets, our results suggest that thermal stability of high-altitude clouds does not follow a simple ``farther is safer'' heuristic: whether a cloud survives at altitude depends as much on the spectral overlap between stellar photons and $Q_{\rm abs}$ as on the overall irradiation level. 
This is a unique characteristic of exoplanetary clouds and suggests the possibility of distinct cloud mineralogy between brown dwarfs and exoplanets.
Stellar spectral properties play a critical role in dictating whether clouds can stably exist, which provides testable predictions for forthcoming surveys of JWST and Ariel.

Our approach omits detailed microphysics such as nucleation, condensation and evaporation, coagulation, and vertical/horizontal transport.
Incorporating thermal non-equilibrium of cloud particles to microphysical models will be of great interest to better understand the thermal stability and spatial distributions of exoplanetary clouds.
Such theoretical effort would be crucial, in particular for understanding inhomogeneous clouds revealed by JWST for multiple exoplanets \citep{Espinoza+24,Murphy+24,Murphy+25,Schlawin+24,Tada+25,Coulombe+25,Mukherjee+25,Fu+25_cloud_assymmetry}.


\section*{Acknowledgements}
We thank the anonymous reviewer for a number of insightful comments that greatly improved the quality of this article.
We thank Hideaki Matsuzaki for his continuous support on this project and Ryo Tazaki, Tetsuo Taki, and Takato Tokuno for heuristic comments.
K.O. thanks Jonathan Fortney for encouraging conversation.
This work was supported by the JSPS KAKENHI Grant number JP23K19072 and JP26K17222.

\section*{Author Contributions}
TMS-N conducted calculations for the main results and wrote the manuscript. KO conceived the research idea and edited the manuscript. Both authors contributed to develop the numerical code.

\appendix

\section{Absorption efficiencies for all condensates}\label{app:qabs}
\begin{figure*}[htbp]
\centering
\includegraphics[width=0.32\textwidth]{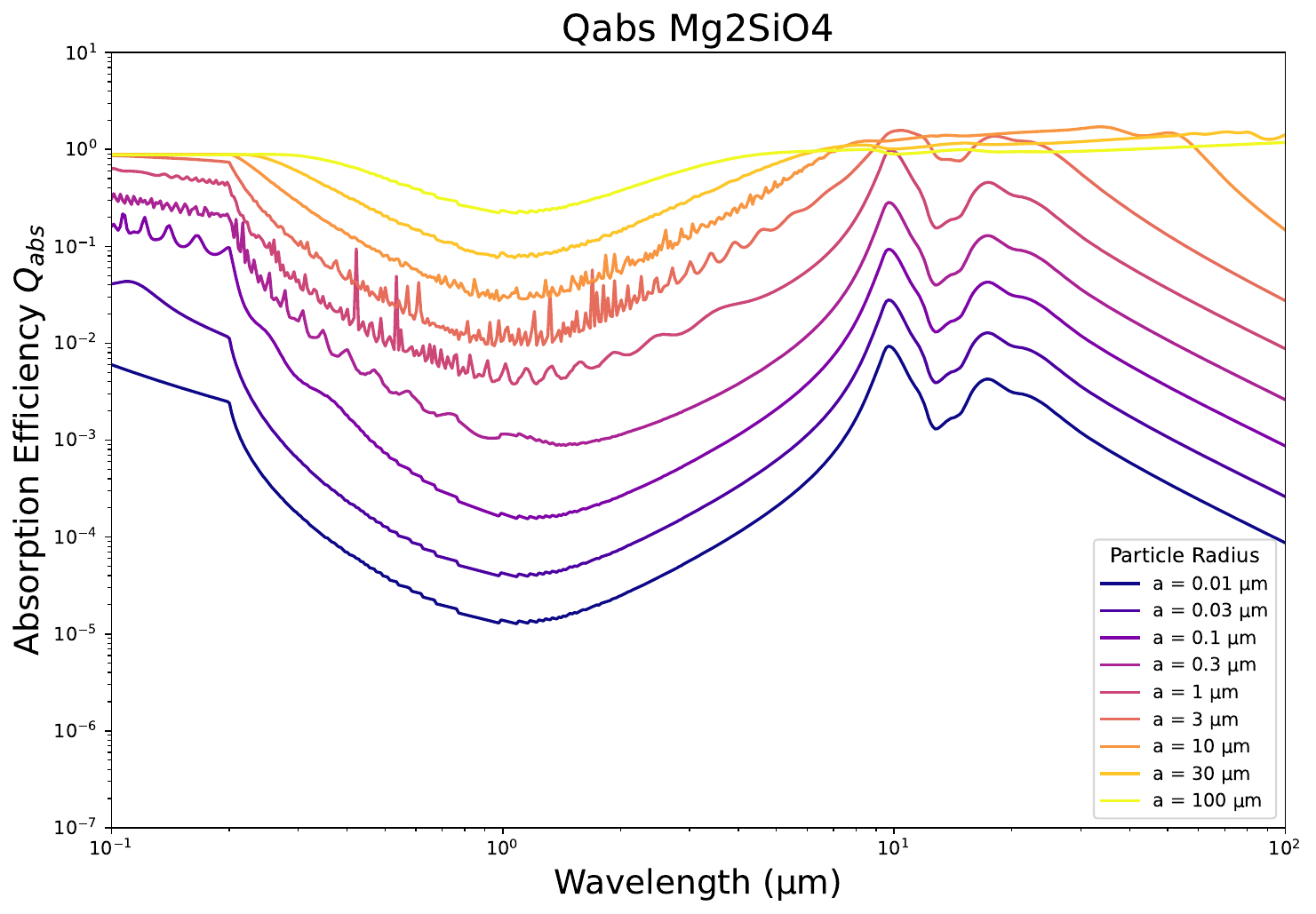}
\includegraphics[width=0.32\textwidth]{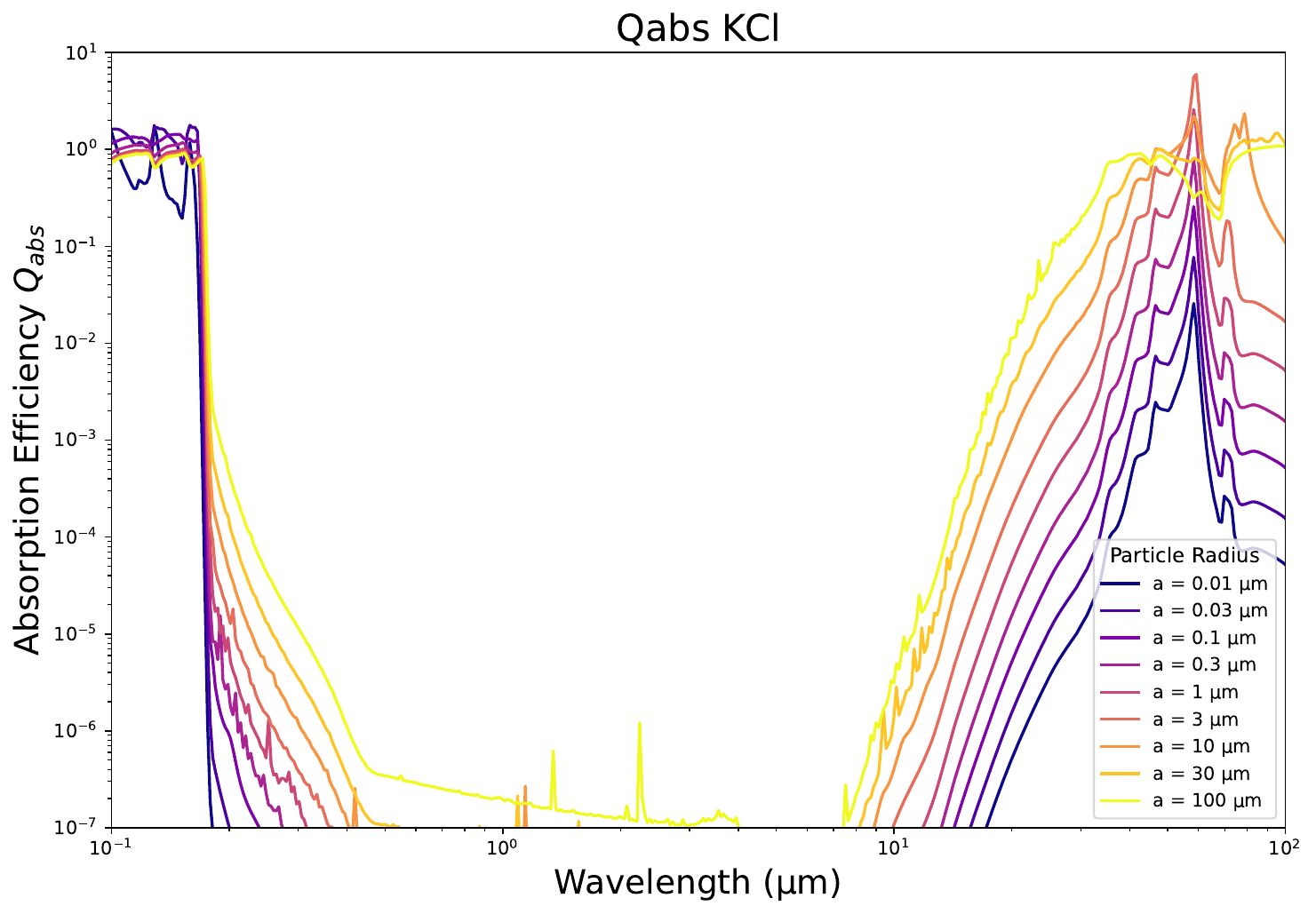}
\includegraphics[width=0.32\textwidth]{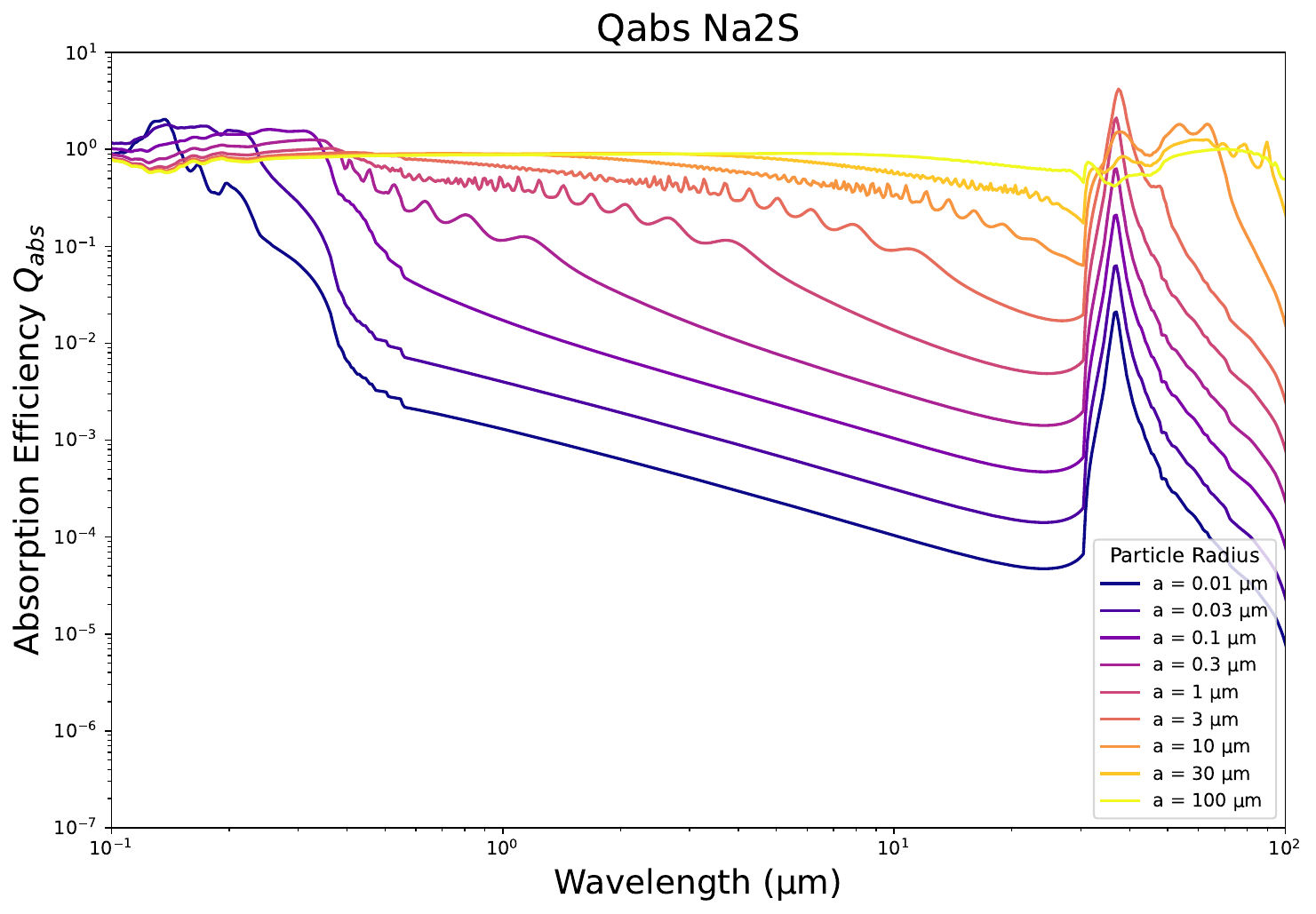}

\includegraphics[width=0.32\textwidth]{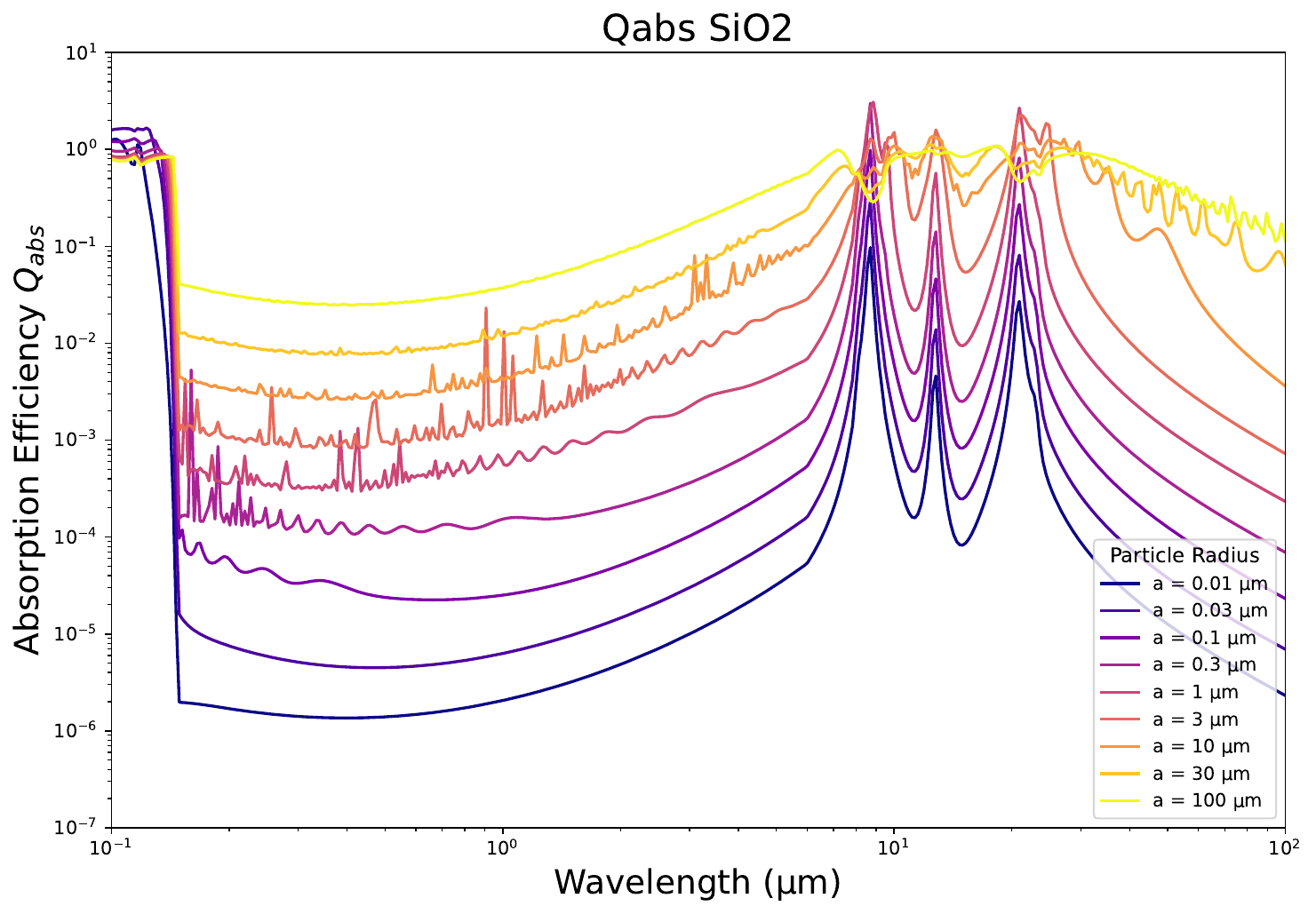}
\includegraphics[width=0.32\textwidth]{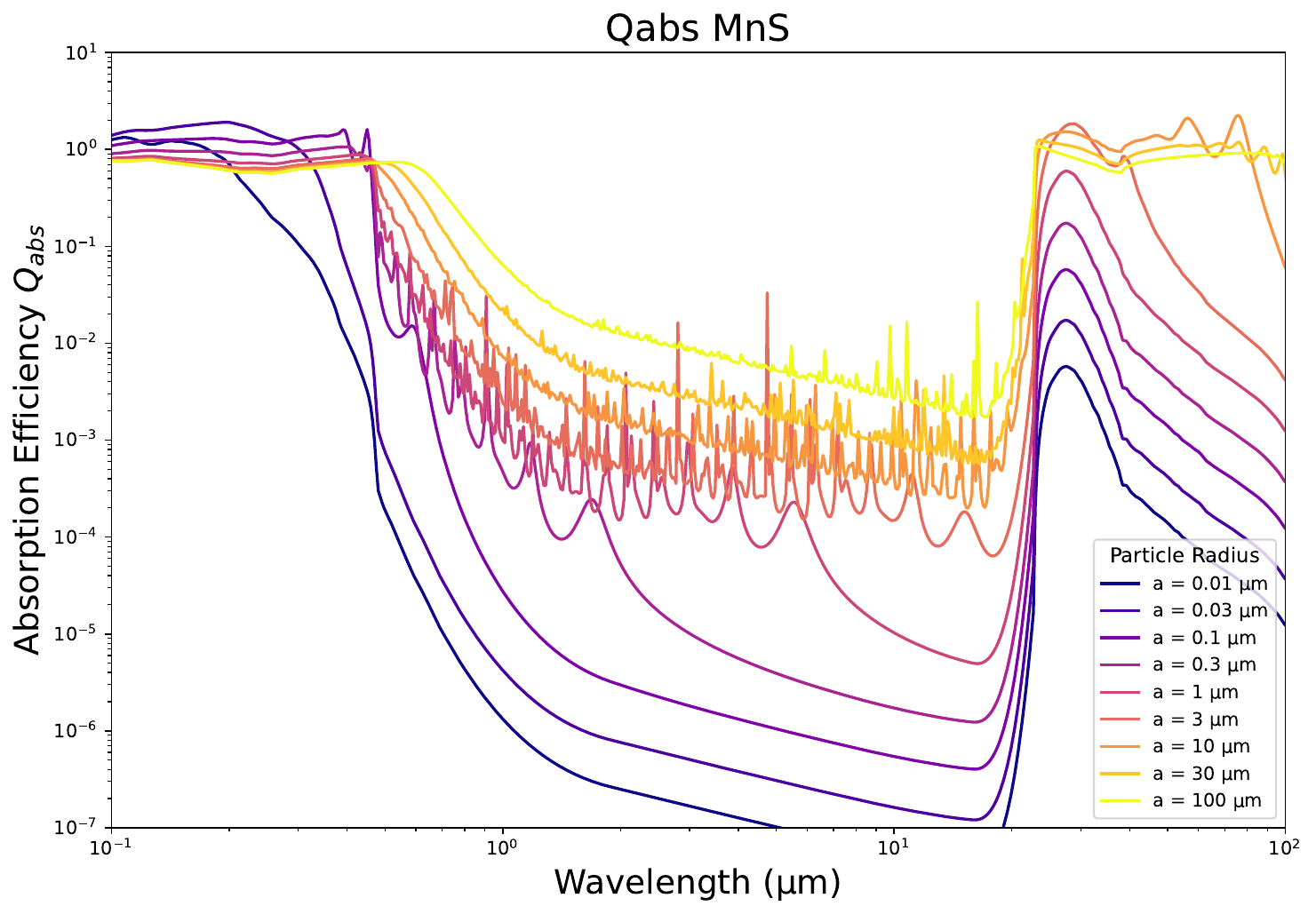}
\includegraphics[width=0.32\textwidth]{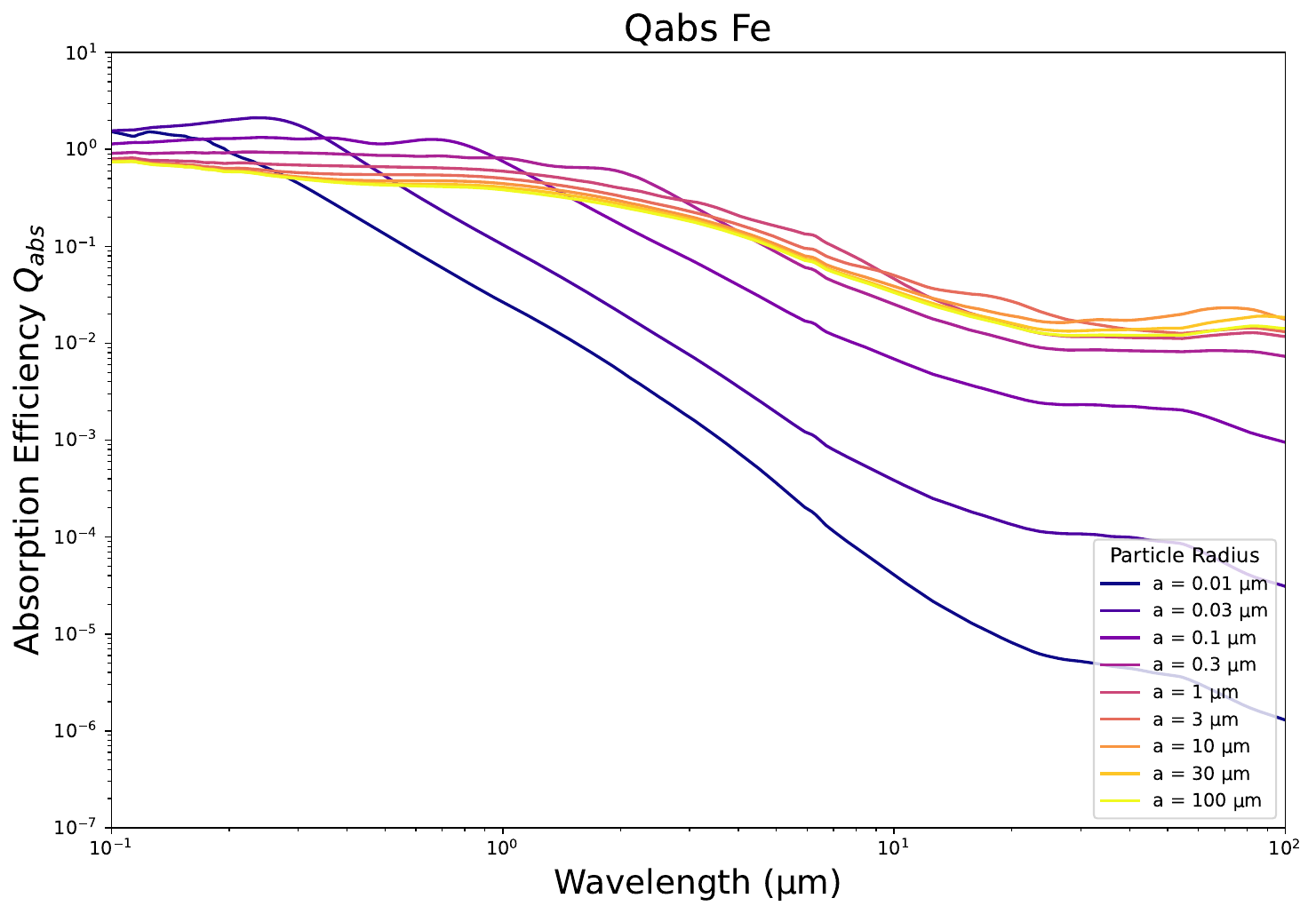}

\includegraphics[width=0.32\textwidth]{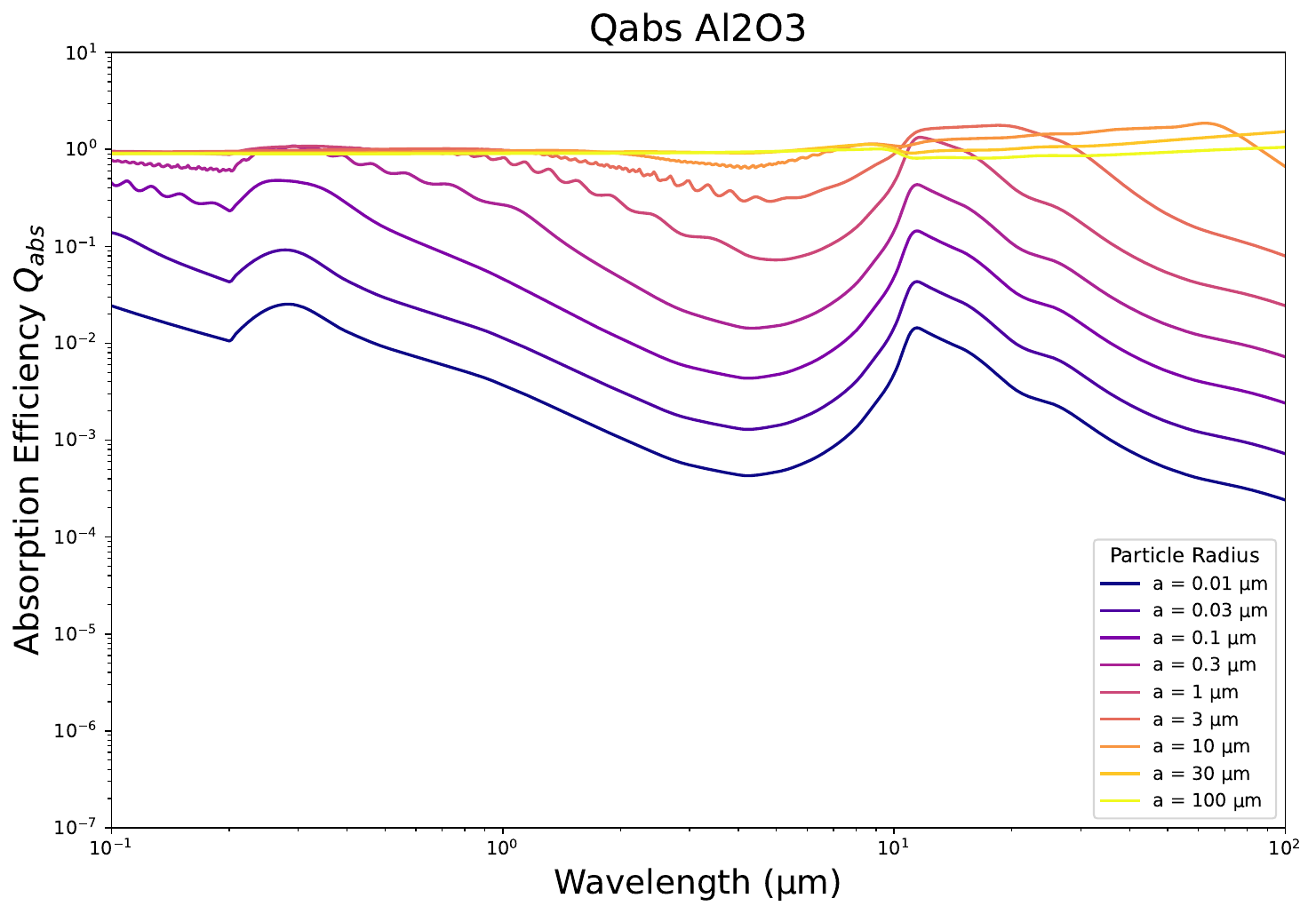}
\includegraphics[width=0.32\textwidth]{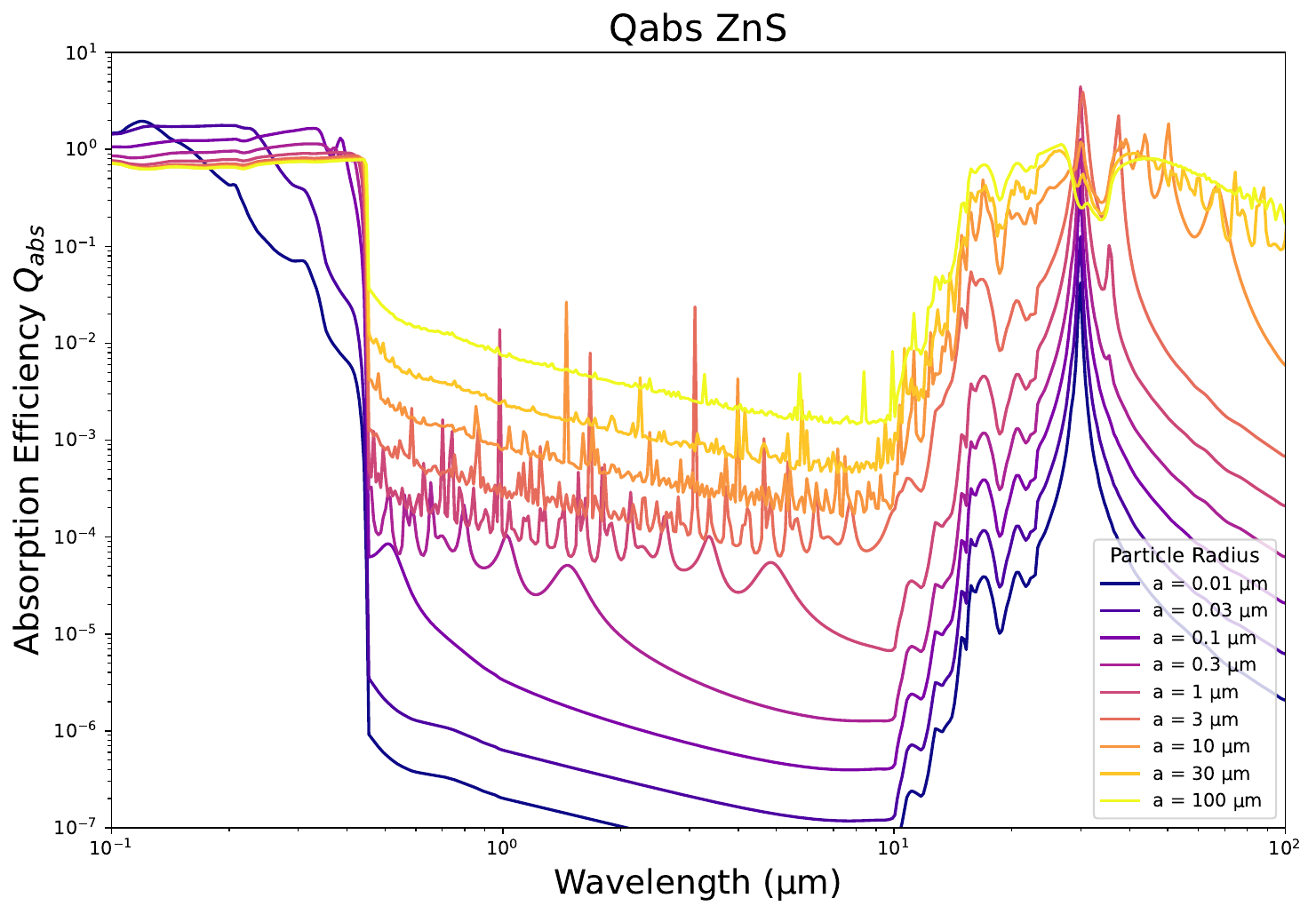}

\caption{
Wavelength-dependent absorption efficiency $Q_{\rm abs}(a,\lambda)$ for the eight condensates
(MnS, KCl, SiO$_2$, Mg$_2$SiO$_4$, Fe, Na$_2$S, Al$_2$O$_3$, and ZnS) considered in this study.
Different colored lines correspond to different particle radii.
}
\label{fig:Qabs_all_full}
\end{figure*}

Figure~\ref{fig:Qabs_all_full} summarizes the wavelength-dependent
absorption efficiencies of all condensates considered in this study at several particle radii.

\bibliographystyle{aa}
\bibliography{references}

@ARTICLE{Kahle+25,
       author = {{Kahle}, K. Angelique and {Blecic}, Jasmina and {Ashtari}, Reza and {Kreidberg}, Laura and {Kawashima}, Yui and {Cubillos}, Patricio E. and {Deming}, Drake and {Jenkins}, James S. and {Molli{\`e}re}, Paul and {Redfield}, Seth and {Tian}, Qiushi Chris and {Vines}, Jose I. and {Wilson}, David J. and {Acu{\~n}a}, Lorena and {Bitsch}, Bertram and {Brande}, Jonathan and {France}, Kevin and {Stevenson}, Kevin B. and {Crossfield}, Ian J.~M. and {Daylan}, Tansu and {Dobbs-Dixon}, Ian and {Evans-Soma}, Thomas M. and {Gapp}, Cyril and {Garc{\'\i}a Mu{\~n}oz}, Antonio and {Heng}, Kevin and {Hu}, Renyu and {Shkolnik}, Evgenya L. and {Stassun}, Keivan G. and {Teske}, Johanna},
        title = "{The SPACE Program: I. The featureless spectrum of HD 86226 c challenges sub-Neptune atmosphere trends}",
      journal = {\aap},
         year = 2025,
        month = sep,
       volume = {701},
          eid = {A184},
        pages = {A184},
          doi = {10.1051/0004-6361/202554916},
archivePrefix = {arXiv},
       eprint = {2507.13439},
 primaryClass = {astro-ph.EP},
       adsurl = {https://ui.adsabs.harvard.edu/abs/2025A&A...701A.184K}
}

@ARTICLE{Gordon+26,
       author = {{Gordon}, Tyler A. and {Batalha}, Natalie M. and {Batalha}, Natasha E. and {Aguichine}, Artyom and {Gagnebin}, Anna and {Kirk}, James and {L{\'o}pez-Morales}, Mercedes and {Meech}, Annabella and {Scarsdale}, Nicholas and {Teske}, Johanna and {Wallack}, Nicole L. and {Wogan}, Nicholas},
        title = "{JWST COMPASS: Insights into the Systematic Noise Properties of NIRSpec/G395H from a Uniform Reanalysis of Seven Transmission Spectra}",
      journal = {\aj},
         year = 2026,
        month = mar,
       volume = {171},
       number = {3},
          eid = {178},
        pages = {178},
          doi = {10.3847/1538-3881/ae3de9},
archivePrefix = {arXiv},
       eprint = {2511.18196},
 primaryClass = {astro-ph.EP},
       adsurl = {https://ui.adsabs.harvard.edu/abs/2026AJ....171..178G}
}

@ARTICLE{Dymont+22,
       author = {{Dymont}, Austin H. and {Yu}, Xinting and {Ohno}, Kazumasa and {Zhang}, Xi and {Fortney}, Jonathan J. and {Thorngren}, Daniel and {Dickinson}, Connor},
        title = "{Cleaning Our Hazy Lens: Exploring Trends in Transmission Spectra of Warm Exoplanets}",
      journal = {\apj},
         year = 2022,
        month = oct,
       volume = {937},
       number = {2},
          eid = {90},
        pages = {90},
          doi = {10.3847/1538-4357/ac7f40},
archivePrefix = {arXiv},
       eprint = {2112.06173},
 primaryClass = {astro-ph.EP},
       adsurl = {https://ui.adsabs.harvard.edu/abs/2022ApJ...937...90D}
}

@ARTICLE{Crossfield&Kreidberg17,
       author = {{Crossfield}, Ian J.~M. and {Kreidberg}, Laura},
        title = "{Trends in Atmospheric Properties of Neptune-size Exoplanets}",
      journal = {\aj},
         year = 2017,
        month = dec,
       volume = {154},
       number = {6},
          eid = {261},
        pages = {261},
          doi = {10.3847/1538-3881/aa9279},
archivePrefix = {arXiv},
       eprint = {1708.00016},
 primaryClass = {astro-ph.EP},
       adsurl = {https://ui.adsabs.harvard.edu/abs/2017AJ....154..261C}
}

@ARTICLE{Benneke+19,
       author = {{Benneke}, Bj{\"o}rn and {Knutson}, Heather A. and {Lothringer}, Joshua and {Crossfield}, Ian J.~M. and {Moses}, Julianne I. and {Morley}, Caroline and {Kreidberg}, Laura and {Fulton}, Benjamin J. and {Dragomir}, Diana and {Howard}, Andrew W. and {Wong}, Ian and {D{\'e}sert}, Jean-Michel and {McCullough}, Peter R. and {Kempton}, Eliza M.-R. and {Fortney}, Jonathan and {Gilliland}, Ronald and {Deming}, Drake and {Kammer}, Joshua},
        title = "{A sub-Neptune exoplanet with a low-metallicity methane-depleted atmosphere and Mie-scattering clouds}",
      journal = {Nature Astronomy},
         year = 2019,
        month = jul,
       volume = {3},
        pages = {813-821},
          doi = {10.1038/s41550-019-0800-5},
archivePrefix = {arXiv},
       eprint = {1907.00449},
 primaryClass = {astro-ph.EP},
       adsurl = {https://ui.adsabs.harvard.edu/abs/2019NatAs...3..813B}
}

@ARTICLE{Knutson+14,
       author = {{Knutson}, Heather A. and {Benneke}, Bj{\"o}rn and {Deming}, Drake and {Homeier}, Derek},
        title = "{A featureless transmission spectrum for the Neptune-mass exoplanet GJ436b}",
      journal = {\nat},
         year = 2014,
        month = jan,
       volume = {505},
       number = {7481},
        pages = {66-68},
          doi = {10.1038/nature12887},
archivePrefix = {arXiv},
       eprint = {1401.3350},
 primaryClass = {astro-ph.EP},
       adsurl = {https://ui.adsabs.harvard.edu/abs/2014Natur.505...66K}
}

@ARTICLE{Kiefer+26,
       author = {{Kiefer}, Sven and {Morley}, Caroline V. and {Rowland}, Melanie J.},
        title = "{Connecting JWST Silicate Cloud Observations to Exoplanet Cloud Microphysics with Nimbus}",
      journal = {\apj},
         year = 2026,
        month = apr,
       volume = {1001},
       number = {1},
          eid = {98},
        pages = {98},
          doi = {10.3847/1538-4357/ae5101},
archivePrefix = {arXiv},
       eprint = {2603.13167},
 primaryClass = {astro-ph.EP},
       adsurl = {https://ui.adsabs.harvard.edu/abs/2026ApJ..1001...98K}
}

@ARTICLE{Bertland+25,
       author = {{Bertrand}, Tanguy and {Lellouch}, Emmanuel and {Holler}, Bryan and {Stansberry}, John and {Wong}, Ian and {Zhang}, Xi and {Lavvas}, Panayotis and {Dufaux}, Elodie and {Merlin}, Frederic and {Villanueva}, Geronimo and {Wan}, Linfeng and {Pinilla-Alonso}, Noem{\'\i} and {de Souza Feliciano}, Ana Carolina and {Murray}, Katherine},
        title = "{Evidence of haze control of Pluto's atmospheric heat balance from JWST/MIRI thermal light curves}",
      journal = {Nature Astronomy},
         year = 2025,
        month = sep,
       volume = {9},
        pages = {1300-1308},
          doi = {10.1038/s41550-025-02573-z},
       adsurl = {https://ui.adsabs.harvard.edu/abs/2025NatAs...9.1300B}
}

@ARTICLE{Zhang+17_Pluto,
       author = {{Zhang}, Xi and {Strobel}, Darrell F. and {Imanaka}, Hiroshi},
        title = "{Haze heats Pluto{\textquoteright}s atmosphere yet explains its cold temperature}",
      journal = {\nat},
         year = 2017,
        month = nov,
       volume = {551},
       number = {7680},
        pages = {352-355},
          doi = {10.1038/nature24465},
       adsurl = {https://ui.adsabs.harvard.edu/abs/2017Natur.551..352Z}
}

@ARTICLE{Morris+24,
       author = {{Morris}, Brett M. and {Heng}, Kevin and {Kitzmann}, Daniel},
        title = "{Observations of scattered light from exoplanet atmospheres}",
      journal = {\aap},
         year = 2024,
        month = may,
       volume = {685},
          eid = {A104},
        pages = {A104},
          doi = {10.1051/0004-6361/202243831},
archivePrefix = {arXiv},
       eprint = {2401.13635},
 primaryClass = {astro-ph.EP},
       adsurl = {https://ui.adsabs.harvard.edu/abs/2024A&A...685A.104M}
}

@ARTICLE{Oreshenko+16,
       author = {{Oreshenko}, Maria and {Heng}, Kevin and {Demory}, Brice-Olivier},
        title = "{Optical phase curves as diagnostics for aerosol composition in exoplanetary atmospheres}",
      journal = {\mnras},
         year = 2016,
        month = apr,
       volume = {457},
       number = {4},
        pages = {3420-3429},
          doi = {10.1093/mnras/stw133},
archivePrefix = {arXiv},
       eprint = {1601.03050},
 primaryClass = {astro-ph.EP},
       adsurl = {https://ui.adsabs.harvard.edu/abs/2016MNRAS.457.3420O}
}

@ARTICLE{Lavvas+24,
       author = {{Lavvas}, Panayotis and {Paraskevaidou}, Sophia and {Arfaux}, Anthony},
        title = "{Atmospheric characterisation of GJ1214b from transit and eclipse observations}",
      journal = {arXiv e-prints},
         year = 2024,
        month = oct,
          eid = {arXiv:2410.09981},
        pages = {arXiv:2410.09981},
          doi = {10.48550/arXiv.2410.09981},
archivePrefix = {arXiv},
       eprint = {2410.09981},
 primaryClass = {astro-ph.EP},
       adsurl = {https://ui.adsabs.harvard.edu/abs/2024arXiv241009981L}
}

@ARTICLE{Mahajan+24,
       author = {{Mahajan}, Alexandra S. and {Eastman}, Jason D. and {Kirk}, James},
        title = "{Using JWST Transits and Occultations to Determine {\ensuremath{\sim}}1\% Stellar Radii and Temperatures of Low-mass Stars}",
      journal = {\apjl},
         year = 2024,
        month = mar,
       volume = {963},
       number = {2},
          eid = {L37},
        pages = {L37},
          doi = {10.3847/2041-8213/ad29f3},
archivePrefix = {arXiv},
       eprint = {2402.05991},
 primaryClass = {astro-ph.SR},
       adsurl = {https://ui.adsabs.harvard.edu/abs/2024ApJ...963L..37M}
}

@ARTICLE{Kempton+12,
       author = {{Miller-Ricci Kempton}, Eliza and {Zahnle}, Kevin and {Fortney}, Jonathan J.},
        title = "{The Atmospheric Chemistry of GJ 1214b: Photochemistry and Clouds}",
      journal = {\apj},
         year = 2012,
        month = jan,
       volume = {745},
       number = {1},
          eid = {3},
        pages = {3},
          doi = {10.1088/0004-637X/745/1/3},
archivePrefix = {arXiv},
       eprint = {1104.5477},
 primaryClass = {astro-ph.EP},
       adsurl = {https://ui.adsabs.harvard.edu/abs/2012ApJ...745....3M}
}

@ARTICLE{Christie+22,
       author = {{Christie}, D.~A. and {Mayne}, N.~J. and {Gillard}, R.~M. and {Manners}, J. and {H{\'e}brard}, E. and {Lines}, S. and {Kohary}, K.},
        title = "{The impact of phase equilibrium cloud models on GCM simulations of GJ 1214b}",
      journal = {\mnras},
         year = 2022,
        month = nov,
       volume = {517},
       number = {1},
        pages = {1407-1421},
          doi = {10.1093/mnras/stac2763},
archivePrefix = {arXiv},
       eprint = {2209.12205},
 primaryClass = {astro-ph.EP},
       adsurl = {https://ui.adsabs.harvard.edu/abs/2022MNRAS.517.1407C}
}

@ARTICLE{Charnay+15,
       author = {{Charnay}, B. and {Meadows}, V. and {Misra}, A. and {Leconte}, J. and {Arney}, G.},
        title = "{3D Modeling of GJ1214b{\textquoteright}s Atmosphere: Formation of Inhomogeneous High Clouds and Observational Implications}",
      journal = {\apjl},
         year = 2015,
        month = nov,
       volume = {813},
       number = {1},
          eid = {L1},
        pages = {L1},
          doi = {10.1088/2041-8205/813/1/L1},
archivePrefix = {arXiv},
       eprint = {1510.01706},
 primaryClass = {astro-ph.EP},
       adsurl = {https://ui.adsabs.harvard.edu/abs/2015ApJ...813L...1C}
}

@ARTICLE{Malsky+25,
       author = {{Malsky}, Isaac and {Rauscher}, Emily and {Stevenson}, Kevin and {Savel}, Arjun B. and {Steinrueck}, Maria E. and {Gao}, Peter and {Kempton}, Eliza M.-R. and {Roman}, Michael T. and {Bean}, Jacob L. and {Zhang}, Michael and {Parmentier}, Vivien and {Piette}, Anjali A.~A. and {Kataria}, Tiffany},
        title = "{Clouds and Hazes in GJ 1214 b's Metal-rich Atmosphere}",
      journal = {\aj},
         year = 2025,
        month = apr,
       volume = {169},
       number = {4},
          eid = {221},
        pages = {221},
          doi = {10.3847/1538-3881/adb7e8},
archivePrefix = {arXiv},
       eprint = {2503.22608},
 primaryClass = {astro-ph.EP},
       adsurl = {https://ui.adsabs.harvard.edu/abs/2025AJ....169..221M}
}

@ARTICLE{Ohno+25,
       author = {{Ohno}, Kazumasa and {Schlawin}, Everett and {Bell}, Taylor J. and {Murphy}, Matthew M. and {Beatty}, Thomas G. and {Welbanks}, Luis and {Greene}, Thomas P. and {Fortney}, Jonathan J. and {Parmentier}, Vivien and {Edelman}, Isaac R. and {Mehta}, Nishil and {Rieke}, Marcia J.},
        title = "{A Possible Metal-dominated Atmosphere below the Thick Aerosols of GJ 1214 b Suggested by Its JWST Panchromatic Transmission Spectrum}",
      journal = {\apjl},
         year = 2025,
        month = jan,
       volume = {979},
       number = {1},
          eid = {L7},
        pages = {L7},
          doi = {10.3847/2041-8213/ada02c},
archivePrefix = {arXiv},
       eprint = {2410.10186},
 primaryClass = {astro-ph.EP},
       adsurl = {https://ui.adsabs.harvard.edu/abs/2025ApJ...979L...7O}
}

@ARTICLE{Gao+23,
       author = {{Gao}, Peter and {Piette}, Anjali A.~A. and {Steinrueck}, Maria E. and {Nixon}, Matthew C. and {Zhang}, Michael and {Kempton}, Eliza M.-R. and {Bean}, Jacob L. and {Rauscher}, Emily and {Parmentier}, Vivien and {Batalha}, Natasha E. and {Savel}, Arjun B. and {Triantafillides}, Anastasia and {Roman}, Michael T. and {Malsky}, Isaac and {Taylor}, Jake},
        title = "{The Hazy and Metal-rich Atmosphere of GJ 1214 b Constrained by Near- and Mid-infrared Transmission Spectroscopy}",
      journal = {\apj},
         year = 2023,
        month = jan,
       volume = {951},
       number = {2},
          eid = {96},
        pages = {96},
          doi = {10.3847/1538-4357/acd16f},
archivePrefix = {arXiv},
       eprint = {2305.05697},
 primaryClass = {astro-ph.EP},
       adsurl = {https://ui.adsabs.harvard.edu/abs/2023ApJ...951...96G}
}

@ARTICLE{Lavvas+19,
       author = {{Lavvas}, Panayotis and {Koskinen}, Tommi and {Steinrueck}, Maria E. and {Garc{\'\i}a Mu{\~n}oz}, Antonio and {Showman}, Adam P.},
        title = "{Photochemical Hazes in Sub-Neptunian Atmospheres with a Focus on GJ 1214b}",
      journal = {\apj},
         year = 2019,
        month = jun,
       volume = {878},
       number = {2},
          eid = {118},
        pages = {118},
          doi = {10.3847/1538-4357/ab204e},
archivePrefix = {arXiv},
       eprint = {1905.02976},
 primaryClass = {astro-ph.EP},
       adsurl = {https://ui.adsabs.harvard.edu/abs/2019ApJ...878..118L}
}

@ARTICLE{Kawashima&Ikoma18,
       author = {{Kawashima}, Yui and {Ikoma}, Masahiro},
        title = "{Theoretical Transmission Spectra of Exoplanet Atmospheres with Hydrocarbon Haze: Effect of Creation, Growth, and Settling of Haze Particles. I. Model Description and First Results}",
      journal = {\apj},
         year = 2018,
        month = jan,
       volume = {853},
       number = {1},
          eid = {7},
        pages = {7},
          doi = {10.3847/1538-4357/aaa0c5},
archivePrefix = {arXiv},
       eprint = {1712.02808},
 primaryClass = {astro-ph.EP},
       adsurl = {https://ui.adsabs.harvard.edu/abs/2018ApJ...853....7K}
}

@ARTICLE{Lee23,
       author = {{Lee}, Elspeth K.~H.},
        title = "{Modelling dynamically driven global cloud formation microphysics in the HAT-P-1b atmosphere}",
      journal = {\mnras},
         year = 2023,
        month = sep,
       volume = {524},
       number = {2},
        pages = {2918-2933},
          doi = {10.1093/mnras/stad2037},
archivePrefix = {arXiv},
       eprint = {2307.02268},
 primaryClass = {astro-ph.EP},
       adsurl = {https://ui.adsabs.harvard.edu/abs/2023MNRAS.524.2918L}
}

@ARTICLE{Lee&Ohno25,
       author = {{Lee}, Elspeth K.~H. and {Ohno}, Kazumasa},
        title = "{Three-dimensional dynamical evolution of cloud particle microphysics in sub-stellar atmospheres: I. Description and exploring Y-dwarf atmospheric variability}",
      journal = {\aap},
         year = 2025,
        month = mar,
       volume = {695},
          eid = {A111},
        pages = {A111},
          doi = {10.1051/0004-6361/202452922},
archivePrefix = {arXiv},
       eprint = {2411.10305},
 primaryClass = {astro-ph.EP},
       adsurl = {https://ui.adsabs.harvard.edu/abs/2025A&A...695A.111L}
}

@ARTICLE{Lines+18,
       author = {{Lines}, S. and {Mayne}, N.~J. and {Boutle}, I.~A. and {Manners}, J. and {Lee}, E.~K.~H. and {Helling}, Ch. and {Drummond}, B. and {Amundsen}, D.~S. and {Goyal}, J. and {Acreman}, D.~M. and {Tremblin}, P. and {Kerslake}, M.},
        title = "{Simulating the cloudy atmospheres of HD 209458 b and HD 189733 b with the 3D Met Office Unified Model}",
      journal = {\aap},
         year = 2018,
        month = jul,
       volume = {615},
          eid = {A97},
        pages = {A97},
          doi = {10.1051/0004-6361/201732278},
archivePrefix = {arXiv},
       eprint = {1803.00226},
 primaryClass = {astro-ph.EP},
       adsurl = {https://ui.adsabs.harvard.edu/abs/2018A&A...615A..97L}
}

@ARTICLE{Lee+16,
       author = {{Lee}, E. and {Dobbs-Dixon}, I. and {Helling}, Ch. and {Bognar}, K. and {Woitke}, P.},
        title = "{Dynamic mineral clouds on HD 189733b. I. 3D RHD with kinetic, non-equilibrium cloud formation}",
      journal = {\aap},
         year = 2016,
        month = oct,
       volume = {594},
          eid = {A48},
        pages = {A48},
          doi = {10.1051/0004-6361/201628606},
archivePrefix = {arXiv},
       eprint = {1603.09098},
 primaryClass = {astro-ph.EP},
       adsurl = {https://ui.adsabs.harvard.edu/abs/2016A&A...594A..48L}
}

@ARTICLE{Powell&Zhang24,
       author = {{Powell}, Diana and {Zhang}, Xi},
        title = "{Two-dimensional Models of Microphysical Clouds on Hot Jupiters. I. Cloud Properties}",
      journal = {\apj},
         year = 2024,
        month = jul,
       volume = {969},
       number = {1},
          eid = {5},
        pages = {5},
          doi = {10.3847/1538-4357/ad3de4},
archivePrefix = {arXiv},
       eprint = {2404.08759},
 primaryClass = {astro-ph.EP},
       adsurl = {https://ui.adsabs.harvard.edu/abs/2024ApJ...969....5P}
}

@ARTICLE{Stevenson16,
       author = {{Stevenson}, Kevin B.},
        title = "{Quantifying and Predicting the Presence of Clouds in Exoplanet Atmospheres}",
      journal = {\apjl},
         year = 2016,
        month = feb,
       volume = {817},
       number = {2},
          eid = {L16},
        pages = {L16},
          doi = {10.3847/2041-8205/817/2/L16},
archivePrefix = {arXiv},
       eprint = {1601.03492},
 primaryClass = {astro-ph.EP},
       adsurl = {https://ui.adsabs.harvard.edu/abs/2016ApJ...817L..16S}
}

@ARTICLE{Fu+17,
       author = {{Fu}, Guangwei and {Deming}, Drake and {Knutson}, Heather and {Madhusudhan}, Nikku and {Mandell}, Avi and {Fraine}, Jonathan},
        title = "{Statistical Analysis of Hubble/WFC3 Transit Spectroscopy of Extrasolar Planets}",
      journal = {\apjl},
         year = 2017,
        month = oct,
       volume = {847},
       number = {2},
          eid = {L22},
        pages = {L22},
          doi = {10.3847/2041-8213/aa8e40},
archivePrefix = {arXiv},
       eprint = {1709.07385},
 primaryClass = {astro-ph.EP},
       adsurl = {https://ui.adsabs.harvard.edu/abs/2017ApJ...847L..22F}
}

@software{prahl_miepython_2026,
  author  = {Scott Prahl},
  title   = {{miepython}: A Python library for Mie scattering calculations},
  url     = {https://github.com/scottprahl/miepython},
  doi     = {10.5281/zenodo.7949263},
  year    = {2026},
  version = {3.2.0},
  publisher = {Zenodo}
}

@ARTICLE{Batalha+26,
       author = {{Batalha}, Natasha E. and {Rooney}, Caoimhe M. and {Visscher}, Channon and {Moran}, Sarah E. and {Marley}, Mark S. and {Sengupta}, Aditya R. and {Kiefer}, Sven and {Lodge}, Matt G. and {Mang}, James and {Morley}, Caroline V. and {Mukherjee}, Sagnick and {Fortney}, Jonathan J. and {Gao}, Peter and {Lewis}, Nikole K. and {Mayorga}, L.~C. and {Pearce}, Logan A. and {Wakeford}, Hannah R.},
        title = "{Condensation Clouds in Substellar Atmospheres with Virga}",
      journal = {\aj},
         year = 2026,
        month = feb,
       volume = {171},
       number = {2},
          eid = {98},
        pages = {98},
          doi = {10.3847/1538-3881/ae29e5},
archivePrefix = {arXiv},
       eprint = {2508.15102},
 primaryClass = {astro-ph.EP},
       adsurl = {https://ui.adsabs.harvard.edu/abs/2026AJ....171...98B}
}

@ARTICLE{Coulombe+25,
       author = {{Coulombe}, Louis-Philippe and {Radica}, Michael and {Benneke}, Bj{\"o}rn and {D'Aoust}, {\'E}lyse and {Dang}, Lisa and {Cowan}, Nicolas B. and {Parmentier}, Vivien and {Albert}, Lo{\"\i}c and {Lafreni{\`e}re}, David and {Taylor}, Jake and {Roy}, Pierre-Alexis and {Pelletier}, Stefan and {Allart}, Romain and {Artigau}, {\'E}tienne and {Doyon}, Ren{\'e} and {Jayawardhana}, Ray and {Johnstone}, Doug and {Kaltenegger}, Lisa and {Langeveld}, Adam B. and {MacDonald}, Ryan J. and {Rowe}, Jason F. and {Turner}, Jake D.},
        title = "{Highly reflective white clouds on the western dayside of an exo-Neptune}",
      journal = {Nature Astronomy},
         year = 2025,
        month = apr,
       volume = {9},
        pages = {512-525},
          doi = {10.1038/s41550-025-02488-9},
archivePrefix = {arXiv},
       eprint = {2501.14016},
 primaryClass = {astro-ph.EP},
       adsurl = {https://ui.adsabs.harvard.edu/abs/2025NatAs...9..512C}
}

@ARTICLE{Fu+25_cloud_assymmetry,
       author = {{Fu}, Guangwei and {Mukherjee}, Sagnick and {Stevenson}, Kevin B. and {Sing}, David K. and {Ashtari}, Reza and {Mayne}, Nathan and {Lothringer}, Joshua D. and {Zamyatina}, Maria and {Schmidt}, Stephen P. and {Gasc{\'o}n}, Carlos and {Allen}, Natalie H. and {Bennett}, Katherine A. and {L{\'o}pez-Morales}, Mercedes},
        title = "{Overcast Mornings and Clear Evenings in Hot Jupiter Exoplanet Atmospheres}",
      journal = {\apjl},
         year = 2025,
        month = aug,
       volume = {989},
       number = {1},
          eid = {L17},
        pages = {L17},
          doi = {10.3847/2041-8213/adf20f},
archivePrefix = {arXiv},
       eprint = {2507.15854},
 primaryClass = {astro-ph.EP},
       adsurl = {https://ui.adsabs.harvard.edu/abs/2025ApJ...989L..17F}
}

@ARTICLE{Mukherjee+25,
       author = {{Mukherjee}, Sagnick and {Sing}, David K. and {Fu}, Guangwei and {Stevenson}, Kevin B. and {Schmidt}, Stephen P. and {Baskett}, Harry and {McCreery}, Patrick and {Allen}, Natalie H. and {Bennett}, Katherine A. and {Christie}, Duncan A. and {Gasc{\'o}n}, Carlos and {Goyal}, Jayesh and {H{\'e}brard}, {\'E}ric and {Lothringer}, Joshua D. and {L{\'o}pez-Morales}, Mercedes and {Lustig-Yaeger}, Jacob and {May}, Erin M. and {Mayorga}, L.~C. and {Mayne}, Nathan and {Ramos Rosado}, Lakeisha M. and {Reggiani}, Henrique and {Rustamkulov}, Zafar and {Schlaufman}, Kevin C. and {Sotzen}, K.~S. and {Thorngren}, Daniel and {Wang}, Le-Chris and {Zamyatina}, Maria},
        title = "{Cloudy mornings and clear evenings on a giant extrasolar world}",
      journal = {arXiv e-prints},
         year = 2025,
        month = may,
          eid = {arXiv:2505.10910},
        pages = {arXiv:2505.10910},
          doi = {10.48550/arXiv.2505.10910},
archivePrefix = {arXiv},
       eprint = {2505.10910},
 primaryClass = {astro-ph.EP},
       adsurl = {https://ui.adsabs.harvard.edu/abs/2025arXiv250510910M}
}

@ARTICLE{Schlawin+24,
       author = {{Schlawin}, Everett and {Mukherjee}, Sagnick and {Ohno}, Kazumasa and {Bell}, Taylor J. and {Beatty}, Thomas G. and {Greene}, Thomas P. and {Line}, Michael and {Challener}, Ryan C. and {Parmentier}, Vivien and {Fortney}, Jonathan J. and {Rauscher}, Emily and {Wiser}, Lindsey and {Welbanks}, Luis and {Murphy}, Matthew and {Edelman}, Isaac and {Batalha}, Natasha and {Moran}, Sarah E. and {Mehta}, Nishil and {Rieke}, Marcia},
        title = "{Multiple Clues for Dayside Aerosols and Temperature Gradients in WASP-69 b from a Panchromatic JWST Emission Spectrum}",
      journal = {\aj},
         year = 2024,
        month = sep,
       volume = {168},
       number = {3},
          eid = {104},
        pages = {104},
          doi = {10.3847/1538-3881/ad58e0},
archivePrefix = {arXiv},
       eprint = {2406.15543},
 primaryClass = {astro-ph.EP},
       adsurl = {https://ui.adsabs.harvard.edu/abs/2024AJ....168..104S}
}

@ARTICLE{Tada+25,
       author = {{Tada}, Shotaro and {Kawahara}, Hajime and {Kawashima}, Yui and {Kotani}, Takayuki and {Masuda}, Kento},
        title = "{Probing Two-dimensional Asymmetries of an Exoplanet Atmosphere from Chromatic Transit Variation}",
      journal = {\aj},
         year = 2025,
        month = may,
       volume = {169},
       number = {5},
          eid = {255},
        pages = {255},
          doi = {10.3847/1538-3881/adbe65},
archivePrefix = {arXiv},
       eprint = {2503.08988},
 primaryClass = {astro-ph.EP},
       adsurl = {https://ui.adsabs.harvard.edu/abs/2025AJ....169..255T}
}

@ARTICLE{Murphy+25,
       author = {{Murphy}, Matthew M. and {Beatty}, Thomas G. and {Schlawin}, Everett and {Bell}, Taylor J. and {Radica}, Michael and {Kennedy}, Thomas D. and {Mehta}, Nishil and {Welbanks}, Luis and {Line}, Michael R. and {Parmentier}, Vivien and {Greene}, Thomas P. and {Mukherjee}, Sagnick and {Fortney}, Jonathan J. and {Ohno}, Kazumasa and {Wiser}, Lindsey and {Arnold}, Kenneth and {Rauscher}, Emily and {Edelman}, Isaac R. and {Rieke}, Marcia J.},
        title = "{A Panchromatic Characterization of the Evening and Morning Atmosphere of WASP-107 b: Composition and Cloud Variations, and Insight into the Effect of Stellar Contamination}",
      journal = {\aj},
         year = 2025,
        month = jul,
       volume = {170},
       number = {1},
          eid = {61},
        pages = {61},
          doi = {10.3847/1538-3881/addf38},
archivePrefix = {arXiv},
       eprint = {2505.13602},
 primaryClass = {astro-ph.EP},
       adsurl = {https://ui.adsabs.harvard.edu/abs/2025AJ....170...61M}
}

@ARTICLE{Murphy+24,
       author = {{Murphy}, Matthew M. and {Beatty}, Thomas G. and {Schlawin}, Everett and {Bell}, Taylor J. and {Line}, Michael R. and {Greene}, Thomas P. and {Parmentier}, Vivien and {Rauscher}, Emily and {Welbanks}, Luis and {Fortney}, Jonathan J. and {Rieke}, Marcia},
        title = "{Evidence for morning-to-evening limb asymmetry on the cool low-density exoplanet WASP-107 b}",
      journal = {Nature Astronomy},
         year = 2024,
        month = dec,
       volume = {8},
        pages = {1562-1574},
          doi = {10.1038/s41550-024-02367-9},
archivePrefix = {arXiv},
       eprint = {2406.09863},
 primaryClass = {astro-ph.EP},
       adsurl = {https://ui.adsabs.harvard.edu/abs/2024NatAs...8.1562M}
}

@ARTICLE{Espinoza+24,
       author = {{Espinoza}, N{\'e}stor and {Steinrueck}, Maria E. and {Kirk}, James and {MacDonald}, Ryan J. and {Savel}, Arjun B. and {Arnold}, Kenneth and {Kempton}, Eliza M.-R. and {Murphy}, Matthew M. and {Carone}, Ludmila and {Zamyatina}, Maria and {Lewis}, David A. and {Samra}, Dominic and {Kiefer}, Sven and {Rauscher}, Emily and {Christie}, Duncan and {Mayne}, Nathan and {Helling}, Christiane and {Rustamkulov}, Zafar and {Parmentier}, Vivien and {May}, Erin M. and {Carter}, Aarynn L. and {Zhang}, Xi and {L{\'o}pez-Morales}, Mercedes and {Allen}, Natalie and {Blecic}, Jasmina and {Decin}, Leen and {Mancini}, Luigi and {Molaverdikhani}, Karan and {Rackham}, Benjamin V. and {Palle}, Enric and {Tsai}, Shang-Min and {Ahrer}, Eva-Maria and {Bean}, Jacob L. and {Crossfield}, Ian J.~M. and {Haegele}, David and {H{\'e}brard}, Eric and {Kreidberg}, Laura and {Powell}, Diana and {Schneider}, Aaron D. and {Welbanks}, Luis and {Wheatley}, Peter and {Brahm}, Rafael and {Crouzet}, Nicolas},
        title = "{Inhomogeneous terminators on the exoplanet WASP-39 b}",
      journal = {\nat},
         year = 2024,
        month = aug,
       volume = {632},
       number = {8027},
        pages = {1017-1020},
          doi = {10.1038/s41586-024-07768-4},
archivePrefix = {arXiv},
       eprint = {2407.10294},
 primaryClass = {astro-ph.EP},
       adsurl = {https://ui.adsabs.harvard.edu/abs/2024Natur.632.1017E}
}

@ARTICLE{Haberle+25,
       author = {{Haberle}, Robert M. and {Kahre}, Melinda A. and {Bertrand}, Tanguy and {Wolff}, Michael J.},
        title = "{Modeling studies of dust/gas non-thermal equilibrium in the Martian atmosphere}",
      journal = {\icarus},
         year = 2025,
        month = mar,
       volume = {429},
          eid = {116452},
        pages = {116452},
          doi = {10.1016/j.icarus.2024.116452},
       adsurl = {https://ui.adsabs.harvard.edu/abs/2025Icar..42916452H}
}

@ARTICLE{Goldenson+08,
       author = {{Goldenson}, Naomi and {Desch}, Steven and {Christensen}, Philip},
        title = "{Non-equilibrium between dust and gas temperatures in the Mars atmosphere}",
      journal = {\grl},
         year = 2008,
        month = apr,
       volume = {35},
       number = {8},
          eid = {L08813},
        pages = {L08813},
          doi = {10.1029/2007GL032907},
       adsurl = {https://ui.adsabs.harvard.edu/abs/2008GeoRL..35.8813G}
}

@ARTICLE{Fiocco+76,
       author = {{Fiocco}, G. and {Mugnai}, A. and {Grams}, G.},
        title = "{Energy exchange and temperature of aerosols in the earth's atmosphere /0-60 km/}",
      journal = {Journal of the Atmospheric Sciences},
         year = 1976,
        month = dec,
       volume = {33},
        pages = {2415-2424},
          doi = {10.1175/1520-0469(1976)033<2415:EEATOA>2.0.CO;2},
       adsurl = {https://ui.adsabs.harvard.edu/abs/1976JAtS...33.2415F}
}

@ARTICLE{Fiocco+75,
       author = {{Fiocco}, G. and {Grams}, G. and {Visconti}, G.},
        title = "{Equilibrium temperatures of small particles in the earth's upper atmosphere /50-110 km/}",
      journal = {Journal of Atmospheric and Terrestrial Physics},
         year = 1975,
        month = oct,
       volume = {37},
        pages = {1327-1337},
          doi = {10.1016/0021-9169(75)90125-7},
       adsurl = {https://ui.adsabs.harvard.edu/abs/1975JATP...37.1327F}
}

@ARTICLE{Anderson+10,
       author = {{Anderson}, D.~R. and {Hellier}, C. and {Gillon}, M. and {Triaud}, A.~H.~M.~J. and {Smalley}, B. and {Hebb}, L. and {Collier Cameron}, A. and {Maxted}, P.~F.~L. and {Queloz}, D. and {West}, R.~G. and {Bentley}, S.~J. and {Enoch}, B. and {Horne}, K. and {Lister}, T.~A. and {Mayor}, M. and {Parley}, N.~R. and {Pepe}, F. and {Pollacco}, D. and {S{\'e}gransan}, D. and {Udry}, S. and {Wilson}, D.~M.},
        title = "{WASP-17b: An Ultra-Low Density Planet in a Probable Retrograde Orbit}",
      journal = {\apj},
         year = 2010,
        month = jan,
       volume = {709},
       number = {1},
        pages = {159-167},
          doi = {10.1088/0004-637X/709/1/159},
archivePrefix = {arXiv},
       eprint = {0908.1553},
 primaryClass = {astro-ph.EP},
       adsurl = {https://ui.adsabs.harvard.edu/abs/2010ApJ...709..159A}
}

@ARTICLE{Stassun+17,
       author = {{Stassun}, Keivan G. and {Collins}, Karen A. and {Gaudi}, B. Scott},
        title = "{Accurate Empirical Radii and Masses of Planets and Their Host Stars with Gaia Parallaxes}",
      journal = {\aj},
         year = 2017,
        month = mar,
       volume = {153},
       number = {3},
          eid = {136},
        pages = {136},
          doi = {10.3847/1538-3881/aa5df3},
archivePrefix = {arXiv},
       eprint = {1609.04389},
 primaryClass = {astro-ph.EP},
       adsurl = {https://ui.adsabs.harvard.edu/abs/2017AJ....153..136S}
}

@ARTICLE{Hamil+25,
       author = {{Hamill}, Colin D. and {Johnson}, Alexandria V. and {Lodge}, Matt and {Gao}, Peter and {Nag}, Rowan and {Batalha}, Natasha and {Christie}, Duncan A. and {Wakeford}, Hannah R.},
        title = "{The Effects of Cuboid Particle Scattering on Reflected Light Phase Curves: Insights from Laboratory Data and Theory}",
      journal = {\apj},
         year = 2025,
        month = jul,
       volume = {987},
       number = {2},
          eid = {176},
        pages = {176},
          doi = {10.3847/1538-4357/add7cf},
archivePrefix = {arXiv},
       eprint = {2507.05485},
 primaryClass = {astro-ph.EP},
       adsurl = {https://ui.adsabs.harvard.edu/abs/2025ApJ...987..176H}
}

@ARTICLE{Hamil+24,
       author = {{Hamill}, Colin D. and {Johnson}, Alexandria V. and {Gao}, Peter},
        title = "{Light Scattering Measurements of KCl Particles as an Exoplanet Cloud Analog}",
      journal = {\psj},
         year = 2024,
        month = aug,
       volume = {5},
       number = {8},
          eid = {186},
        pages = {186},
          doi = {10.3847/PSJ/ad6569},
archivePrefix = {arXiv},
       eprint = {2411.00952},
 primaryClass = {astro-ph.EP},
       adsurl = {https://ui.adsabs.harvard.edu/abs/2024PSJ.....5..186H}
}

@ARTICLE{Lodge+25,
       author = {{Lodge}, Matt G. and {Moran}, Sarah E. and {Wakeford}, Hannah R. and {Leinhardt}, Zoe M. and {Marley}, Mark S.},
        title = "{Fractal Aggregate Aerosols in the Virga Cloud Code II: Exploring the Effects of Key Cloud Parameters in Warm Neptune, Hot Jupiter and Brown Dwarf Atmospheres}",
      journal = {arXiv e-prints},
         year = 2025,
        month = dec,
          eid = {arXiv:2512.04186},
        pages = {arXiv:2512.04186},
          doi = {10.48550/arXiv.2512.04186},
archivePrefix = {arXiv},
       eprint = {2512.04186},
 primaryClass = {astro-ph.EP},
       adsurl = {https://ui.adsabs.harvard.edu/abs/2025arXiv251204186L}
}

@ARTICLE{Lavvas2021,
       author = {{Lavvas}, P. and {Arfaux}, A.},
        title = "{Impact of photochemical hazes and gases on exoplanet atmospheric thermal structure}",
      journal = {\mnras},
         year = 2021,
        month = apr,
       volume = {502},
       number = {4},
        pages = {5643-5657},
          doi = {10.1093/mnras/stab456},
archivePrefix = {arXiv},
       eprint = {2102.05763},
 primaryClass = {astro-ph.EP},
       adsurl = {https://ui.adsabs.harvard.edu/abs/2021MNRAS.502.5643L}
}

@ARTICLE{Steinrueck+23,
       author = {{Steinrueck}, Maria E. and {Koskinen}, Tommi and {Lavvas}, Panayotis and {Parmentier}, Vivien and {Zieba}, Sebastian and {Tan}, Xianyu and {Zhang}, Xi and {Kreidberg}, Laura},
        title = "{Photochemical Hazes Dramatically Alter Temperature Structure and Atmospheric Circulation in 3D Simulations of Hot Jupiters}",
      journal = {\apj},
         year = 2023,
        month = jul,
       volume = {951},
       number = {2},
          eid = {117},
        pages = {117},
          doi = {10.3847/1538-4357/acd4bb},
archivePrefix = {arXiv},
       eprint = {2305.09654},
 primaryClass = {astro-ph.EP},
       adsurl = {https://ui.adsabs.harvard.edu/abs/2023ApJ...951..117S}
}

@ARTICLE{Steinrueck+25,
       author = {{Steinrueck}, Maria E. and {Parmentier}, Vivien and {Kreidberg}, Laura and {Gao}, Peter and {Kempton}, Eliza M.-R. and {Zhang}, Michael and {Stevenson}, Kevin B. and {Malsky}, Isaac and {Roman}, Michael T. and {Rauscher}, Emily and {Malik}, Matej and {Lupu}, Roxana and {Kataria}, Tiffany and {Piette}, Anjali A.~A. and {Bean}, Jacob L. and {Nixon}, Matthew C.},
        title = "{The Radiative Effects of Photochemical Hazes on the Atmospheric Circulation and Phase Curves of Sub-Neptunes}",
      journal = {\apj},
         year = 2025,
        month = may,
       volume = {985},
       number = {1},
          eid = {98},
        pages = {98},
          doi = {10.3847/1538-4357/adc99a},
archivePrefix = {arXiv},
       eprint = {2503.22642},
 primaryClass = {astro-ph.EP},
       adsurl = {https://ui.adsabs.harvard.edu/abs/2025ApJ...985...98S}
}

@ARTICLE{Molaverdikhani+20,
       author = {{Molaverdikhani}, Karan and {Henning}, Thomas and {Molli{\`e}re}, Paul},
        title = "{The Role of Clouds on the Depletion of Methane and Water Dominance in the Transmission Spectra of Irradiated Exoplanets}",
      journal = {\apj},
         year = 2020,
        month = aug,
       volume = {899},
       number = {1},
          eid = {53},
        pages = {53},
          doi = {10.3847/1538-4357/aba52b},
archivePrefix = {arXiv},
       eprint = {2007.06562},
 primaryClass = {astro-ph.EP},
       adsurl = {https://ui.adsabs.harvard.edu/abs/2020ApJ...899...53M}
}

@ARTICLE{Kitzmann&Heng18,
       author = {{Kitzmann}, Daniel and {Heng}, Kevin},
        title = "{Optical properties of potential condensates in exoplanetary atmospheres}",
      journal = {\mnras},
         year = 2018,
        month = mar,
       volume = {475},
       number = {1},
        pages = {94-107},
          doi = {10.1093/mnras/stx3141},
archivePrefix = {arXiv},
       eprint = {1710.04946},
 primaryClass = {astro-ph.EP},
       adsurl = {https://ui.adsabs.harvard.edu/abs/2018MNRAS.475...94K}
}

@ARTICLE{Koike+06,
       author = {{Koike}, C. and {Mutschke}, H. and {Suto}, H. and {Naoi}, T. and {Chihara}, H. and {Henning}, Th. and {J{\"a}ger}, C. and {Tsuchiyama}, A. and {Dorschner}, J. and {Okuda}, H.},
        title = "{Temperature effects on the mid-and far-infrared spectra of olivine particles}",
      journal = {\aap},
         year = 2006,
        month = apr,
       volume = {449},
       number = {2},
        pages = {583-596},
          doi = {10.1051/0004-6361:20053256},
       adsurl = {https://ui.adsabs.harvard.edu/abs/2006A&A...449..583K}
}

@ARTICLE{Zeidler+13,
       author = {{Zeidler}, S. and {Posch}, Th. and {Mutschke}, H.},
        title = "{Optical constants of refractory oxides at high temperatures. Mid-infrared properties of corundum, spinel, and {\ensuremath{\alpha}}-quartz, potential carriers of the 13 {\ensuremath{\mu}}m feature}",
      journal = {\aap},
         year = 2013,
        month = may,
       volume = {553},
          eid = {A81},
        pages = {A81},
          doi = {10.1051/0004-6361/201220459},
       adsurl = {https://ui.adsabs.harvard.edu/abs/2013A&A...553A..81Z}
}

@ARTICLE{Zeidler+15,
       author = {{Zeidler}, S. and {Mutschke}, H. and {Posch}, Th.},
        title = "{Temperature-dependent Infrared Optical Constants of Olivine and Enstatite}",
      journal = {\apj},
         year = 2015,
        month = jan,
       volume = {798},
       number = {2},
          eid = {125},
        pages = {125},
          doi = {10.1088/0004-637X/798/2/125},
       adsurl = {https://ui.adsabs.harvard.edu/abs/2015ApJ...798..125Z}
}

@ARTICLE{Lodge+24,
       author = {{Lodge}, M.~G. and {Wakeford}, H.~R. and {Leinhardt}, Z.~M.},
        title = "{Aerosols are not spherical cows: using discrete dipole approximation to model the properties of fractal particles}",
      journal = {\mnras},
         year = 2024,
        month = feb,
       volume = {527},
       number = {4},
        pages = {11113-11137},
          doi = {10.1093/mnras/stad3743},
archivePrefix = {arXiv},
       eprint = {2312.02301},
 primaryClass = {astro-ph.EP},
       adsurl = {https://ui.adsabs.harvard.edu/abs/2024MNRAS.52711113L}
}

@ARTICLE{Lin+25,
       author = {{Lin}, Zhe-Yu Daniel and {Weinberger}, Alycia J. and {Zubko}, Evgenij and {Arnold}, Jessica A. and {Videen}, Gorden},
        title = "{glitterin: Towards Replacing the Role of Lorenz-Mie Theory in Astronomy Using Neural Networks Trained on Light Scattering of Irregularly Shaped Grains}",
      journal = {arXiv e-prints},
         year = 2025,
        month = nov,
          eid = {arXiv:2511.09668},
        pages = {arXiv:2511.09668},
          doi = {10.48550/arXiv.2511.09668},
archivePrefix = {arXiv},
       eprint = {2511.09668},
 primaryClass = {astro-ph.EP},
       adsurl = {https://ui.adsabs.harvard.edu/abs/2025arXiv251109668L}
}

@ARTICLE{Kataoka+14,
       author = {{Kataoka}, Akimasa and {Okuzumi}, Satoshi and {Tanaka}, Hidekazu and {Nomura}, Hideko},
        title = "{Opacity of fluffy dust aggregates}",
      journal = {\aap},
         year = 2014,
        month = aug,
       volume = {568},
          eid = {A42},
        pages = {A42},
          doi = {10.1051/0004-6361/201323199},
archivePrefix = {arXiv},
       eprint = {1312.1459},
 primaryClass = {astro-ph.EP},
       adsurl = {https://ui.adsabs.harvard.edu/abs/2014A&A...568A..42K}
}

@ARTICLE{Moran+25,
       author = {{Moran}, Sarah E. and {Lodge}, Matt G. and {Batalha}, Natasha E. and {Ohno}, Kazumasa and {Vahidinia}, Sanaz and {Marley}, Mark S. and {Wakeford}, Hannah R. and {Leinhardt}, Zo{\"e} M.},
        title = "{Fractal Aggregate Aerosols in the Virga Cloud Code. I. Model Description and Application to a Benchmark Cloudy Exoplanet}",
      journal = {\apj},
         year = 2025,
        month = nov,
       volume = {994},
       number = {1},
          eid = {116},
        pages = {116},
          doi = {10.3847/1538-4357/ae0583},
archivePrefix = {arXiv},
       eprint = {2509.06708},
 primaryClass = {astro-ph.EP},
       adsurl = {https://ui.adsabs.harvard.edu/abs/2025ApJ...994..116M}
}

@ARTICLE{Vahidinia+24,
       author = {{Vahidinia}, Sanaz and {Moran}, Sarah E. and {Marley}, Mark S. and {Cuzzi}, Jeffrey N.},
        title = "{Aggregate Cloud Particle Effects in Exoplanet Atmospheres}",
      journal = {\pasp},
         year = 2024,
        month = aug,
       volume = {136},
       number = {8},
          eid = {084404},
        pages = {084404},
          doi = {10.1088/1538-3873/ad6cf2},
archivePrefix = {arXiv},
       eprint = {2408.11215},
 primaryClass = {astro-ph.EP},
       adsurl = {https://ui.adsabs.harvard.edu/abs/2024PASP..136h4404V}
}

@ARTICLE{Samra+20,
       author = {{Samra}, D. and {Helling}, Ch. and {Min}, M.},
        title = "{Mineral snowflakes on exoplanets and brown dwarfs. Effects of micro-porosity, size distributions, and particle shape}",
      journal = {\aap},
         year = 2020,
        month = jul,
       volume = {639},
          eid = {A107},
        pages = {A107},
          doi = {10.1051/0004-6361/202037553},
archivePrefix = {arXiv},
       eprint = {2004.13502},
 primaryClass = {astro-ph.EP},
       adsurl = {https://ui.adsabs.harvard.edu/abs/2020A&A...639A.107S}
}

@ARTICLE{Adams+19,
       author = {{Adams}, Danica and {Gao}, Peter and {de Pater}, Imke and {Morley}, Caroline V.},
        title = "{Aggregate Hazes in Exoplanet Atmospheres}",
      journal = {\apj},
         year = 2019,
        month = mar,
       volume = {874},
       number = {1},
          eid = {61},
        pages = {61},
          doi = {10.3847/1538-4357/ab074c},
archivePrefix = {arXiv},
       eprint = {1902.05231},
 primaryClass = {astro-ph.EP},
       adsurl = {https://ui.adsabs.harvard.edu/abs/2019ApJ...874...61A}
}

@ARTICLE{Paredes2021,
       author = {{Paredes}, Leonardo A. and {Henry}, Todd J. and {Quinn}, Samuel N. and {Gies}, Douglas R. and {Hinojosa-Go{\~n}i}, Rodrigo and {James}, Hodari-Sadiki and {Jao}, Wei-Chun and {White}, Russel J.},
        title = "{The Solar Neighborhood XLVIII: Nine Giant Planets Orbiting Nearby K Dwarfs, and the CHIRON Spectrograph's Radial Velocity Performance}",
      journal = {\aj},
         year = 2021,
        month = nov,
       volume = {162},
       number = {5},
          eid = {176},
        pages = {176},
          doi = {10.3847/1538-3881/ac082a},
archivePrefix = {arXiv},
       eprint = {2111.15028},
 primaryClass = {astro-ph.EP},
       adsurl = {https://ui.adsabs.harvard.edu/abs/2021AJ....162..176P}
}

@ARTICLE{Addison2019,
       author = {{Addison}, Brett and {Wright}, Duncan J. and {Wittenmyer}, Robert A. and {Horner}, Jonathan and {Mengel}, Matthew W. and {Johns}, Daniel and {Marti}, Connor and {Nicholson}, Belinda and {Soutter}, Jack and {Bowler}, Brendan and {Crossfield}, Ian and {Kane}, Stephen R. and {Kielkopf}, John and {Plavchan}, Peter and {Tinney}, C.~G. and {Zhang}, Hui and {Clark}, Jake T. and {Clerte}, Mathieu and {Eastman}, Jason D. and {Swift}, Jon and {Bottom}, Michael and {Muirhead}, Philip and {McCrady}, Nate and {Herzig}, Erich and {Hogstrom}, Kristina and {Wilson}, Maurice and {Sliski}, David and {Johnson}, Samson A. and {Wright}, Jason T. and {Johnson}, John Asher and {Blake}, Cullen and {Riddle}, Reed and {Lin}, Brian and {Cornachione}, Matthew and {Bedding}, Timothy R. and {Stello}, Dennis and {Huber}, Daniel and {Marsden}, Stephen and {Carter}, Bradley D.},
        title = "{Minerva-Australis. I. Design, Commissioning, and First Photometric Results}",
      journal = {\pasp},
         year = 2019,
        month = nov,
       volume = {131},
       number = {1005},
        pages = {115003},
          doi = {10.1088/1538-3873/ab03aa},
archivePrefix = {arXiv},
       eprint = {1901.11231},
 primaryClass = {astro-ph.IM},
       adsurl = {https://ui.adsabs.harvard.edu/abs/2019PASP..131k5003A}
}

@ARTICLE{Huang+24,
       author = {{Huang}, Helong and {Ormel}, Chris W. and {Min}, Michiel},
        title = "{ExoLyn: A golden mean approach to multispecies cloud modeling in atmospheric retrieval}",
      journal = {\aap},
         year = 2024,
        month = nov,
       volume = {691},
          eid = {A291},
        pages = {A291},
          doi = {10.1051/0004-6361/202451112},
archivePrefix = {arXiv},
       eprint = {2409.18181},
 primaryClass = {astro-ph.EP},
       adsurl = {https://ui.adsabs.harvard.edu/abs/2024A&A...691A.291H}
}

@ARTICLE{Helling+06,
       author = {{Helling}, Ch. and {Thi}, W.-F. and {Woitke}, P. and {Fridlund}, M.},
        title = "{Detectability of dirty dust grains in brown dwarf atmospheres}",
      journal = {\aap},
         year = 2006,
        month = may,
       volume = {451},
       number = {2},
        pages = {L9-L12},
          doi = {10.1051/0004-6361:20064944},
archivePrefix = {arXiv},
       eprint = {astro-ph/0603341},
 primaryClass = {astro-ph},
       adsurl = {https://ui.adsabs.harvard.edu/abs/2006A&A...451L...9H}
}

@ARTICLE{Inglis+24,
       author = {{Inglis}, Julie and {Batalha}, Natasha E. and {Lewis}, Nikole K. and {Kataria}, Tiffany and {Knutson}, Heather A. and {Kilpatrick}, Brian M. and {Gagnebin}, Anna and {Mukherjee}, Sagnick and {Pettyjohn}, Maria M. and {Crossfield}, Ian J.~M. and {Foote}, Trevor O. and {Grant}, David and {Henry}, Gregory W. and {Lally}, Maura and {McKemmish}, Laura K. and {Sing}, David K. and {Wakeford}, Hannah R. and {Zapata Trujillo}, Juan C. and {Zellem}, Robert T.},
        title = "{Quartz Clouds in the Dayside Atmosphere of the Quintessential Hot Jupiter HD 189733 b}",
      journal = {\apjl},
         year = 2024,
        month = oct,
       volume = {973},
       number = {2},
          eid = {L41},
        pages = {L41},
          doi = {10.3847/2041-8213/ad725e},
archivePrefix = {arXiv},
       eprint = {2409.11395},
 primaryClass = {astro-ph.EP},
       adsurl = {https://ui.adsabs.harvard.edu/abs/2024ApJ...973L..41I}
}

@BOOK{Draine+11,
       author = {{Draine}, Bruce T.},
        title = "{Physics of the Interstellar and Intergalactic Medium}",
         year = 2011,
       adsurl = {https://ui.adsabs.harvard.edu/abs/2011piim.book.....D}
}

@ARTICLE{Moran+24,
       author = {{Moran}, Sarah E. and {Marley}, Mark S. and {Crossley}, Samuel D.},
        title = "{Neglected Silicon Dioxide Polymorphs as Clouds in Substellar Atmospheres}",
      journal = {\apjl},
         year = 2024,
        month = sep,
       volume = {973},
       number = {1},
          eid = {L3},
        pages = {L3},
          doi = {10.3847/2041-8213/ad72e7},
archivePrefix = {arXiv},
       eprint = {2408.00698},
 primaryClass = {astro-ph.EP},
       adsurl = {https://ui.adsabs.harvard.edu/abs/2024ApJ...973L...3M}
}

@ARTICLE{Grant+23,
       author = {{Grant}, David and {Lewis}, Nikole K. and {Wakeford}, Hannah R. and {Batalha}, Natasha E. and {Glidden}, Ana and {Goyal}, Jayesh and {Mullens}, Elijah and {MacDonald}, Ryan J. and {May}, Erin M. and {Seager}, Sara and {Stevenson}, Kevin B. and {Valenti}, Jeff A. and {Visscher}, Channon and {Alderson}, Lili and {Allen}, Natalie H. and {Ca{\~n}as}, Caleb I. and {Col{\'o}n}, Knicole and {Clampin}, Mark and {Espinoza}, N{\'e}stor and {Gressier}, Am{\'e}lie and {Huang}, Jingcheng and {Lin}, Zifan and {Long}, Douglas and {Louie}, Dana R. and {Pe{\~n}a-Guerrero}, Maria and {Ranjan}, Sukrit and {Sotzen}, Kristin S. and {Valentine}, Daniel and {Anderson}, Jay and {Balmer}, William O. and {Bellini}, Andrea and {Hoch}, Kielan K.~W. and {Kammerer}, Jens and {Libralato}, Mattia and {Mountain}, C. Matt and {Perrin}, Marshall D. and {Pueyo}, Laurent and {Rickman}, Emily and {Rebollido}, Isabel and {Sohn}, Sangmo Tony and {van der Marel}, Roeland P. and {Watkins}, Laura L.},
        title = "{JWST-TST DREAMS: Quartz Clouds in the Atmosphere of WASP-17b}",
      journal = {\apjl},
         year = 2023,
        month = oct,
       volume = {956},
       number = {2},
          eid = {L32},
        pages = {L32},
          doi = {10.3847/2041-8213/acfc3b10.3847/2041-8213/acfdab},
archivePrefix = {arXiv},
       eprint = {2310.08637},
 primaryClass = {astro-ph.EP},
       adsurl = {https://ui.adsabs.harvard.edu/abs/2023ApJ...956L..32G}
}

@ARTICLE{Ohno24,
       author = {{Ohno}, Kazumasa},
        title = "{Photochemical Hazes in Exoplanetary Skies with Diamonds: Microphysical Modeling of Haze Composition Evolution via Chemical Vapor Deposition}",
      journal = {\apj},
         year = 2024,
        month = dec,
       volume = {977},
       number = {2},
          eid = {188},
        pages = {188},
          doi = {10.3847/1538-4357/ad8e67},
archivePrefix = {arXiv},
       eprint = {2410.10197},
 primaryClass = {astro-ph.EP},
       adsurl = {https://ui.adsabs.harvard.edu/abs/2024ApJ...977..188O}
}

@ARTICLE{Visscher+06,
       author = {{Visscher}, Channon and {Lodders}, Katharina and {Fegley}, Jr., Bruce},
        title = "{Atmospheric Chemistry in Giant Planets, Brown Dwarfs, and Low-Mass Dwarf Stars. II. Sulfur and Phosphorus}",
      journal = {\apj},
         year = 2006,
        month = sep,
       volume = {648},
       number = {2},
        pages = {1181-1195},
          doi = {10.1086/506245},
archivePrefix = {arXiv},
       eprint = {astro-ph/0511136},
 primaryClass = {astro-ph},
       adsurl = {https://ui.adsabs.harvard.edu/abs/2006ApJ...648.1181V}
}

@ARTICLE{Gao+20,
       author = {{Gao}, Peter and {Thorngren}, Daniel P. and {Lee}, Elspeth K.~H. and {Fortney}, Jonathan J. and {Morley}, Caroline V. and {Wakeford}, Hannah R. and {Powell}, Diana K. and {Stevenson}, Kevin B. and {Zhang}, Xi},
        title = "{Aerosol composition of hot giant exoplanets dominated by silicates and hydrocarbon hazes}",
      journal = {Nature Astronomy},
         year = 2020,
        month = may,
       volume = {4},
        pages = {951-956},
          doi = {10.1038/s41550-020-1114-3},
archivePrefix = {arXiv},
       eprint = {2005.11939},
 primaryClass = {astro-ph.EP},
       adsurl = {https://ui.adsabs.harvard.edu/abs/2020NatAs...4..951G}
}

@ARTICLE{Lee+25,
       author = {{Lee}, Elspeth K.~H. and {Werlen}, Aaron and {Dorn}, Caroline},
        title = "{Mineral Cloud Formation above Magma Oceans in Sub-Neptune Atmospheres}",
      journal = {\apjl},
         year = 2025,
        month = sep,
       volume = {990},
       number = {2},
          eid = {L43},
        pages = {L43},
          doi = {10.3847/2041-8213/adfe62},
archivePrefix = {arXiv},
       eprint = {2508.15097},
 primaryClass = {astro-ph.EP},
       adsurl = {https://ui.adsabs.harvard.edu/abs/2025ApJ...990L..43L}
}

@ARTICLE{Mehta+25,
       author = {{Mehta}, Nishil and {Parmentier}, Vivien and {Tan}, Xianyu and {Lee}, Elspeth K.~H. and {Guillot}, Tristan and {Wiser}, Lindsey S. and {Bell}, Taylor J. and {Schlawin}, Everett and {Arnold}, Kenneth and {Mukherjee}, Sagnick and {Greene}, Thomas P. and {Beatty}, Thomas G. and {Welbanks}, Luis and {Line}, Michael R. and {Murphy}, Matthew M. and {Fortney}, Jonathan J. and {Ohno}, Kazumasa},
        title = "{How clear are the skies of WASP-80b?: 3D Cloud feedback on the atmosphere and spectra of the warm Jupiter}",
      journal = {arXiv e-prints},
         year = 2025,
        month = sep,
          eid = {arXiv:2509.23406},
        pages = {arXiv:2509.23406},
          doi = {10.48550/arXiv.2509.23406},
archivePrefix = {arXiv},
       eprint = {2509.23406},
 primaryClass = {astro-ph.EP},
       adsurl = {https://ui.adsabs.harvard.edu/abs/2025arXiv250923406M}
}

@ARTICLE{Roman&Rauscher19,
       author = {{Roman}, Michael and {Rauscher}, Emily},
        title = "{Modeled Temperature-dependent Clouds with Radiative Feedback in Hot Jupiter Atmospheres}",
      journal = {\apj},
         year = 2019,
        month = feb,
       volume = {872},
       number = {1},
          eid = {1},
        pages = {1},
          doi = {10.3847/1538-4357/aafdb5},
archivePrefix = {arXiv},
       eprint = {1807.08890},
 primaryClass = {astro-ph.EP},
       adsurl = {https://ui.adsabs.harvard.edu/abs/2019ApJ...872....1R}
}

@ARTICLE{Parmentier+16,
       author = {{Parmentier}, Vivien and {Fortney}, Jonathan J. and {Showman}, Adam P. and {Morley}, Caroline and {Marley}, Mark S.},
        title = "{Transitions in the Cloud Composition of Hot Jupiters}",
      journal = {\apj},
         year = 2016,
        month = sep,
       volume = {828},
       number = {1},
          eid = {22},
        pages = {22},
          doi = {10.3847/0004-637X/828/1/22},
archivePrefix = {arXiv},
       eprint = {1602.03088},
 primaryClass = {astro-ph.EP},
       adsurl = {https://ui.adsabs.harvard.edu/abs/2016ApJ...828...22P}
}

@ARTICLE{Ormel&Min19,
       author = {{Ormel}, Chris W. and {Min}, Michiel},
        title = "{ARCiS framework for exoplanet atmospheres. The cloud transport model}",
      journal = {\aap},
         year = 2019,
        month = feb,
       volume = {622},
          eid = {A121},
        pages = {A121},
          doi = {10.1051/0004-6361/201833678},
archivePrefix = {arXiv},
       eprint = {1812.05053},
 primaryClass = {astro-ph.EP},
       adsurl = {https://ui.adsabs.harvard.edu/abs/2019A&A...622A.121O}
}

@ARTICLE{Ohno&Okuzumi18,
       author = {{Ohno}, Kazumasa and {Okuzumi}, Satoshi},
        title = "{Microphysical Modeling of Mineral Clouds in GJ1214 b and GJ436 b: Predicting Upper Limits on the Cloud-top Height}",
      journal = {\apj},
         year = 2018,
        month = may,
       volume = {859},
       number = {1},
          eid = {34},
        pages = {34},
          doi = {10.3847/1538-4357/aabee3},
archivePrefix = {arXiv},
       eprint = {1804.05708},
 primaryClass = {astro-ph.EP},
       adsurl = {https://ui.adsabs.harvard.edu/abs/2018ApJ...859...34O}
}

@ARTICLE{Gao&Benneke18,
       author = {{Gao}, Peter and {Benneke}, Bj{\"o}rn},
        title = "{Microphysics of KCl and ZnS Clouds on GJ 1214 b}",
      journal = {\apj},
         year = 2018,
        month = aug,
       volume = {863},
       number = {2},
          eid = {165},
        pages = {165},
          doi = {10.3847/1538-4357/aad461},
archivePrefix = {arXiv},
       eprint = {1807.04924},
 primaryClass = {astro-ph.EP},
       adsurl = {https://ui.adsabs.harvard.edu/abs/2018ApJ...863..165G}
}

@ARTICLE{Powell+18,
       author = {{Powell}, Diana and {Zhang}, Xi and {Gao}, Peter and {Parmentier}, Vivien},
        title = "{Formation of Silicate and Titanium Clouds on Hot Jupiters}",
      journal = {\apj},
         year = 2018,
        month = jun,
       volume = {860},
       number = {1},
          eid = {18},
        pages = {18},
          doi = {10.3847/1538-4357/aac215},
archivePrefix = {arXiv},
       eprint = {1805.01468},
 primaryClass = {astro-ph.EP},
       adsurl = {https://ui.adsabs.harvard.edu/abs/2018ApJ...860...18P}
}

@ARTICLE{Wakeford+17,
       author = {{Wakeford}, H.~R. and {Visscher}, C. and {Lewis}, N.~K. and {Kataria}, T. and {Marley}, M.~S. and {Fortney}, J.~J. and {Mandell}, A.~M.},
        title = "{High-temperature condensate clouds in super-hot Jupiter atmospheres}",
      journal = {\mnras},
         year = 2017,
        month = feb,
       volume = {464},
       number = {4},
        pages = {4247-4254},
          doi = {10.1093/mnras/stw2639},
archivePrefix = {arXiv},
       eprint = {1610.03325},
 primaryClass = {astro-ph.EP},
       adsurl = {https://ui.adsabs.harvard.edu/abs/2017MNRAS.464.4247W}
}

@ARTICLE{Mbarek&Kempton16,
       author = {{Mbarek}, Rostom and {Kempton}, Eliza M. -R.},
        title = "{Clouds in Super-Earth Atmospheres: Chemical Equilibrium Calculations}",
      journal = {\apj},
         year = 2016,
        month = aug,
       volume = {827},
       number = {2},
          eid = {121},
        pages = {121},
          doi = {10.3847/0004-637X/827/2/121},
archivePrefix = {arXiv},
       eprint = {1602.02759},
 primaryClass = {astro-ph.EP},
       adsurl = {https://ui.adsabs.harvard.edu/abs/2016ApJ...827..121M}
}

@ARTICLE{Burrows&Sharp99,
       author = {{Burrows}, Adam and {Sharp}, C.~M.},
        title = "{Chemical Equilibrium Abundances in Brown Dwarf and Extrasolar Giant Planet Atmospheres}",
      journal = {\apj},
         year = 1999,
        month = feb,
       volume = {512},
       number = {2},
        pages = {843-863},
          doi = {10.1086/306811},
archivePrefix = {arXiv},
       eprint = {astro-ph/9807055},
 primaryClass = {astro-ph},
       adsurl = {https://ui.adsabs.harvard.edu/abs/1999ApJ...512..843B}
}

@ARTICLE{Visscher+10,
       author = {{Visscher}, Channon and {Lodders}, Katharina and {Fegley}, Jr., Bruce},
        title = "{Atmospheric Chemistry in Giant Planets, Brown Dwarfs, and Low-mass Dwarf Stars. III. Iron, Magnesium, and Silicon}",
      journal = {\apj},
         year = 2010,
        month = jun,
       volume = {716},
       number = {2},
        pages = {1060-1075},
          doi = {10.1088/0004-637X/716/2/1060},
archivePrefix = {arXiv},
       eprint = {1001.3639},
 primaryClass = {astro-ph.EP},
       adsurl = {https://ui.adsabs.harvard.edu/abs/2010ApJ...716.1060V}
}

@ARTICLE{Morley+15,
       author = {{Morley}, Caroline V. and {Fortney}, Jonathan J. and {Marley}, Mark S. and {Zahnle}, Kevin and {Line}, Michael and {Kempton}, Eliza and {Lewis}, Nikole and {Cahoy}, Kerri},
        title = "{Thermal Emission and Reflected Light Spectra of Super Earths with Flat Transmission Spectra}",
      journal = {\apj},
         year = 2015,
        month = dec,
       volume = {815},
       number = {2},
          eid = {110},
        pages = {110},
          doi = {10.1088/0004-637X/815/2/110},
archivePrefix = {arXiv},
       eprint = {1511.01492},
 primaryClass = {astro-ph.EP},
       adsurl = {https://ui.adsabs.harvard.edu/abs/2015ApJ...815..110M}
}

@ARTICLE{Heng+12,
       author = {{Heng}, Kevin and {Hayek}, Wolfgang and {Pont}, Fr{\'e}d{\'e}ric and {Sing}, David K.},
        title = "{On the effects of clouds and hazes in the atmospheres of hot Jupiters: semi-analytical temperature-pressure profiles}",
      journal = {\mnras},
         year = 2012,
        month = feb,
       volume = {420},
       number = {1},
        pages = {20-36},
          doi = {10.1111/j.1365-2966.2011.19943.x},
archivePrefix = {arXiv},
       eprint = {1107.1390},
 primaryClass = {astro-ph.EP},
       adsurl = {https://ui.adsabs.harvard.edu/abs/2012MNRAS.420...20H}
}

@ARTICLE{Brande+24,
       author = {{Brande}, Jonathan and {Crossfield}, Ian J.~M. and {Kreidberg}, Laura and {Morley}, Caroline V. and {Barman}, Travis and {Benneke}, Bj{\"o}rn and {Christiansen}, Jessie L. and {Dragomir}, Diana and {Fortney}, Jonathan J. and {Greene}, Thomas P. and {Hardegree-Ullman}, Kevin K. and {Howard}, Andrew W. and {Knutson}, Heather A. and {Lothringer}, Joshua D. and {Mikal-Evans}, Thomas},
        title = "{Clouds and Clarity: Revisiting Atmospheric Feature Trends in Neptune-size Exoplanets}",
      journal = {\apjl},
         year = 2024,
        month = jan,
       volume = {961},
       number = {1},
          eid = {L23},
        pages = {L23},
          doi = {10.3847/2041-8213/ad1b5c},
archivePrefix = {arXiv},
       eprint = {2310.07714},
 primaryClass = {astro-ph.EP},
       adsurl = {https://ui.adsabs.harvard.edu/abs/2024ApJ...961L..23B}
}

@ARTICLE{2014Natur.505...69K,
       author = {{Kreidberg}, Laura and {Bean}, Jacob L. and {D{\'e}sert}, Jean-Michel and {Benneke}, Bj{\"o}rn and {Deming}, Drake and {Stevenson}, Kevin B. and {Seager}, Sara and {Berta-Thompson}, Zachory and {Seifahrt}, Andreas and {Homeier}, Derek},
        title = "{Clouds in the atmosphere of the super-Earth exoplanet GJ1214b}",
      journal = {\nat},
         year = 2014,
        month = jan,
       volume = {505},
       number = {7481},
        pages = {69-72},
          doi = {10.1038/nature12888},
archivePrefix = {arXiv},
       eprint = {1401.0022},
 primaryClass = {astro-ph.EP},
       adsurl = {https://ui.adsabs.harvard.edu/abs/2014Natur.505...69K}
}

@ARTICLE{2016Natur.529...59S,
       author = {{Sing}, David K. and {Fortney}, Jonathan J. and {Nikolov}, Nikolay and {Wakeford}, Hannah R. and {Kataria}, Tiffany and {Evans}, Thomas M. and {Aigrain}, Suzanne and {Ballester}, Gilda E. and {Burrows}, Adam S. and {Deming}, Drake and {D{\'e}sert}, Jean-Michel and {Gibson}, Neale P. and {Henry}, Gregory W. and {Huitson}, Catherine M. and {Knutson}, Heather A. and {Lecavelier Des Etangs}, Alain and {Pont}, Frederic and {Showman}, Adam P. and {Vidal-Madjar}, Alfred and {Williamson}, Michael H. and {Wilson}, Paul A.},
        title = "{A continuum from clear to cloudy hot-Jupiter exoplanets without primordial water depletion}",
      journal = {\nat},
         year = 2016,
        month = jan,
       volume = {529},
       number = {7584},
        pages = {59-62},
          doi = {10.1038/nature16068},
archivePrefix = {arXiv},
       eprint = {1512.04341},
 primaryClass = {astro-ph.EP},
       adsurl = {https://ui.adsabs.harvard.edu/abs/2016Natur.529...59S}
}

@ARTICLE{Morley12,
       author = {{Morley}, Caroline V. and {Fortney}, Jonathan J. and {Marley}, Mark S. and {Visscher}, Channon and {Saumon}, Didier and {Leggett}, S.~K.},
        title = "{Neglected Clouds in T and Y Dwarf Atmospheres}",
      journal = {\apj},
         year = 2012,
        month = sep,
       volume = {756},
       number = {2},
          eid = {172},
        pages = {172},
          doi = {10.1088/0004-637X/756/2/172},
archivePrefix = {arXiv},
       eprint = {1206.4313},
 primaryClass = {astro-ph.SR},
       adsurl = {https://ui.adsabs.harvard.edu/abs/2012ApJ...756..172M}
}

@ARTICLE{Jones22,
       author = {{Jones}, A.~P.},
        title = "{Nano-diamonds in proto-planetary discs. Life on the edge}",
      journal = {\aap},
         year = 2022,
        month = sep,
       volume = {665},
          eid = {A21},
        pages = {A21},
          doi = {10.1051/0004-6361/202142718},
archivePrefix = {arXiv},
       eprint = {2206.13474},
 primaryClass = {astro-ph.GA},
       adsurl = {https://ui.adsabs.harvard.edu/abs/2022A&A...665A..21J}
}

@ARTICLE{Gao+21,
       author = {{Gao}, Peter and {Wakeford}, Hannah R. and {Moran}, Sarah E. and {Parmentier}, Vivien},
        title = "{Aerosols in Exoplanet Atmospheres}",
      journal = {Journal of Geophysical Research (Planets)},
         year = 2021,
        month = apr,
       volume = {126},
       number = {4},
          eid = {e06655},
        pages = {e06655},
          doi = {10.1029/2020JE006655},
archivePrefix = {arXiv},
       eprint = {2102.03480},
 primaryClass = {astro-ph.EP},
       adsurl = {https://ui.adsabs.harvard.edu/abs/2021JGRE..12606655G}
}

@ARTICLE{Fu+25,
       author = {{Fu}, Guangwei and {Stevenson}, Kevin B. and {Sing}, David K. and {Mukherjee}, Sagnick and {Welbanks}, Luis and {Thorngren}, Daniel and {Tsai}, Shang-Min and {Gao}, Peter and {Lothringer}, Joshua and {Beatty}, Thomas G. and {Gapp}, Cyril and {Evans-Soma}, Thomas M. and {Allart}, Romain and {Pelletier}, Stefan and {Thao}, Pa Chia and {Mann}, Andrew W.},
        title = "{Statistical trends in JWST transiting exoplanet atmospheres}",
      journal = {arXiv e-prints},
         year = 2025,
        month = jan,
          eid = {arXiv:2501.02081},
        pages = {arXiv:2501.02081},
          doi = {10.48550/arXiv.2501.02081},
archivePrefix = {arXiv},
       eprint = {2501.02081},
 primaryClass = {astro-ph.EP},
       adsurl = {https://ui.adsabs.harvard.edu/abs/2025arXiv250102081F}
}

@ARTICLE{Louie+24,
       author = {{Louie}, Dana R. and {Mullens}, Elijah and {Alderson}, Lili and {Glidden}, Ana and {Lewis}, Nikole K. and {Wakeford}, Hannah R. and {Batalha}, Natasha E. and {Col{\'o}n}, Knicole D. and {Gressier}, Am{\'e}lie and {Long}, Douglas and {Radica}, Michael and {Espinoza}, N{\'e}stor and {Goyal}, Jayesh and {MacDonald}, Ryan J. and {May}, Erin M. and {Seager}, Sara and {Stevenson}, Kevin B. and {Valenti}, Jeff A. and {Allen}, Natalie H. and {Ca{\~n}as}, Caleb I. and {Challener}, Ryan C. and {Grant}, David and {Huang}, Jingcheng and {Lin}, Zifan and {Valentine}, Daniel and {Clampin}, Mark and {Perrin}, Marshall and {Pueyo}, Laurent and {van der Marel}, Roeland P. and {Mountain}, C. Matt},
        title = "{JWST-TST DREAMS: A Precise Water Abundance for Hot Jupiter WASP-17b from the NIRISS SOSS Transmission Spectrum}",
      journal = {\aj},
         year = 2025,
        month = feb,
       volume = {169},
       number = {2},
          eid = {86},
        pages = {86},
          doi = {10.3847/1538-3881/ad9688},
archivePrefix = {arXiv},
       eprint = {2412.03675},
 primaryClass = {astro-ph.EP},
       adsurl = {https://ui.adsabs.harvard.edu/abs/2025AJ....169...86L}
}

@ARTICLE{Bonomo+17,
       author = {{Bonomo}, A.~S. and {Desidera}, S. and {Benatti}, S. and {Borsa}, F. and {Crespi}, S. and {Damasso}, M. and {Lanza}, A.~F. and {Sozzetti}, A. and {Lodato}, G. and {Marzari}, F. and {Boccato}, C. and {Claudi}, R.~U. and {Cosentino}, R. and {Covino}, E. and {Gratton}, R. and {Maggio}, A. and {Micela}, G. and {Molinari}, E. and {Pagano}, I. and {Piotto}, G. and {Poretti}, E. and {Smareglia}, R. and {Affer}, L. and {Biazzo}, K. and {Bignamini}, A. and {Esposito}, M. and {Giacobbe}, P. and {H{\'e}brard}, G. and {Malavolta}, L. and {Maldonado}, J. and {Mancini}, L. and {Martinez Fiorenzano}, A. and {Masiero}, S. and {Nascimbeni}, V. and {Pedani}, M. and {Rainer}, M. and {Scandariato}, G.},
        title = "{The GAPS Programme with HARPS-N at TNG . XIV. Investigating giant planet migration history via improved eccentricity and mass determination for 231 transiting planets}",
      journal = {\aap},
         year = 2017,
        month = jun,
       volume = {602},
          eid = {A107},
        pages = {A107},
          doi = {10.1051/0004-6361/201629882},
archivePrefix = {arXiv},
       eprint = {1704.00373},
 primaryClass = {astro-ph.EP},
       adsurl = {https://ui.adsabs.harvard.edu/abs/2017A&A...602A.107B}
}

@ARTICLE{Kreidberg+14,
       author = {{Kreidberg}, Laura and {Bean}, Jacob L. and {D{\'e}sert}, Jean-Michel and {Line}, Michael R. and {Fortney}, Jonathan J. and {Madhusudhan}, Nikku and {Stevenson}, Kevin B. and {Showman}, Adam P. and {Charbonneau}, David and {McCullough}, Peter R. and {Seager}, Sara and {Burrows}, Adam and {Henry}, Gregory W. and {Williamson}, Michael and {Kataria}, Tiffany and {Homeier}, Derek},
        title = "{A Precise Water Abundance Measurement for the Hot Jupiter WASP-43b}",
      journal = {\apjl},
         year = 2014,
        month = oct,
       volume = {793},
       number = {2},
          eid = {L27},
        pages = {L27},
          doi = {10.1088/2041-8205/793/2/L27},
archivePrefix = {arXiv},
       eprint = {1410.2255},
 primaryClass = {astro-ph.EP},
       adsurl = {https://ui.adsabs.harvard.edu/abs/2014ApJ...793L..27K}
}

@ARTICLE{Feinstein+23,
       author = {{Feinstein}, Adina D. and {Radica}, Michael and {Welbanks}, Luis and {Murray}, Catriona Anne and {Ohno}, Kazumasa and {Coulombe}, Louis-Philippe and {Espinoza}, N{\'e}stor and {Bean}, Jacob L. and {Teske}, Johanna K. and {Benneke}, Bj{\"o}rn and {Line}, Michael R. and {Rustamkulov}, Zafar and {Saba}, Arianna and {Tsiaras}, Angelos and {Barstow}, Joanna K. and {Fortney}, Jonathan J. and {Gao}, Peter and {Knutson}, Heather A. and {MacDonald}, Ryan J. and {Mikal-Evans}, Thomas and {Rackham}, Benjamin V. and {Taylor}, Jake and {Parmentier}, Vivien and {Batalha}, Natalie M. and {Berta-Thompson}, Zachory K. and {Carter}, Aarynn L. and {Changeat}, Quentin and {dos Santos}, Leonardo A. and {Gibson}, Neale P. and {Goyal}, Jayesh M. and {Kreidberg}, Laura and {L{\'o}pez-Morales}, Mercedes and {Lothringer}, Joshua D. and {Miguel}, Yamila and {Molaverdikhani}, Karan and {Moran}, Sarah E. and {Morello}, Giuseppe and {Mukherjee}, Sagnick and {Sing}, David K. and {Stevenson}, Kevin B. and {Wakeford}, Hannah R. and {Ahrer}, Eva-Maria and {Alam}, Munazza K. and {Alderson}, Lili and {Allen}, Natalie H. and {Batalha}, Natasha E. and {Bell}, Taylor J. and {Blecic}, Jasmina and {Brande}, Jonathan and {Caceres}, Claudio and {Casewell}, S.~L. and {Chubb}, Katy L. and {Crossfield}, Ian J.~M. and {Crouzet}, Nicolas and {Cubillos}, Patricio E. and {Decin}, Leen and {D{\'e}sert}, Jean-Michel and {Harrington}, Joseph and {Heng}, Kevin and {Henning}, Thomas and {Iro}, Nicolas and {Kempton}, Eliza M. -R. and {Kendrew}, Sarah and {Kirk}, James and {Krick}, Jessica and {Lagage}, Pierre-Olivier and {Lendl}, Monika and {Mancini}, Luigi and {Mansfield}, Megan and {May}, E.~M. and {Mayne}, N.~J. and {Nikolov}, Nikolay K. and {Palle}, Enric and {Petit dit de la Roche}, Dominique J.~M. and {Piaulet}, Caroline and {Powell}, Diana and {Redfield}, Seth and {Rogers}, Laura K. and {Roman}, Michael T. and {Roy}, Pierre-Alexis and {Nixon}, Matthew C. and {Schlawin}, Everett and {Tan}, Xianyu and {Tremblin}, P. and {Turner}, Jake D. and {Venot}, Olivia and {Waalkes}, William C. and {Wheatley}, Peter J. and {Zhang}, Xi},
        title = "{Early Release Science of the exoplanet WASP-39b with JWST NIRISS}",
      journal = {\nat},
         year = 2023,
        month = feb,
       volume = {614},
       number = {7949},
        pages = {670-675},
          doi = {10.1038/s41586-022-05674-1},
archivePrefix = {arXiv},
       eprint = {2211.10493},
 primaryClass = {astro-ph.EP},
       adsurl = {https://ui.adsabs.harvard.edu/abs/2023Natur.614..670F}
}

@ARTICLE{Kempton+23,
       author = {{Kempton}, Eliza M. -R. and {Zhang}, Michael and {Bean}, Jacob L. and {Steinrueck}, Maria E. and {Piette}, Anjali A.~A. and {Parmentier}, Vivien and {Malsky}, Isaac and {Roman}, Michael T. and {Rauscher}, Emily and {Gao}, Peter and {Bell}, Taylor J. and {Xue}, Qiao and {Taylor}, Jake and {Savel}, Arjun B. and {Arnold}, Kenneth E. and {Nixon}, Matthew C. and {Stevenson}, Kevin B. and {Mansfield}, Megan and {Kendrew}, Sarah and {Zieba}, Sebastian and {Ducrot}, Elsa and {Dyrek}, Achr{\`e}ne and {Lagage}, Pierre-Olivier and {Stassun}, Keivan G. and {Henry}, Gregory W. and {Barman}, Travis and {Lupu}, Roxana and {Malik}, Matej and {Kataria}, Tiffany and {Ih}, Jegug and {Fu}, Guangwei and {Welbanks}, Luis and {McGill}, Peter},
        title = "{A reflective, metal-rich atmosphere for GJ 1214b from its JWST phase curve}",
      journal = {arXiv e-prints},
         year = 2023,
        month = may,
          eid = {arXiv:2305.06240},
        pages = {arXiv:2305.06240},
          doi = {10.48550/arXiv.2305.06240},
archivePrefix = {arXiv},
       eprint = {2305.06240},
 primaryClass = {astro-ph.EP},
       adsurl = {https://ui.adsabs.harvard.edu/abs/2023arXiv230506240K}
}

@ARTICLE{Ohno+20,
       author = {{Ohno}, Kazumasa and {Okuzumi}, Satoshi and {Tazaki}, Ryo},
        title = "{Clouds of Fluffy Aggregates: How They Form in Exoplanetary Atmospheres and Influence Transmission Spectra}",
      journal = {\apj},
         year = 2020,
        month = mar,
       volume = {891},
       number = {2},
          eid = {131},
        pages = {131},
          doi = {10.3847/1538-4357/ab44bd},
archivePrefix = {arXiv},
       eprint = {1908.02201},
 primaryClass = {astro-ph.EP},
       adsurl = {https://ui.adsabs.harvard.edu/abs/2020ApJ...891..131O}
}

\end{document}